\documentclass[11pt,a4paper]{article}
\usepackage{amsmath,amssymb,amsfonts,a4wide,graphicx,bm,times,bbm}
\usepackage{mcite}
\usepackage{comment}
\usepackage{tikz}
\usepackage{gensymb} 

\makeatletter \let\old@startsection=\@startsection
\renewcommand{\@startsection}[6]
{\old@startsection{#1}{#2}{#3}{#4}{#5}{#6\mathversion{bold}}}
\makeatother

\makeatletter

\@addtoreset{equation}{section}
\makeatother
\let\refOld\ref
\renewcommand{\ref}[1]{(\refOld{#1})}

 \DeclareMathOperator*{\tr}{tr} 
 \newcommand{\superp}[2]{\genfrac{}{}{0pt}{}{#1}{#2}}
 \newcommand{\superpsim}[2]{\genfrac{}{}{0pt}{2}{#1}{#2}}
 
 \def\Abox{\tikz[scale=0.007cm] \draw (0,0) rectangle (1,1);}
 \def\sAbox{\tikz[scale=0.004cm] \draw (0,0) rectangle (1,1);}
 \def\AAbox{\begin{tikzpicture}[scale=0.007cm] 
 \draw (0,0) rectangle (1,1);
 \draw (1,0) rectangle (2,1);
 \end{tikzpicture}}
 \def\sAAbox{\begin{tikzpicture}[scale=0.004cm] 
 \draw (0,0) rectangle (1,1);
 \draw (1,0) rectangle (2,1);
 \end{tikzpicture}}
 \def\Bbox{\begin{tikzpicture}[scale=0.007cm] 
 \draw (0,0) rectangle (1,1);
 \draw (0,1) rectangle (1,2);
 \end{tikzpicture}}
 \def\sBbox{\begin{tikzpicture}[scale=0.004cm] 
 \draw (0,0) rectangle (1,1);
 \draw (0,1) rectangle (1,2);
 \end{tikzpicture}}
 \def\Cbox{\begin{tikzpicture}[scale=0.007cm] 
 \draw (0,0) rectangle (1,1);
 \draw (0,1) rectangle (1,2);
 \draw (0,2) rectangle (1,3);
 \end{tikzpicture}}
 \def\sCbox{\begin{tikzpicture}[scale=0.004cm] 
 \draw (0,0) rectangle (1,1);
 \draw (0,1) rectangle (1,2);
 \draw (0,2) rectangle (1,3);
 \end{tikzpicture}}
 \def\d{\delta}

 \def\p{\partial}
 
 \def\a{\alpha}
 \def\b{\beta}
 \def\g{\gamma}
 \def\d{\delta}

 \def\k{\kappa}
 \def\l{\lambda}
 
 \def\n{\nu}

 \def\s{\sigma}

 \def\z{\zeta }
 
 \def\G{\Gamma}
 \def\D{\Delta}

 \def\L{\Lambda}

\def\CE{{\mathcal{E}}}

\def\CG{{\mathcal{G}}}

\def\CN{{\mathcal{N}}}

\def\CR{{\mathcal{R}}}
\def\CS{{\mathcal{S}}}
\def\CT{{\mathcal{T}}}

\def\CX{{\mathcal{X}}}
\def\CY{{\mathcal{Y}}}

\def\la{\left\langle}
\def\ra{\right\rangle}

\def\Op{\mathcal{O}}

\def\implies{\quad\Rightarrow\quad}
\def\equivalent{\quad\Longleftrightarrow\quad}

\def\vphi{\varphi}

\def\Zv{\mathcal{Z}_\text{vect.}}
\def\Zbf{\mathcal{Z}_\text{bfd.}}

\def\Zf{\mathcal{Z}_\text{fund.}}
\def\Zaf{\mathcal{Z}_\text{a.f.}}
\def\Zinst{\mathcal{Z}_\text{inst.}}
\def\ZCS{\mathcal{Z}_\text{CS}}

\def\res{\mathop{\text{Res}}}

\def\qf{\mathfrak{q}}

\def\pr{\mathrm{P}}
\def\tf{\tilde{f}}

\def\pf{p_\text{fund.}}
\def\paf{p_\text{a.f.}}

\def\vac{\emptyset}

\def\tV{\tilde{V}}
\def\hg{\hat\gamma}
\def\rag{\right\rangle_\text{gauge}}

\def\ZZ{$\mathbb{Z}^{\otimes 2}$}

\begin{document}
\begin{titlepage}
\renewcommand{\thefootnote}{\fnsymbol{footnote}}
\vspace*{-2cm}
\begin{flushright}
UT-17-12
\end{flushright}

\vspace*{1cm}

\begin{center}
{\huge {\bf   $(p,q)$-webs of DIM representations, 5d $\CN=1$ instanton partition functions and qq-characters}}

\vspace{10mm}
{\Large J.-E. Bourgine$^\dagger$ $^\ast$, 
M. Fukuda$^\diamond$, K. Harada$^\diamond$, Y. Matsuo$^\diamond$, R.-D. Zhu$^\diamond$}
\\[.6cm]
{\em {}$^\dagger$ Korea Institute for Advanced Study (KIAS)}\\
{\em Quantum Universe Center (QUC)}\\
{\em 85 Hoegiro, Dongdaemun-gu, Seoul, South Korea}\\
[.3cm]
{\em {}$^\ast$Sezione INFN di Bologna, Dipartimento di Fisica e Astronomia,
Universit\`a di Bologna} \\
{\em Via Irnerio 46, 40126 Bologna, Italy}
\\[.3cm]
{\em {}$^\diamond$ Department of Physics, The University of Tokyo}\\
{\em Bunkyo-ku, Tokyo, Japan}
\\[.4cm]
\texttt{bourgine\,@\,kias.re.kr,$\quad$ fukuda\,@\,hep-th.phys.s.u-tokyo.ac.jp,$\quad$ harada\,@\,hep-th.phys.s.u-tokyo.ac.jp, $\quad$  matsuo\,@\,phys.s.u-tokyo.ac.jp,$\quad$ nick\_zrd\,@\,hep-th.phys.s.u-tokyo.ac.jp}

\end{center}

\vspace{0.7cm}

\begin{abstract}
\noindent

Instanton partition functions of $\CN=1$ 5d Super Yang-Mills reduced on $S^1$ can be engineered in type IIB string theory from the $(p,q)$-branes web diagram. To this diagram is superimposed a web of representations of the Ding-Iohara-Miki (DIM) algebra that acts on the partition function. In this correspondence, each segment is associated to a representation, and the (topological string) vertex is identified with the intertwiner operator constructed by Awata, Feigin and Shiraishi. We define a new intertwiner acting on the representation spaces of levels $(1,n)\otimes(0,m)\to(1,n+m)$, thereby generalizing to higher rank $m$ the original construction. It allows us to use a folded version of the usual $(p,q)$-web diagram, bringing great simplifications to actual computations. As a result, the characterization of Gaiotto states and vertical intertwiners, previously obtained by some of the authors, is uplifted to operator relations acting in the Fock space of horizontal representations. We further develop a method to build qq-characters of linear quivers based on the horizontal action of DIM elements. While fundamental qq-characters can be built using the coproduct, higher ones require the introduction of a (quantum) Weyl reflection acting on tensor products of DIM generators.

\vspace{0.5cm}
\end{abstract}

\vfill

\end{titlepage}
\vfil\eject

\setcounter{footnote}{0}

\section{Introduction}
Duality has been one of the most fundamental issues in string/gauge theories, and it has been studied from many different viewpoints and in various contexts. One of the standard approaches to the problem is to assign a brane configuration to the gauge dynamics, and interpret the duality in graphical ways. Such an approach has been taken in $\CN=2$ super-Yang-Mills in 4 dimensions (and $\CN=1$ in 5d). The corresponding graphical object, the Seiberg-Witten curve, is expressed through various configurations of D- and NS-branes. For example, supersymmetric gauge theories with $\CN=1$ supercharges in five dimensions can be engineered in type IIB string theory using the methods developed in \cite{Hanany1996}. Linear quiver gauge theories with $U(m)$ gauge groups are obtained from webs of $(p,q)$-branes that are bound states of $p$ D5 branes and $q$ NS5 branes \cite{Aharony1997,Aharony1997a}.

A useful algebraic tool to analyze brane configurations is the topological vertex \cite{aganagic2005topological}. It has been introduced to reproduce the topological string amplitude on toric Calabi-Yau manifolds. In fact, the toric diagram of Calabi-Yau threefold can be identified with the $(p,q)$-branes web diagrams of type IIB string theory \cite{Leung1997}. From this identification, it is possible to build the instanton partition function of the gauge theory using the machinery of topological string theory. In order to recover the Nekrasov partition function \cite{Nekrasov2004} in a general Omega-background, it is necessary to refine the definition of the topological vertex to include the  gravi-photon background \cite{iqbal2009refined}. The importance of this representation of gauge theories in the context of the BPS/CFT correspondence was first realized in \cite{Morozov2015,Mironov2016a} where the connection with the decomposition of conformal blocks was also investigated.

In \cite{Bourgine2015c,Bourgine2016}, Nekrasov partition functions have been studied from a different perspective, namely through the representation of underlying quantum algebras: the \underline{s}pherical double affine \underline{H}ecke algebra with \underline{c}entral charges (SH$^c$) \cite{SHc} for the 4d gauge theory, and its quantum deformation, the Ding-Iohara-Miki (DIM) algebra \cite{Ding1997,Miki2007,Feigin2009} for the 5d gauge theory. The representations of these algebras coincide with those of (quantum) $W_N$-algebra \cite{feigin2011quantum,Feigin:2010qea,SHc,fukuda2015sh} while the basis of the representation coincides with the set of fixed points which represent the equivariant cohomology of the instanton moduli space. This observation was essential in the proof of the AGT conjecture elaborated in \cite{SHc}. In addition, the presence of these algebras reflects the integrable nature of the BPS sector of the gauge theory. It led to the construction of $\CR$ and $\CT$ matrices satisfying the standard $\CR\CT\CT$ relation of quantum integrable systems \cite{Maulik2012,Smirnov2013,Awata2016,Awata2016b}.

The main focus of our previous works \cite{Bourgine2015c,Bourgine2016} was the derivation of the qq-characters from SH$^c$/DIM algebras. These particular correlators of the gauge theory were first introduced in \cite{NPS}, and further studied in \cite{Nekrasov2015,Nekrasov2016,Nekrasov2016b,Nekrasov2016a,Kim2016}. They have the essential property to be polynomials, thus defining a resolvent for the matrix model representing the localized gauge partition function. These quantities generalize the q-characters of quantum groups defined in \cite{Knight1995,Frenkel1998}, and naturally associated to the $\CT$-operators of integrable systems.  In \cite{Bourgine2015c,Bourgine2016}, we have shown that the representation theoretical properties of the Gaiotto state and the intertwiner associated with bifundamental matter are directly translated into the regularity property of qq-characters. 

A different construction of qq-characters has been presented by Kimura and Pestun (KP) in \cite{Kimura2015} (see also \cite{Kimura2016a}). While both constructions are based on quantum W-algebras, the action of these algebras is seemingly different. In our approach, a copy of the DIM algebra is associated to each node of the quiver diagram, so that a quantum $W_m$ algebra is attached to each gauge group $U(m)$. On the other hand, the quiver W-algebras constructed in \cite{Kimura2015} is based on a Lie algebra whose Dynkin diagram coincides with the gauge theory quiver. In a sense, the two approaches are S-dual to each-other: in our approach the rank is the number of D-branes while it is the number of NS5-branes in KP's work. On the algebraic level, the S-duality is believed to be realized by Miki's automorphism \cite{Miki2007} that exchanges the labels $(l_1,l_2)$ of DIM representations, here identified with the $(p,q)$ charge of the branes.\footnote{Due to a different choice of conventions for DIM representations, the usual representation of $(p,q)$-branes webs is rotated by 90 degrees, with the NS charge $q$ in the horizontal direction and the Ramond (D-brane) charge $p$ in the vertical one. Thus, a brane with charge $(p,q)$ corresponds to a representation of label $(q,p)$.}

It was realized in \cite{Mironov2016,Awata2016a} that the two different pictures can be better understood using the refined topological vertex. Indeed, in \cite{Awata2011}, Awata, Feigin and Shiraishi (AFS) have reconstructed this object using the generators of the DIM algebra where it plays the role of an intertwiner between vertical $(0,m)$ (associated to $m$ D-branes) and horizontal $(1,n)$ (associated to a NS5-brane bound to $n$ D-branes) representations \cite{Awata2011}. Hence, like a string junction, it interpolates between the representations associated to different brane charges. In this way, different representations of DIM algebra can be combined to form a \textit{representation web} \cite{Mironov2016,Awata2016a} that can be identified with the $(p,q)$-web diagram engineering the gauge theory.\footnote{In \cite{Mironov2016,Awata2016a} this structure was called a \textit{network matrix model}. However, since we do not use the matrix model presentation of partition functions, we prefer not to employ this terminology here.} This presentation clarifies the two approaches for the construction of qq-characters: KP employed DIM generators in horizontal representations \cite{Feigin2009a} while we used similar generators but in vertical representations \cite{feigin2011quantum,Mironov2016}.

The purpose of this paper is to propose a unified method to build qq-characters and prove their regularity property. As suggested in \cite{Mironov2016}, the method is based on insertions of DIM operators in the horizontal representations. However, it also makes use of the commutation of vertical actions which was instrumental in our previous derivation \cite{Bourgine2016}. The link is made by a set of lemmas that intertwines horizontal and vertical actions on AFS intertwiners. The AFS lemmas can be regarded as an uplift in the horizontal representation space of the relation characterizing Gaiotto state and vertical intertwiners obtained in \cite{Bourgine2016}. With this new method, we derive all the (higher) qq-character of linear quiver theories with $U(m)$ gauge groups, thus largely extending our previous results that were restricted to fundamental qq-characters. In order to achieve this general treatment of quivers, several new insights were necessary. First, we generalized the AFS intertwiner to higher level $m$ of vertical representations, allowing us to treat gauge groups of arbitrary rank. Most of the previous considerations, including results on integrability, were restricted to gauge groups of rank one or two. Then, in order to build higher qq-characters, a Weyl reflection acting on (tensor products of) DIM generators has been introduced. Called \textit{quantum Weyl transformation}, it keeps the qq-character invariant.
In practice, it is used to build operators commuting with a $\CT$-operator, the vacuum expectation value (vev) of which reproduces the instanton partition function. This commutation property is directly related to the regularity property of the qq-character, thus providing another link with the manifestation of integrability in supersymmetric gauge theories.

This paper is organized as follows. The second section provides the main properties of DIM algebra and its vertical and horizontal representations. We put some emphasis on the various duality properties, including the SL(2,$\mathbb{Z}$) automorphisms. This section also includes a brief reminder on $\CN = 1$ 5d gauge theories.  The third section starts from the definition of the AFS intertwiners and proposes a generalization obtained from horizontal composition.  The generalized intertwiners simplify the computation of amplitudes associated to the brane-web. In sections 3.3 and 3.4, these intertwiners are used to reconstruct the Gaiotto state and the vertical intertwiner built in \cite{Bourgine2016}. In the fourth section, the horizontal intertwiner is defined by taking vertical contractions of generalized AFS vertices.  It is shown that it commutes with the co-product of DIM generators. The quantum Weyl transformation is defined in the section 4.2 as an operation on the tensor product of generators.　It leads to a systematic method to construct qq-character which is the main result of the paper. Finally, the details of computations can be found in the appendix, along with several useful identities.



\section{DIM algebra and representations}
\subsection{DIM algebra}
\begin{figure}
\begin{center}
\includegraphics[width=4cm]{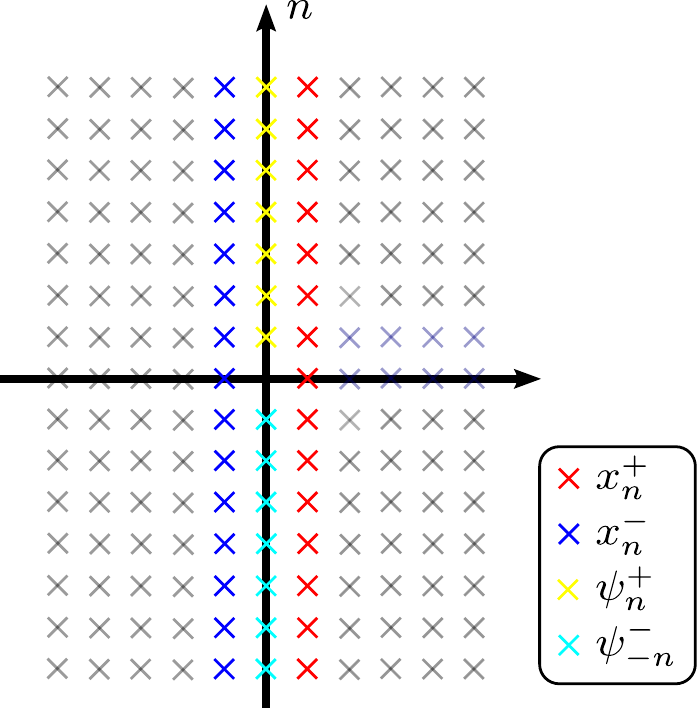}
\end{center}
\caption{DIM generators represented on a \ZZ  lattice where the $\CS$-transformation of the modular group SL(2,$\mathbb{Z}$) acts a 90 degrees rotation.}
\label{fig_DIM}
\end{figure}

The Ding-Iohara-Miki algebra $\CE$ \cite{Ding1997,Miki2007} can be presented in terms of the Drinfeld currents
\begin{equation}
x^\pm(z)=\sum_{k\in\mathbb{Z}}z^{-k}x^\pm_k,\quad \psi^+(z)=\sum_{k\geq0}z^{-k}\psi_k^+,\quad \psi^-(z)=\sum_{k\geq0}z^k\psi^-_{-k}.
\end{equation}
The modes $x_k^\pm$ and $\psi^\pm_{\pm k}$ are usually associated to points of a \ZZ-lattice representing the elements of the algebra (see figure \refOld{fig_DIM}). We assign \ZZ-degree for generators as $\mbox{deg}(x^\pm_n)=(\pm 1, n)$, $\mbox{deg}(\psi^\pm_n)=(0,n)$. The notations and conventions used here are mostly borrowed from \cite{Awata2016}, up to minor differences in the normalization of operators. The q-commutation relations satisfied by the currents read
\begin{align}
\begin{split}\label{def_DIM}
&[\psi^\pm(z),\psi^\pm(w)]=0,\quad \psi^+(z)\psi^-(w)=\dfrac{g(\hg w/z)}{g(\hg^{-1}w/z)}\psi^-(w)\psi^+(z)\\
&\psi^+(z)x^\pm(w)=g(\hg^{\mp1/2}w/z)^{\mp1}x^\pm(w)\psi^+(z),\quad \psi^-(z)x^\pm(w)=g(\hg^{\mp1/2}z/w)^{\pm1}x^\pm(w)\psi^-(z)\\
&x^\pm(z)x^\pm(w)=g(z/w)^{\pm1}x^\pm(w)x^\pm(z)\\
&[x^+(z),x^-(w)]=\dfrac{(1-q_1)(1-q_2)}{(1-q_1q_2)}\left(\d(\hg^{-1}z/w)\psi^+(\hg^{1/2}w)-\d(\hg z/w)\psi^-(\hg^{-1/2}w)\right),
\end{split}
\end{align}
with $\hg$ a central element. This algebra has two independent parameters encoded in the $q_\a$ with $\a=1,2,3$ under the relation $q_1q_2q_3=1$. It is sometimes more convenient to use instead the parameters $q=q_2$ and $t=q_1^{-1}$, in particular in the context of representations over Macdonald polynomials. We will also introduce the notation $\g=q_3^{1/2}=t^{1/2}q^{-1/2}$. These parameters appear through the functions
\begin{equation}
g(z)=\prod_{\a=1,2,3}\dfrac{1-q_\a z}{1-q_\a^{-1}z}=\dfrac{S(z)}{S(q_3z)},\quad S(z)=\dfrac{(1-q_1z)(1-q_2z)}{(1-z)(1-q_3^{-1}z)},
\end{equation}
that obey the unitarity properties $g(z)g(1/z)=1$ and $S(\g z)=S(\g/z)$ necessary to the consistency of the algebraic relations \ref{def_DIM}. Since these functions arise from the exchange of two operators, they are sometimes called \textit{scattering factors}.

In this paper, representations of level $(l_1,l_2)\in\mathbb{Z}\times\mathbb{Z}$ with a weight $u$ will be denoted $\rho_u^{(l_1,l_2)}$ or sometimes simply $(l_1,l_2)_u$. The levels are defined through the representations of the central element $\hg$ and the zero modes $\psi_0^\pm$,\begin{equation}
\rho_u^{(l_1,l_2)}(\hg)=q_3^{l_1/2},\quad \dfrac{\rho_u^{(l_1,l_2)}(\psi_0^-)}{\rho_u^{(l_1,l_2)}(\psi_0^+)}=q_3^{l_2}.
\end{equation}
Here, we will focus on the so-called vertical representations $(0,m)$ and horizontal representations $(1,n)$. The intertwiner defined in the next section relates three different representations, it will be portrayed as a three-legged vertex. Products and tensor products of such operators can be described using diagrams resembling the $(p,q)$-web diagrams of brane configurations in type IIB theory, they will be called \textit{representation webs}. Note however, that here diagrams are rotated by 90\degree, such that vertical lines $(0,m)$ are associated to $m$ D5 branes and the horizontal line $(1,0)$ to a NS5 brane. Note also that we will take no care of the precise slope of horizontal lines: vertical lines in the diagrams will always refer to a vertical representation $(0,m)$, while horizontal and inclined lines can represent any of the horizontal representations $(1,n)$. To avoid confusion, the representation space associated to each line will be explicitly written on every figure. 

The DIM algebra is a Hopf algebra with the following coproduct:
\begin{align}
\begin{split}\label{AFS_coproduct}
&\D(x^+(z))=x^+(z)\otimes 1+\psi^-(\hg_{(1)}^{1/2}z)\otimes x^+(\hg_{(1)}z)\\
&\D(x^-(z))=x^-(\hg_{(2)} z)\otimes \psi^+(\hg_{(2)}^{1/2}z)+1\otimes x^-(z)\\
&\D(\psi^+(z))=\psi^+(\hg_{(2)}^{1/2}z)\otimes\psi^+(\hg_{(1)}^{-1/2}z)\\
&\D(\psi^-(z))=\psi^-(\hg_{(2)}^{-1/2}z)\otimes\psi^-(\hg_{(1)}^{1/2}z)
\end{split}
\end{align}
with $\D(\hg)=\hg\otimes\hg$, $\hg_{(1)}=\hg\otimes1$ and $\hg_{(2)}=1\otimes\hg$.

\subsection{Quantum torus and DIM}
Before we start the detailed explanation of DIM representations, 
it is illuminating to mention the simplest representation with level $(0,0)$
which may be identified with the symmetry of a quantum mechanical system.
We consider the noncommutative torus generated by the two operators $U,V$
satisfying
\begin{eqnarray}
VU=q_1 UV.
\end{eqnarray}
where $q_1$ is not a root of unity. The enveloping algebra of $U,V$ is generated
by the elements $w_{rs}=U^r V^s$ identified as the degree $(r,s)$ generators defining the algebra
\begin{eqnarray}\label{winfty}
\left[w_{r_1s_1},w_{r_2s_2}\right]=(q_1^{s_1r_2}-q_1^{s_2 r_1})w_{r_1+s_1,r_2+s_2}\,.
\end{eqnarray}

This algebra has a $SL(2,\mathbb{Z})$ duality realized as 
the redefinition of the basis,
$U'=A U^a V^b$, $V'=B U^c V^d$ ($a,b,c,d\in \mathbb{Z}$) which
satisfies $V' U'=q_1 U' V'$ as long as $ad-bc=1$.\footnote{
	A different but similar duality structure is realized by writing $q_1=:e^{2\pi i \tau_1}$
	and defining a SL(2,$\mathbb{Z}$) modular transformation for $\tau_1$:
	$\tau_1'=\frac{a\tau_1+b}{c\tau_1+d}$. Then, there exists two
	generators $\tilde{U},\tilde{V}$ they satisfy the quantum torus algebra
	with $q_1'=e^{2\pi i \tau_1'}$ and that commute with the original generators
	$[U^n V^m, \tilde U^{r}\tilde V^s]=0$. This duality is known as the \textit{Morita equivalence}, it plays a fundamental role
	in non-commutative geometry \cite{rieffel1981c,rieffel1988projective}, and is also relevant in more recent works such as \cite{grassi2016topological}. }
In particular, the $\CS$-transformation is realized as $\mathcal{S}: (U,V)\rightarrow (V,U^{-1})$.

In this simple set-up, the DIM algebra with $(l_1,l_2)=(0,0)$ is realized as
\begin{eqnarray}
&&\rho^{(0,0)}_u(x^{+}(z))= \frac{1}{1-q_1} U\delta(V/z),\quad
\rho_u^{(0,0)}(x^-(z))= \sum_n x^-_n z^{-n}=\frac{-1}{1-q_1^{-1}}\delta(V/z)U^{-1},\nonumber\\
&&\rho_u^{(0,0)}(\psi^\pm(z)) =\frac{(1-(q_2V/z)^{\pm 1})(1-(q_3V/z)^{\pm 1})}{(1-(V/z)^{\pm1})(1-(q_1^{-1} Vt/z)^{\pm1})}
= \exp\left(\sum_{r=1}^\infty \frac{1}{r}(1-q_2^{\pm r})(1-q_3^{\pm r}) V^{\pm r} z^{\mp r}\right)
\end{eqnarray}
with $\delta(z)=\sum_{n\in\mathbb{Z}} z^n$. In this representation, the expression \ref{AFS_coproduct} of the coproduct simplifies as $\rho_u^{(0,0}(\hat\g)=1$. In the \textit{vector representation}, the generators $U,V$ are represented on a basis labeled by a single integer,
\begin{eqnarray}
\rho_u^{(0,0)}(U)|u,i\rangle = |u,i+1\rangle,\quad
\rho_u^{(0,0)}(V)|u,i\rangle = u q_1^i |u,i\rangle\,,
\end{eqnarray}
where $u$ is the weight of the representation.



It is of some interest to compare the DIM algebra with the loop algebra $\hat{\mathfrak{g}}$ of a Lie algebra $\mathfrak{g}$. The generators of $\hat{\mathfrak{g}}$ (without the central extension) are defined in terms of the generator $t_a$ of the Lie algebra as $J^a_n = t^a U^{n}$ where $U$ is a formal variable. DIM algebra is a natural generalization of this setting in which two formal variables are introduced. It sometimes referred to as a {\it two-loop symmetry}. The algebra \ref{winfty} depending on a single deformation parameter $q_1$ can be extended by two central charges, $c_1, c_2$ \cite{Miki2007},
\begin{eqnarray}
\left[w_{r_1s_1},w_{r_2s_2}\right]=(q_1^{s_1r_2}-q_1^{s_2 r_1})w_{r_1+s_1,r_2+s_2}+\delta_{r_1+r_2}\delta_{s_1+s_2}(r_1 c_1+s_1 c_2)q_1^{r_1 s_1}\,.
\end{eqnarray}
In this formulation, the SL(2,$\mathbb{Z}$) symmetry is manifest, and the $\CS$-transformation is realized on the central charges as $\mathcal{S}: c_1\rightarrow c_2, c_2\rightarrow -c_1$. The introduction of a second quantization parameter $q_2$ leads to the DIM algebra \cite{Miki2007}.  One may identify the levels $(l_1,l_2)$ with the representation of the two central charges, $\rho_u^{(l_1,l_2)}(c_1)=q_3^{l_2}, \rho_u^{(l_1,l_2)}(c_2)=q_3^{l_1}$.

\subsection{Vertical $(0,m)$ representation}
The vertical representation $(0,1)$ has been formulated in \cite{Feigin2009,feigin2011quantum}, it is equivalent to the rank $m$ representation studied in \cite{Bourgine2016} with $m=1$ and up to a normalization. Here we employ conventions similar to the ones used in \cite{Bourgine2016}, but with different states normalization (the change of the convention is summarized in appendix \refOld{AppA}).

The $(0,m)$ representations depend on a $m$-vector of weights $\vec v=(v_1,\cdots v_m)$ and act on a space spanned by states in one-to-one correspondence with $m$-tuple Young diagrams $\vec\l=(\l^{(1)},\cdots,\l^{(m)})$,
\begin{align}
\begin{split}\label{def_vert}
&\rho^{(0,m)}_{\vec v}(x^+(z))|\vec v,\vec\l\rangle\rangle=\g^{m-1}z^{-(m-1)}\sum_{x\in A(\vec\l)}\d(z/\chi_x)\res_{z=\chi_x}\dfrac1{z\CY_{\vec\l}(z)}|\vec v,\vec\l+x\rangle\rangle,\\
&\rho^{(0,m)}_{\vec v}(x^-(z))|\vec v,\vec\l\rangle\rangle=\g^{-2m+1}z^{m-1}\sum_{x\in R(\vec\l)}\d(z/\chi_x)\res_{z=\chi_x}z^{-1}\CY_{\vec\l}(zq_3^{-1})|\vec v,\vec\l-x\rangle\rangle,\\
&\rho^{(0,m)}_{\vec v}(\psi^\pm(z))|\vec v,\vec\l\rangle\rangle=\g^{-m}\left[\Psi_{\vec\l}(z)\right]_\pm|\vec v,\vec\l\rangle\rangle.
\end{split}
\end{align}
where $A(\vec\l)$ and $R(\vec\l)$ denote respectively the set of boxes that can be added to or removed from the set of the Young diagrams $\l_1,\cdots,\l_m$.
These expressions involve the functions $\Psi_{\vec\l}(z)$ and $\CY_{\vec \lambda}(z)$ that depend on a $m$-tuple Young diagram. Their expression can be factorized in terms of individual Young diagram contributions,
\begin{eqnarray}\label{def_CY}
\CY_{\vec\l}(z)&=&\prod_{l=1}^m \CY_{\l_l}(z),\quad
\Psi_{\vec\lambda}(z)
=\prod_{l=1}^m\Psi_{\lambda^{(l)}}(z),\\
\Psi_\l(z)&=&\dfrac{\CY_\l(zq_3^{-1})}{\CY_\l(z)},\quad
\CY_\l(z)=\left(1-\dfrac{v}{z}\right)\prod_{x\in\l}S(\chi_x/z)=\dfrac{\prod_{x\in A(\l)}z-\chi_x}{\prod_{x\in R(\l)}z-q_3^{-1}\chi_x},\\
\end{eqnarray}
Here, each box $x\in\vec\l$ is defined by three integer labels $(l,i,j)$ such that $(i,j)$ indicates the position of the box in $\l^{(l)}$. The associated box coordinate reads $\chi_x=v_lq_1^{i-1}q_2^{j-1}$. 

As in \cite{Bourgine2016}, it will be important to add a set of diagonal operators
$\CY^{\pm}(z)$ such that
\begin{eqnarray}
\CY^\pm(z) |\vec v,\vec\l\rangle\rangle=\left[\CY_{\vec\l}(z)\right]_\pm |\vec v,\vec\l\rangle\rangle\,.
\end{eqnarray}
The notation $[\cdots]_\pm$ refers to an expansion in powers of $z^{\mp1}$ of the quantity inside the brackets. 
They will be used to define the qq-character.

The action on the bra states will be referred to as the \textit{dual vertical representation}. In this representation, the roles of $x^+$ and $x^-$ are exchanged:
\begin{align}
\begin{split}\label{vert_repres_dual}
&\langle\langle\vec v,\vec\l|\hat\rho^{(0,m)}_{\vec v}(x^+(z))=-\g^{-1}\sum_{x\in R(\vec\l)}\d(z/\chi_x)\res_{z=\chi_x}z^{-1}\CY_{\vec\l}(zq_3^{-1})\langle\langle\vec v,\vec\l-x|,\\
&\langle\langle\vec v,\vec\l|\hat\rho^{(0,m)}_{\vec v}(x^-(z))=-\g^{-m+1}\sum_{x\in A(\vec\l)}\d(z/\chi_x)\res_{z=\chi_x}\dfrac1{z\CY_{\vec\l}(z)}\langle\langle\vec v,\vec\l+x|,\\
&\langle\langle\vec v,\vec\l|\hat\rho^{(0,m)}_{\vec v}(\psi^\pm(z))=\g^{-m}\left[\Psi_{\vec\l}(z)\right]_\pm\langle\langle\vec v,\vec\l|.
\end{split}
\end{align}
and $\hat\rho_v^{(0,m)}(\hg)=1$. Strictly speaking, this is not a representation of the DIM algebra because some of the q-commutation relations are no longer satisfied. Instead, it should be seen as a representation on the dual states $\langle\langle\vec v,\vec\l|$ that are orthogonal to the basis $|\vec v,\vec\l\rangle\rangle$,\footnote{Due to the change of states normalization performed in appendix \refOld{AppA}, and since the original states were orthonormal, the coefficient $a_{\vec\l}$ is expected to be proportional to $\CN(\vec\l)^{-2}$. The additional factors are chosen to simplify the formulation of the AFS lemmas below.}
\begin{equation}\label{scalar_vert}
\langle\langle\vec v,\vec\l|\vec v,\vec\l'\rangle\rangle=\d_{\vec\l,\vec\l'}\ a_{\vec\l}^{-1},
\end{equation} 
such that we have the property
\begin{equation}\label{dual_vert_repres}
\left(\langle\langle\vec v,\vec\l|\hat\rho_{\vec v}^{(m,0)}(e)\right)|\vec v,\vec\l'\rangle\rangle=\langle\langle\vec v,\vec\l|\left(\rho_{\vec v}^{(m,0)}(e)|\vec v,\vec\l'\rangle\rangle\right)
\end{equation} 
for any element $e$ of the DIM algebra. The norm of the states involves the coefficients $a_{\vec\l}$ which will play an important role in the construction of instanton partition functions. They are defined in terms of the vector multiplet contribution to the instanton partition function $\Zv(\vec v,\vec\l)$ as follows,
\begin{equation}\label{def_al}
a_{\vec\l}=\Zv(\vec v,\vec\l)\prod_{l=1}^m(-\g v_l)^{-|\vec\l|}\prod_{x\in\vec\l}\chi_x.
\end{equation} 
The vector contribution $\Zv(\vec v,\vec\l)$ will be defined in the section 2.5 below, it is expressed in terms of the Nekrasov factor \ref{Zbf} as a result of localization. As shown in \cite{Kanno:2013aha,Bourgine2016}, the Nekrasov factor obeys a set of \textit{discrete Ward identities}. Consequently, the coefficients $a_{\vec\l}$ also obey similar identities. They can be written in terms of the function $\CY_{\vec\l}(z)$ as
\begin{align}
\begin{split}\label{prop_al}
&\dfrac{a_{\vec\l+x}}{a_{\vec\l}}=\dfrac{1-q_1q_2}{(1-q_1)(1-q_2)}\g^m\chi_x^{-m}\res_{z=\chi_x}\dfrac1{\CY_{\vec\l}(z)\CY_{\vec\l}(zq_3^{-1})},\\
&\dfrac{a_{\vec\l-x}}{a_{\vec\l}}=-\dfrac{1-q_1q_2}{(1-q_1)(1-q_2)}\g^{-m}\chi_x^{m-2}\res_{z=\chi_x}\CY_{\vec\l}(z)\CY_{\vec\l}(zq_3^{-1}).
\end{split}
\end{align}

\subsection{Horizontal $(1,n)$ representations}
Horizontal representations \cite{Feigin2009a} of level $(1,n)$ act as a vertex operator algebra in the Fock space of $q$-bosonic modes with the commutation relations\footnote{Here we use parameters $q,t$
instead of $q_\alpha$ to follow the convention of \cite{Awata2011}: $q=q_2, t=q_1^{-1}$.
The oscillator modes can be represented on symmetric polynomials as follows:
\begin{equation}
\rho_\text{Macdonald}(a_{-k})=p_k,\quad \rho_\text{Macdonald}(a_k)=k\dfrac{1-q^k}{1-t^k}\dfrac{\p}{\p p_k},
\end{equation}
where $p_k$ denotes the power-sum symmetric polynomials.}
\begin{equation}\label{com_an}
[a_k,a_{-l}]=k\dfrac{1-q^k}{1-t^k}\d_{k,l},\quad k,l>0.
\end{equation}
The representations involve the positive/negative modes of the vertex operator
\begin{equation}\label{def_V}
V_\pm(z)=\exp\left(\mp\sum_{k=1}^\infty\dfrac{1-t^{\pm k}}{k}z^{\mp k}a_{\pm k}\right),
\end{equation}
and can be defined in terms of the following operators:
\begin{equation}
\eta(z)=V_-(z)V_+(z),\quad \xi(z)=V_-(\g z)^{-1}V_+(z/\g)^{-1},\quad \vphi^\pm(z)=V_\pm(\g^{\pm1/2}z)V_\pm(\g^{\mp3/2}z)^{-1}.
\end{equation}
Explicitly, we have
\begin{align}
\begin{split}
&\eta(z)=\exp\left(\sum_{k=1}^\infty\dfrac{1-t^{-k}}{k}z^ka_{-k}\right)\exp\left(-\sum_{k=1}^\infty\dfrac{1-t^k}{k}z^{-k}a_k\right),\\
&\xi(z)=\exp\left(-\sum_{k=1}^\infty\dfrac{1-t^{-k}}{k}\g^kz^ka_{-k}\right)\exp\left(\sum_{k=1}^\infty\dfrac{1-t^k}{k}\g^kz^{-k}a_k\right),\\
&\vphi^+(z)=\exp\left(-\sum_{k=1}^\infty\dfrac{1-t^k}{k}(1-\g^{2k})\g^{-k/2}z^{-k}a_k\right),\\
&\vphi^-(z)=\exp\left(\sum_{k=1}^\infty\dfrac{1-t^{-k}}{k}(1-\g^{2k})\g^{-k/2}z^ka_{-k}\right).
\end{split}
\end{align}
Useful commutation relations involving these operators are presented in appendix \refOld{AppB}. The horizontal representation $(1,n)_u$ reads
\begin{align}
\begin{split}
&\rho_u^{(1,n)}(x^+(z))=u\g^nz^{-n}\eta(z),\quad\rho_u^{(1,n)}(x^-(z))=u^{-1}\g^{-n}z^{n}\xi(z),\\
&\rho_u^{(1,n)}(\psi^+(z))=\g^{-n}\vphi^+(z),\quad\rho_u^{(1,n)}(\psi^-(z))=\g^n\vphi^-(z),
\end{split}
\end{align}
and $\rho_u^{(1,n)}(\hg)=\g$. Similarly, it is possible to define the representation $(-1,n)_u$ using the same vertex algebra,
\begin{align}
\begin{split}
&\rho_u^{(-1,n)}(x^+(z))=u^{-1}\g^nz^{n}\xi(z^{-1}),\quad\rho_u^{(-1,n)}(x^-(z))=u\g^{-n}z^{-n}\eta(z^{-1}),\\
&\rho_u^{(-1,n)}(\psi^+(z))=\g^{-n}\vphi^-(z^{-1}),\quad\rho_u^{(-1,n)}(\psi^-(z))=\g^n\vphi^+(z^{-1}),
\end{split}
\end{align}
with $\rho_u^{(-1,n)}(\hg)=\g^{-1}$.

By definition, the vacuum state $|\vac\rangle_{(1,n)_u}$ is annihilated by the positive modes $a_k$, and $\vphi^+(z)|\vac\rangle_{(1,n)_u}=|\vac\rangle_{(1,n)_u}$. The dual vacuum state $_{(1,n)_u\!}\langle\vac|$ is annihilated by negative modes, and $_{(1,n)_u\!}\langle\vac|\vphi^-(z)=\ _{(1,n)_u\!}\langle\vac|$. The normal ordering, denoted $:\cdots:$, corresponds to write all the positive modes on the right, and all the negative modes on the left. Correlators of operators $\Op_i(z_i)$ acting in the Fock space are defined as the vacuum expectation values 
\begin{equation}
\la\Op_1(z_1)\cdots\Op_N(z_N)\ra=\ _{(1,n)_u\!}\langle\vac|\Op_1(z_1)\cdots\Op_N(z_N)|\vac\rangle_{(1,n)_u},
\end{equation} 
with the radial ordering $|z_1|>|z_2|>\cdots>|z_N|$.

\subsection{Reminder on 5d $\CN=1$ instanton partition functions}
The quiver Super Yang-Mills gauge theories with $\CN=1$ in 5d reduced on $S^1$ are characterized by a simply laced Dynkin diagram $\G$. Each node $i\in\G$ is associated to a vector multiplet with gauge group $U(m_i)$, and an exponentiated gauge coupling $\qf_i$. To each edge $<ij>\in\G$ corresponds a bifundamental matter multiplet of mass $\mu_{ij}$ that transforms under the gauge group $U(m_i)\times U(m_j)$. In addition, a Chern-Simons term of level $\k_i$ coupled to the gauge group $U(m_i)$ can be introduced at each node $i$. Thus, each node bears two integer labels $(m_i,\k_i)$ with $m_i>0$ that will later be related to the levels $(l_1,l_2)$ of DIM representations. Extra matter fields in the fundamental/antifundamental representation of the gauge group $U(m_i)$ can also be attached to each node, and the corresponding masses will be denoted $\mu^{(\text{f})}_{i,j}$ with $j=1\cdots f_i$, and  $\mu^{(\text{af})}_{i,j}$ with $j=1\cdots \tilde{f}_i$ respectively.

The expression of the instanton contribution to the (K-theoretic) partition function reflects the particle content of the theory. It is written as a sum over $m_i$-tuple Young diagrams $\vec\l_i$, and each term is factorized into vector, Chern-Simons, (anti)fundamental and bifundamental contributions:
\begin{equation}\label{def_Zinst}
\Zinst[\G]=\sum_{\{\vec\l_i\}}\prod_{i\in\G}\qf_i^{|\vec\l_i|}\Zv(\vec v_i,\vec \l_i)\ZCS(\k_i,\vec \l_i)\Zf(\vec \mu^{(\text{f})}_i,\vec \l_i)\Zaf(\vec \mu^{(\text{af})}_i,\vec \l_i)\ \prod_{<ij>\in\G}\Zbf(\vec v_i\vec\l_i,\vec v_j,\vec\l_j|\mu_{ij}),
\end{equation} 
where the (exponentiated) Coulomb branch vevs $\vec v_i$ are related to the vacuum expectation value of the scalar field in the gauge multiplets. From this expression, it is possible to define a normalized trace of functions depending on the realization of the set of (tuple) Young diagrams $\{\vec\l_i\}$ as follows,
\begin{equation}\label{def_vev_gauge}\small
\la F[\{\vec\l_i\}]\rag\!\!\!\!\!\!=\dfrac1{\Zinst[\G]}\sum_{\{\vec\l_i\}}F[\{\vec\l_i\}]\prod_{i\in\G}\qf_i^{|\vec\l_i|}\Zv(\vec v_i,\vec \l_i)\ZCS(\k_i,\vec \l_i)\Zf(\vec \mu^{(\text{f})}_i,\vec \l_i)\Zaf(\vec \mu^{(\text{af})}_i,\vec \l_i)\prod_{<ij>\in\G}\Zbf(\vec v_i\vec\l_i,\vec v_j,\vec\l_j|\mu_{ij}).
\end{equation} 
This operation will be very useful in order to express the qq-characters of the gauge theory.

The bifundamental contribution with $U(m)\times U(m')$ gauge group can be decomposed as a product 
of \textit{Nekrasov factors},\footnote{The Nekrasov factors enjoy the property 
\begin{equation}
N(v_2,\l_2,v_1q_3^{-1},\l_1)=(-v_1)^{-|\l_2|}(-q_3v_2)^{|\l_1|}\prod_{x\in\l_1}\chi_x^{-1}\prod_{x\in\l_2}\chi_x\ N(v_1,\l_1,v_2,\l_2).
\end{equation}}
\begin{equation}\label{def_Zbf}
\Zbf(\vec v,\vec\l,\vec v',\vec\l'|\mu)=\prod_{l=1}^{m}\prod_{l'=1}^{m'}N(v_l,\l^{(l)},\mu v_{l'}',\l^{(l')\prime}).
\end{equation} 
Various expressions of the Nekrasov factors have been written, the one given here has been obtained by solving the discrete Ward identities derived in \cite{Kanno:2013aha,Bourgine2016},
\begin{equation}
N(v_1,\l_1,v_2,\l_2)=\prod_{\superp{x\in\l_1}{y\in\l_2}}S\left(\dfrac{\chi_x}{\chi_y}\right)\times\prod_{x\in\l_1}\left(1-\dfrac{\chi_x}{q_3 v_2}\right)\times\prod_{x\in\l_2}\left(1-\dfrac{v_1}{\chi_x}\right).
\end{equation} 
The vector multiplet contribution is expressed in terms of the Nekrasov factors as follows:
\begin{equation}
\label{Zbf}
\Zv(\vec v,\vec\l)=\prod_{l,l'=1}^m\dfrac1{N(v_l,\l^{(l)},v_{l'},\l^{(l')})}.
\end{equation}
Finally, the Chern-Simons and fundamental/antifundamental contributions are expressed in terms of a simple product over all boxes in the Young diagrams,
\begin{equation}\label{ZCS}
\ZCS(\k,\vec\l)=\prod_{x\in\vec\l}\left(\chi_x\right)^\k,\quad \Zf(\vec \mu^{(\text{f})}_i,\vec \l_i)=\prod_{x\in\vec\l}\prod_{j=1}^{f_i}\left(1-\chi_xq_3^{-1}(\mu^{(\text{f})}_{i,j})^{-1}\right),\quad 
\Zaf(\vec \mu^{(\text{af})}_i,\vec \l_i)=\prod_{x\in\vec\l}\prod_{j=1}^{\tilde{f}_i}(1-\mu^{(\text{af})}_{i,j}\chi_x^{-1}).
\end{equation}

The instanton partition functions defined in \ref{def_Zinst} are invariant under the rescaling $\vec v_i\to\a_i\vec v_i$, $\qf_i\to \a_i^{-\k_i}\qf_i$, $\vec\mu^{(\text{f})}_i\to\a_i\vec\mu^{(\text{f})}_i$, $\vec\mu^{(\text{af})}_i\to\a_i\vec\mu^{(\text{af})}_i$ and $\mu_{ij}\to (\a_i/\a_j)\mu_{ij}$. This invariance can be used to set the bifundamental masses to a specific value which simplifies the algebraic formulation developed here. Thus, from now on, all bifundamental masses will be set to $\mu_{ij}=\g^{-1}$.

These $\CN=1$ supersymmetric gauge theories can be engineered in type IIB string theory \cite{Hanany1996}. Linear quiver gauge theories with $U(m)$ gauge groups are obtained from webs of $(p,q)$-branes that are bound states of $p$ D5 branes and $q$ NS5 branes \cite{Aharony1997,Aharony1997a}. The branes occupy the dimensions 01234 corresponding to the space-time of the 5d gauge theory, plus an extra one-dimensional object (line) in the 56-planes. In order to preserve supersymmetry, the lines have the slope $\D x^6/\D x^5=p/q$, so that the world-volume of D5-branes with charge $(1,0)$ occupy the dimensions 012346, i.e. they are vertical segments in the 56-plane. On the other hand, NS5 branes of charge $(0,1)$ are extended in the 012345 directions, and correspond to an horizontal line in the 56-plane. A representation of DIM algebra has been associated to each brane of the $(p,q)$-web diagram \cite{Mironov2016,Awata2016a}. Representations of level $(l_1,l_2)$ correspond to $(l_2,l_1)$-branes so that horizontal $(1,0)$ representations are associated to NS5 branes and vertical $(0,1)$-representations to D5 branes. The topological vertex play the role of creation/annihilation operators for the $(p,q)$-branes, they will be identified with the generalized AFS intertwiners in the next section.

\subsection{Discrete symmetries of DIM algebra}\label{sec_DIM_sym}
In \cite{Miki2007}, Miki has introduced an automorphism of the DIM algebra that he denoted $\Psi$. Since it can be identified with the action of S-duality on the $(p,q)$-branes, it will be denoted by $\CS$ here. This automorphism leaves the DIM algebra invariant, but map degree $(r,l)$ generator into degree $(l,-r)$ and representations of different levels:
\begin{equation}
\CS\cdot(l_1,l_2)=(-l_2,l_1).
\end{equation}
Although the explicit transformation of the modes is rather complicated (it can be found in \cite{Miki2007}), the square of the automorphism takes a rather simple form: $\psi_n^+\leftrightarrow\psi_{-n}^-$, $x_n^\pm\leftrightarrow x_{-n}^\mp$ and $\hg\leftrightarrow\hg^{-1}$, or in terms of generating series, $\psi^+(z)\leftrightarrow\psi^-(1/z)$ and $x^\pm(z)\leftrightarrow x^\mp(1/z)$. The action of $\CS^2$ transforms horizontal representations $(1,n)_u$ into the representations $(-1,-n)_{u}$, so that $\rho_u^{(1,n)}(\CS^2\cdot e)=\rho_u^{(-1,-n)}(e)$.

By examination of the commutation relations, it is possible to define another transformation $\CT$ acting on the Drinfeld currents as
\begin{equation}
\CT\cdot x^\pm(z)=\left(\dfrac{\hg}{z}\right)^{\pm1}x^\pm(z),\quad \CT\cdot\psi^\pm(z)=\hg^{\mp1}\psi^\pm(z),
\end{equation}
or, in terms of modes,
\begin{equation}
\CT\cdot x^\pm_k(z)=\hg^{\pm1}x_{k\mp1}^\pm,\quad \CT\cdot\psi^\pm_k=\hg^{\mp1}\psi^\pm_k.
\end{equation}
Again, the DIM algebra is invariant, but representations of different levels are mapped to each-other,
\begin{equation}
\CT\cdot(l_1,l_2)=(l_1,l_1+l_2).
\end{equation}
Vertical representations are invariant under the action of $\CT$, and horizontal representations of level $n$ are mapped to horizontal representations of level $n+1$. The operations $\CS$ and $\CT$ obey the properties $\CS^4=1$ and $(\CS\CT)^3=1$, so that they generate a group of $SL(2,\mathbb{Z})$ transformations. To some extent, this group can be identified with the modular group of type IIB string theory. In particular, the Miki automorphism $\CS$ would correspond to the S-duality that rotates the $(p,q)$-web diagrams by 90\degree, exchanging NS5 and D5 branes. 

The second duality symmetry in DIM algebra is permutation of three parameters
$q_1,q_2, q_3$, which is manifest at the level of algebra.  This $S_3$ symmetry is sometimes referred to as a ``triality" \cite{fukuda2015sh} in connection with higher spin gravity \cite{Gaberdiel:2012ku}.  We note that the representations of DIM are constructed with the reference to $q_1,q_2$. In this sense, the exchange between $q_1$ and $q_2$ is manifest.  In 2D CFT language, such permutation is realized as $\beta\leftrightarrow 1/\beta$ where $\beta$ parametrizes the central charge $c=(n-1)(1-Q^2n(n+1))$ with $Q=\sqrt{\b}-\sqrt{\b}^{-1}$.   In terms of the vertical representation basis, it is realized by taking the transpose of each Young diagram, $\lambda\leftrightarrow \lambda'$.  The other transformations, such as $q_1\leftrightarrow q_3$, are less obvious.  When the parameters are suitably chosen, they are identified with the level-rank duality \cite{Altschuler:1990th,kuniba1991ferro,Gaberdiel:2012ku} where the correspondence between basis is also more involved \cite{fukuda2015sh}. While the SL(2,$\mathbb{Z}$) transformation may be regarded as a M-theoretical target space duality since it interchanges D-brane and NS-brane, the $S_3$ duality may be interpreted as a worldsheet symmetry since it acts on the Hilbert space of equivalent 2D conformal field theories.  From the viewpoint of super Yang-Mills, $q_1,q_2$ represent the graviphoton background in the Euclidean planes $(01)$ and $(23)$.  In this sense, we sometime denotes the symmetry $q_1\leftrightarrow q_2$ as $\s_{(01)(23)}$. On the other hand $q_1\leftrightarrow q_3$ does not have an immediate woldvolume interpretation. 

Another reflection symmetry $\s_5$ is obtained by replacing the parameters $q_\a$ by their inverse, effectively exchanging $S(z)$ with $S(q_3z)$, and $g(z)$ with $g(z)^{-1}=g(z^{-1})$.\footnote{Obviously, the two reflections $\s_{(01)(23)}$ and $\s_5$ commute. The composition $\s_{(01)(23)}\s_5$ acts on the DIM parameters as the exchange $q_1\leftrightarrow q_2^{-1}$, or $t\leftrightarrow q$. This is a well-known symmetry in the context of Macdonald polynomials, see for instance \cite{Macdonald1995}.} The transformation of DIM generators resemble the action of $\CS^2$, except that $x^+$ and $x^-$ are not exchanged: $\psi^\pm(z)\leftrightarrow\psi^\mp(1/z)$, $x^\pm(z)\leftrightarrow x^\pm(1/z)$, and the central parameter $\hg$ remains invariant. Thus, representations of level $(l_1,l_2)$ are mapped to representations of level $(-l_1,l_2)$ and vertical representations are left invariant.The $\s_5$ reflection of vertical $(0,m)$ representations follows from the transformation of the functions
\begin{equation}
\CY_{\vec\l}(z)\to z^{-m}\prod_{l=1}^m(-v_l)^{-1}\ \CY_{\vec\l}(z^{-1}),\quad \Psi_{\vec\l}(z)\to q_3^m\Psi_{\vec\l}(z^{-1}),
\end{equation} 
provided that the weights transform as $v_l\to1/v_l$ so that $\chi_x\to1/\chi_x$ for $x\in\vec\l$. On the other hand, $(1,n)_u$ representations are sent to $(-1,n)_u$ representations so that $\s_5\cdot\rho_u^{(1,n)}(e)=\rho_u^{(-1,n)}(\s_5\cdot e)$ where the transformation $\s_5$ sends the background parameters $q_\a\to q_\a^{-1}$ together with the modes $a_k\to t^k\g^{|k|}a_k$ and the weights $u\to \g^{2n}u^{-1}$. In this manner, $\eta(z)$ and $\xi(z)$ are exchanged but $\vphi^\pm(z)$ remain invariant.

The action of the $\s_5$ symmetry on instanton partition functions is closely related to the reflection symmetry studied in \cite{Bourgine2017,Bourgine2017a}, where it is interpreted as a quantum version of the exchange of the two sheets of the Seiberg-Witten curve. However, here the Coulomb branch vevs behave differently, since $v_l\to 1/v_l$. Vector and Chern-Simons contributions transform as
\begin{equation}
\Zv(\vec v,\vec \l)\to q_3^{-m|\vec\l|}\Zv(\vec v,\vec \l),\quad \ZCS(\k,\vec \l)\to \ZCS(-\k,\vec \l),
\end{equation} 
and the $A_1$ pure $U(m)$ partition function is invariant provided that the sign of the Chern-Simons level is flipped, and the extra $q_3$-factor is absorbed in the transformation $\qf\to q_3^{m}\qf$. On the other hand, the bifundamental contribution transforms into itself, but with the two nodes exchanged:
\begin{equation}
\Zbf(\vec v_1,\vec \l_1,\vec v_2,\vec \l_2|\mu)\to\Zbf(\vec v_2,\vec \l_2,\vec v_1,\vec \l_1|q_3^{-1}\mu')
\end{equation} 
where $\mu'$ is the reflection of $\mu$ ($\mu'=\g$ when $\mu=\g^{-1}$). As a result, the $\s_5$ symmetry for the instanton partition function of linear $A_r$ quiver consists in reflecting the order of the nodes $123\cdots r\to r(r-1)\cdots 1$. Hence, this $S_2$-symmetry can be interpreted as the reflection symmetry of the $(p,q)$-web diagram with respect to the horizontal ($x^5$) axis. In fact, the symmetry $\s_5$ also acts as a reflection along the horizontal axis in the graphical representation of the DIM modes $x_n^\pm$ and $\psi_n^\pm$ (see figure \refOld{fig_DIM}).

\section{Generalized AFS intertwiners}
\subsection{Definition of the generalized intertwiners}
\begin{figure}
\begin{center}
\begin{tikzpicture}
\node[below,scale=0.7] at (1,-1) {$(0,m)_{\vec v}$};
\node[above,scale=0.7] at (0,0) {$(1,n)_u$};
\node[above,scale=0.7] at (1.7,0.7) {$(1,n+m)_{u'}$};
\node[below right,scale=0.7] at (1,0) {$\Phi^{(n,m)}[u,\vec v]$};
\draw (0,0) -- (1,0) -- (1.7,0.7);
\draw (1,0) -- (1,-1);
\end{tikzpicture}
\hspace{1cm}
\begin{tikzpicture}
\node[below,scale=0.7] at (-0.7,-1.7) {$(1,n+m)_{u'}$};
\node[below,scale=0.7] at (1,-1) {$(1,n)_u$};
\node[above,scale=0.7] at (0,0) {$(0,m)_{\vec v}$};
\node[above left,scale=0.7] at (0,-1) {$\Phi^{(n,m)\ast}[u,\vec v]$};
\draw (0,0) -- (0,-1) -- (1,-1);
\draw (0,-1) -- (-0.7,-1.7);
\end{tikzpicture}
\hspace{1cm}
\begin{tikzpicture}
\tikzset{>=latex}
\draw[<->] (0,1) -- (0,0) -- (1,0);
\node[right,scale=0.7] at (1,0) {$x^5$ (NS5)};
\node[above,scale=0.7] at (0,1) {$x^6$ (D5)};
\end{tikzpicture}
\end{center}
\caption{$\Phi^{(n,m)}[u,\vec v]$ and $\Phi^{(n,m)\ast}[u,\vec v]$}
\label{fig_vertex}
\end{figure}
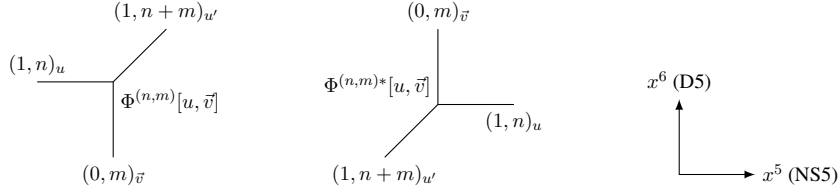

The AFS intertwiner operator has been introduced in \cite{Awata2011}, it generalizes the free fermion presentation \cite{Okounkov2003a} of the topological string vertex to the refined case. It is built over bosonic fields that coincide with those introduced in the horizontal representation of DIM algebra, thus providing a direct link with the representation theory. The original intertwiner acts in the tensor product of the representation spaces $(0,1)_v$ and $(1,n)_u$, and takes values in the space $(1,n+1)_{-uv}$. The vertical space $(0,1)_v$ is spanned by states in one to one correspondence with Young diagrams $\l$. Hence, the intertwiner is a vector in this space with index $\l$,
\begin{equation}
\Phi^{(n)}[u,v]=\left(\Phi_\l^{(n)}[u,v]\right)_\l,\quad \Phi_\l^{(n)}[u,v]:(1,n)_u\to(1,n+1)_{-uv}.
\end{equation}
Both horizontal spaces $(1,n)_u$ and $(1,n+1)_{-uv}$ can be identified with the same Fock space of $q$-bosonic modes, and the elements $\Phi_\l^{(n)}$ are expressed in terms of the modes as follows:
\begin{equation}
\Phi_\l^{(n)}[u,v]=t_n(\l,u,v)\ :\Phi_\vac(v)\prod_{x\in\l}\eta(\chi_x):.
\end{equation}
The vacuum component is defined in terms of a new vertex operator
\begin{equation}
\Phi_\vac(v)=\tV_-(v)\tV_+(v),\quad \tV_\pm(z)=\exp\left(\pm\sum_{k=1}^\infty\dfrac1k\dfrac{z^{\mp k}}{1-q^{\mp k}}a_{\pm k}\right),
\end{equation}
which is related to the previous operator $V_\pm(z)$ defined in \ref{def_V} as
\begin{equation}
V_\pm(z)=\tV_\pm(q_1z)\tV_\pm(q_2z)\tV_\pm(z)^{-1}\tV_\pm(q_3^{-1}z)^{-1}.
\end{equation}
In fact, the vacuum operator can be obtained as a (normal-ordered) product of $\eta(\chi_x)$ factors associated to an infinite Young diagram since
\begin{equation}
\tilde{V}_\pm(v)=\prod_{i,j=1}^\infty V_\pm(v q_1^{i-1}q_2^{j-1})^{-1}\implies \Phi_{\vac}(v)=:\prod_{i,j=1}^\infty\eta(vq_1^{i-1}q_2^{j-1})^{-1}:.
\end{equation} 
Thus, this operator is associated to the perturbative part of the partition function, extending the arguments developed in \cite{Bourgine2017} for the degenerate limit relevant to 4d gauge theories. Indeed, the prefactors obtained from the normal ordering of two vacuum intertwiner and involving the function $\CG(z)$ (defined in appendix \refOld{AppB}) should be interpreted as perturbative (one loop) contributions to the gauge theory partition function. However, to keep our arguments simple, we will simply neglect these factors and no longer refer to this interpretation here.

The normalization factor $t_n(\l,u,v)$ is the vev of the operator $\Phi_\l^{(n)}$, i.e. the correlator $\la\Phi_\l^{(n)}[u,v]\ra$. It is chosen in order to recover the exact form of the AFS relations presented below. Its explicit expression depends on the form of the vertical representation, which is slightly different than the original one employed by AFS,
\begin{equation}
t_n(\l,u,v)=(-uv)^{|\l|}\prod_{x\in\l}(\g/\chi_x)^{n+1}.
\end{equation} 
The reason for this different choice of normalization is that Awata, Feigin and Shiraishi were using the action on Macdonald polynomials to investigate the connection with the refined topological vertex \cite{Awata2008}. On the other hand, here we have chosen to keep a certain symmetry in the way the boxes of Young diagrams enter the formulas. It also makes the connection with our previous results on qq-characters more explicit \cite{Bourgine2016}.

\begin{figure}
\begin{center}
\begin{tikzpicture}
\node[below,scale=0.7] at (1,-1) {$(0,m)_{\vec v}$};
\node[above,scale=0.7] at (0,0) {$(1,n)_u$};
\node[above,scale=0.7] at (1.7,0.7) {$(1,n+m)_{u'}$};
\node[below right,scale=0.7] at (1,0) {$\Phi^{(n,m)}[u,\vec v]$};
\draw (0,0) -- (1,0) -- (1.7,0.7);
\draw (1,0) -- (1,-1);
\node[scale=0.7] at (3,-0.2) {$\equiv$};
\end{tikzpicture}
\begin{tikzpicture}
\node[below,scale=0.7] at (1,-1) {$(0,1)_{v_1}$};
\node[below,scale=0.7] at (3,-1) {$(0,1)_{v_2}$};
\node[below,scale=0.7] at (5,-1) {$(0,1)_{v_m}$};
\node[below,scale=0.7] at (0,0) {$(1,n)_{u_1}$};
\node[below,scale=0.7] at (2,0) {$(1,n+1)_{u_2}$};
\node[above,scale=0.7] at (5.7,0.7) {$(1,n+m)_{u'}$};
\node[above,scale=0.7] at (1,0) {$\Phi^{(n)}[u_1,v_1]$};
\node[above,scale=0.7] at (3,0) {$\Phi^{(n+1)}[u_2,v_2]$};
\node[right, below,scale=0.7] at (6.5,0) {$\Phi^{(n+m-1)}[u_m,v_m]$};
\draw (0,0) -- (3.5,0);
\draw[dotted] (3.5,0) -- (4.5,0);
\draw (4.5,0) -- (5,0) -- (5.7,0.7);
\draw (1,0) -- (1,-1);
\draw (3,0) -- (3,-1);
\draw (5,0) -- (5,-1);
\end{tikzpicture}
\end{center}
\caption{Construction of the generalized intertwiners $\Phi^{(n,m)}[u,\vec v]$}
\label{fig_Phi_nm}
\end{figure}
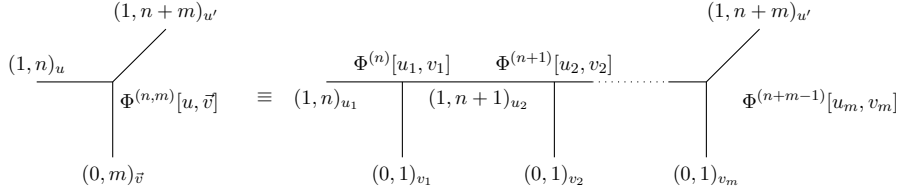

The AFS intertwiner can be generalized to vertical representations of higher level,
\begin{equation}
\Phi^{(n,m)}[u,\vec v]:(0,m)_{\vec v}\otimes(1,n)_u\to(1,n+m)_{u'},\qquad \text{with}\quad u'=u\prod_{l=1}^m(-v_l),
\end{equation} 
where the vector in the vertical space $(0,m)_{\vec v}$ has components labeled by the $m$-tuple $\vec\l$ that reads
\begin{equation}
\Phi_{\vec\l}^{(n,m)}[u,\vec v]=t_{n,m}(\vec\l,u,\vec v):\prod_{l=1}^m\Phi_\vac(v_l)\prod_{x\in\vec\l}\eta(\chi_x):,\quad t_{n,m}(\vec\l,u,\vec v)=(u')^{|\vec\l|}\prod_{x\in\vec\l}(\g/\chi_x)^{n+1}.
\end{equation}
This operator can be constructed as a product of vertical level one intertwiners coupled in the horizontal channel, as represented in the figure \refOld{fig_Phi_nm}. The contraction in the horizontal channel simply corresponds to a product of operators in the $q$-boson Fock space. However, it is only possible if the weights of the intermediate representation spaces coincide. The resulting product can be normal ordered, and commutations produce a bifundamental contribution,
\begin{equation}
\Phi_{\l_2}^{(n+1)}[u_2,v_2]\Phi_{\l_1}^{(n)}[u_1,v_1]=\dfrac{\CG(v_1/\g^2 v_2)}{N(v_1,\l_1,v_2,\l_2)} :\Phi_{\l_1}^{(n)}[u_1,v_1]\Phi_{\l_2}^{(n+1)}[u_2,v_2]:,
\end{equation} 
with the requirement $u_2=-u_1v_1$. The function $\CG(z)$ is defined in appendix, formula \ref{def_CG}. It only depends on the ratio $v_1/v_2$ and thus can be easily discarded. The (vertical) level $m$ intertwiner is obtained by repeating this operation $m$ times,
\begin{equation}\label{Phi_decomposed}
\Phi_{\l_m}^{(n+m-1)}[u_m,v_m]\cdots\Phi_{\l_2}^{(n+1)}[u_2,v_2]\Phi_{\l_1}^{(n)}[u_1,v_1]=\prod_{\superp{l,l'=1}{l<l'}}^m\dfrac{\CG(v_l/\g^2v_{l'})}{N(v_l,\l_l,v_{l'},\l_{l'})}\times\dfrac{\prod_{l=1}^mt_{n+l-1}(\l_l,u_l,v_l)}{t_{n,m}(\vec\l,u,\vec v)}\Phi_{\vec\l}^{(n,m)}[u,\vec v],
\end{equation} 
with for each intermediate horizontal space the weight
\begin{equation}\label{def_ul}
u_l=u\prod_{l'=1}^{l-1}(-v_{l'}).
\end{equation}
The extra factors in \ref{Phi_decomposed} and in \ref{Phi_ast_decomposed} below will be absorbed in the replacement of products of $a_{\l^{(l)}}$ by $a_{\vec\l}$ in the definition of the gauge theory operators (see the next section).

The AFS dual intertwiner can be generalized in the same way. It is defined as the operator\footnote{Note that we have exchanged the role of $u$ and $v$ with respect to the original definition so that $\vec v$ is always the weight in the vertical space.}
\begin{equation}
\Phi^{(n,m)\ast}[u,\vec v]:(1,n+m)_{u'}\to(1,n)_u\otimes(0,m)_{\vec v},\qquad \text{with}\quad u'=u\prod_{l=1}^m(-v_l),
\end{equation}
with vertical components
\begin{align}
\begin{split}
&\Phi_{\vec\l}^{(n,m)\ast}[u,\vec v]=t_{n,m}^\ast(\vec\l,u,\vec v):\prod_{l=1}^m\Phi_\vac^\ast(v_l)\prod_{x\in\vec\l}\xi(\chi_x):,\\
&\text{where:}\quad \Phi_\vac^\ast(v)=\tilde{V}_-(\g v)^{-1}\tilde{V}_+(\g^{-1}v)^{-1},\quad t_{n,m}^\ast(\vec\l,u,\vec v)=(\g u)^{-|\vec\l|}\prod_{x\in\vec\l}(\chi_x/\g)^{n}.
\end{split}
\end{align}
Again, it can be constructed from the original dual intertwiners of vertical level one as a product in the horizontal channel using the relation
\begin{equation}
\Phi_{\l_2}^{(n^\ast)\ast}[u_2,v_2]\Phi_{\l_1}^{(n^\ast+1)\ast}[u_1,v_1]=\CG(v_1/v_2)(-v_2)^{|\l_1|}(-q_3v_1)^{-|\l_2|}\dfrac{\prod_{x\in\l_1}\chi_x^{-1}\prod_{x\in\l_2}\chi_x}{N(v_2,\l_2,v_1,\l_1)} :\Phi_{\l_1}^{(n^\ast+1)\ast}[u_1,v_1]\Phi_{\l_2}^{(n^\ast)\ast}[u_2,v_2]:
\end{equation} 
Repeating the operation $m$ times reproduces the dual intertwiner of level $m$,
\begin{align}\small
\begin{split}\label{Phi_ast_decomposed}
&\Phi_{\l_m}^{(n)\ast}[u_m,v_m]\cdots\Phi_{\l_1}^{(n+m-1)\ast}[u_1,v_1]\\
&=\prod_{\superp{l,l'=1}{l>l'}}^m\dfrac{(-v_l)^{|\l_{l'}|}(-q_3v_{l'})^{-|\l_l|}\CG(v_{l'}/v_{l})}{N(v_l,\l_l,v_{l'},\l_{l'})}\times\prod_{l=1}^m\prod_{x\in\l_l}\chi_x^{-m+2l-1}\times\dfrac{\prod_{l=1}^mt^\ast_{n+m-l}(\l_l,u_l,v_l)}{t_{n,m}^\ast(\vec\l,u,\vec v)}\Phi_{\vec\l}^{(n,m)\ast}[u,\vec v],
\end{split}
\end{align}
with the intermediate horizontal weights $u_l$ such that $u_l=-v_{l+1}u_{l+1}$ and $u_m=u$:
\begin{equation}
u_l=u'\prod_{\superp{l,l'=1}{l'\leq l}}^{m}(-v_l)^{-1}.
\end{equation} 

\paragraph{Representation web} In order to form a particular web of representations relevant to a given gauge theory, the AFS interwiners can be coupled in two different ways: either along the horizontal $(1,n)$ or the vertical $(0,m)$ channels. These two contraction channels will be discussed below. To anticipate, an horizontal contraction corresponds to the operator product of the q-Heisenberg algebra describing the horizontal representation. On the other hand, the vertical contraction will be associated a scalar product that generates the trace of a tensor product. These contractions are represented by joining the vertex drawn in the figure \refOld{fig_vertex} along horizontal/vertical legs.\footnote{We remind the reader that we call ``horizontal'' every segment that is not vertical.} Taking only AFS intertwiners of rank $m=1$, we recover the $(p,q)$-web diagram giving the brane configuration engineering the gauge theory. From the NS5 brane perspective, the operators $\Phi$ and $\Phi^\ast$ are interpreted as creation/annihilation operators of D5 branes since they increase/decrease the horizontal level by one respectively. Branes charge conservation takes the form of representation levels conservation, with horizontal edges oriented from left to right, and vertical edges from bottom to top.

The introduction of intertwiners with higher rank allows us to simplify the diagram, effectively folding it $m$ times in the horizontal direction. In this way, it describes the creation/annihilation of $m$ coinciding D5 branes in one go. The resulting web diagram corresponds to the $(p,q)$-web diagram of the gauge theory where all gauge groups have been replaced by $U(1)$ groups. Although the information contained in this diagram become less visible, calculations are much more efficient in this setting.

\paragraph{Mass-deformed intertwiners} For simplicity, the parameters of the gauge theory have been rescaled in order to set all the bifundamental masses to $\g^{-1}$. It is however possible to keep arbitrary masses upon the introduction of mass-deformed intertwiners. This can be achieved by using the twisted operators $\xi_\mu(z)=V_-(z\mu^{-1})^{-1}V_+(zq_3^{-1}\mu^{-1})^{-1}$ and $\Phi_{\mu,\vac}^\ast(v)=\tilde{V}_-(z\mu^{-1})^{-1}\tilde{V}_+(zq_3^{-1}\mu^{-1})^{-1}$ to construct the dual intertwiner $\Phi^{(n,m)\ast}_{\mu,\vec\l}[u,\vec v]$. Then, the interwiner products of the form $\Phi_1\Phi_2^\ast$ reproduces the bifundamental contribution coupled to the nodes $1$ and $2$ with an arbitrary bifundamental mass $\mu$. Since such a deformation brings only little new insight to our discussion, in the following we will keep all the bifundamental masses set to $\g^{-1}$ to lighten the notations.

\subsection{AFS lemmas}
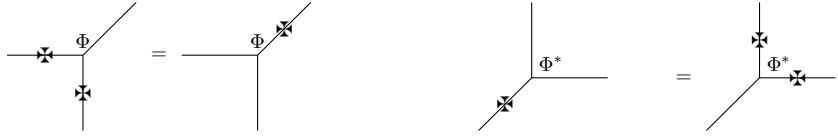
\begin{figure}
\begin{center}
\begin{tikzpicture}
\draw (0,0) -- (1,0) -- (1.7,0.7);
\draw (1,0) -- (1,-1);
\node[above,scale=0.7] at (1,0) {$\Phi$};
\node[scale=0.7] at (0.5,0) {$\maltese$};
\node[scale=0.7] at (1,-0.5) {$\maltese$};
\node[scale=0.7] at (2,0) {$=$};
\end{tikzpicture}
\begin{tikzpicture}
\draw (0,0) -- (1,0) -- (1.7,0.7);
\draw (1,0) -- (1,-1);
\node[above,scale=0.7] at (1,0) {$\Phi$};
\node[scale=0.7] at (1.35,0.35) {$\maltese$};
\end{tikzpicture}
\hspace{2cm}
\begin{tikzpicture}
\node[above right,scale=0.7] at (0,-1) {$\Phi^\ast$};
\draw (0,0) -- (0,-1) -- (1,-1);
\draw (0,-1) -- (-0.7,-1.7);
\node[scale=0.7] at (-0.35,-1.35) {$\maltese$};
\node[scale=0.7] at (2,-1) {$=$};
\end{tikzpicture}
\begin{tikzpicture}
\node[above right,scale=0.7] at (0,-1) {$\Phi^\ast$};
\draw (0,0) -- (0,-1) -- (1,-1);
\draw (0,-1) -- (-0.7,-1.7);
\node[scale=0.7] at (0,-0.5) {$\maltese$};
\node[scale=0.7] at (0.5,-1) {$\maltese$};
\end{tikzpicture}
\end{center}
\caption{Representation of the AFS lemma for the intertwiner and its dual, the symbol $\maltese$ denotes the insertion of an operator $e\in\CE$}
\label{fig_AFS}
\end{figure}

In \cite{Awata2011}, a series of commutation relations were derived, involving the intertwiners $\Phi$, $\Phi^\ast$ and the DIM generators in the appropriate representations. These relations can be extended to the generalized intertwiners defined here. They are expressed formally as
\begin{align}
\begin{split}\label{AFS}
&\rho_{u'}^{(1,n+m)}(e)\Phi^{(n,m)}[u,\vec v]=\Phi^{(n,m)}[u,\vec v]\cdot\left(\rho_{\vec v}^{(0,m)}\otimes\rho_u^{(1,n)}\right)\D(e),\\
&\Phi^{(n,m)\ast}[u,\vec v]\rho_{u'}^{(1,n+m)}(e)=\left(\rho_u^{(1,n)}\otimes\hat\rho_{\vec v}^{(0,m)}\right)\D(e)\cdot\Phi^{(n,m)\ast}[u,\vec v],
\end{split}
\end{align}
for any element $e\in\CE$ of the algebra. A proof is briefly sketched in appendix \refOld{AppC}. To be a little more explicit, the lemmas can be expressed as an action in the $q$-boson Fock space of the horizontal representation of DIM generators on the vertical components $\Phi_{\vec\l}$ of the intertwiner operators:
\begin{align}\small
\begin{split}\label{AFS_Phi}
&\psi^+(\g^{-1/2}z)\Phi_{\vec\l}^{(n,m)}[u,\vec v]-\g^{-m}\Psi_{\vec \l}(z)\ \Phi_{\vec\l}^{(n,m)}[u,\vec v]\psi^+(\g^{-1/2}z)=0\\
&\psi^-(\g^{1/2}z)\Phi_{\vec\l}^{(n,m)}[u,\vec v]-\g^{-m}\Psi_{\vec \l}(z)\ \Phi_{\vec\l}^{(n,m)}[u,\vec v]\psi^-(\g^{1/2}z)=0\\
&x^+(z)\Phi_{\vec\l}^{(n,m)}[u,\vec v]-\g^{-m}\Psi_{\vec\l}(z)\Phi_{\vec\l}^{(n,m)}[u,\vec v]x^+(z)=\left(\dfrac{\g}{z}\right)^{m-1}\sum_{x\in A(\vec\l)}\d(z/\chi_x)\res_{z=\chi_x}\dfrac1{z\CY_{\vec\l}(z)}\ \Phi_{\vec\l+x}^{(n,m)}[u,\vec v],\\
&x^-(\g^{-1}z)\Phi_{\vec\l}^{(n,m)}[u,\vec v]-\Phi_{\vec\l}^{(n,m)}[u,\vec v]x^-(\g^{-1}z)=z^{m-1}\g^{-2m+1}\!\!\sum_{x\in R(\vec\l)}\d(z/\chi_x)\res_{z=\chi_x}z^{-1}\CY_{\vec\l}(zq_3^{-1})\Phi_{\vec\l-x}^{(n,m)}[u,\vec v]\psi^+(\g^{-1/2}z).
\end{split}
\end{align}
In these relations, the representation in the horizontal space of the DIM operators has been omitted: for instance $x^\pm(z)$ reads $\rho_u^{(1,n)}(x^\pm(z))$ on the right of the intertwiner $\Phi_{\vec\l}$, and $\rho_{u'}^{(1,n+m)}(x^\pm(z))$ on the left. For the dual intertwiner, $x^\pm(z)$ is understood to be $\rho_u^{(1,n)}(x^\pm(z))$ on the left and $\rho_{u'}^{(1,n+m)}(x^\pm(z))$ on the right. Symmetric relations can be written for the dual intertwiner,
\begin{align}\small
\begin{split}\label{AFS_Phiast}
&\Phi_{\vec\l}^{(n,m)\ast}[u,\vec v]\psi^+(\g^{1/2}z)-\g^{-m}\Psi_{\vec\l}(z)\ \psi^+(\g^{1/2}z)\Phi_{\vec\l}^{(n,m)\ast}[u,\vec v]=0\\
&\Phi_{\vec\l}^{(n,m)\ast}[u,\vec v]\psi^-(\g^{-1/2}z)-\g^{-m}\Psi_{\vec\l}(z)\ \psi^-(\g^{-1/2}z)\Phi_{\vec\l}^{(n,m)\ast}[u,\vec v]=0,\\
&\Phi_{\vec\l}^{(n,m)\ast}[u,\vec v]x^+(\g^{-1}z)-x^+(\g^{-1}z)\Phi_{\vec\l}^{(n,m)\ast}[u,\vec v]=-\g^{-1}\sum_{x\in R(\vec\l)}\d(z/\chi_x)\res_{z=\chi_x}z^{-1}\CY_{\vec\l}(zq_3^{-1})\ \psi^-(\g^{-1/2}z)\Phi_{\vec\l-x}^{(n,m)\ast}[u,\vec v],\\
&\Phi_{\vec\l}^{(n,m)\ast}[u,\vec v]x^-(z)-\g^{-m}\Psi_{\vec\l}(z)x^-(z)\Phi_{\vec\l}^{(n,m)\ast}[u,\vec v]=-\g^{-m+1}\sum_{x\in A(\vec\l)}\d(z/\chi_x)\res_{z=\chi_x}\dfrac1{z\CY_{\vec\l}(z)}\ \Phi_{\vec\l+x}^{(n,m)\ast}[u,\vec v].
\end{split}
\end{align}
In fact, it is possible to re-write the RHS of the relations involving $x^\pm$ in a slightly more condensed way,
\begin{align}
\begin{split}\label{AFS_xpm}
&x^+(z)\Phi_{\vec\l}^{(n,m)}[u,\vec v]-\g^{-m}\Psi_{\vec\l}(z)\Phi_{\vec\l}^{(n,m)}[u,\vec v]x^+(z)=\rho_{\vec v}^{(0,m)}(x^+(z))\cdot\Phi_{\vec\l}^{(n,m)}[u,\vec v],\\
&x^-(\g^{-1}z)\Phi_{\vec\l}^{(n,m)}[u,\vec v]-\Phi_{\vec\l}^{(n,m)}[u,\vec v]x^-(\g^{-1}z)=\left[\rho_{\vec v}^{(0,m)}(x^-(z))\cdot\Phi_{\vec\l}^{(n,m)}[u,\vec v]\right]\psi^+(\g^{-1/2}z),\\
&\Phi_{\vec\l}^{(n,m)\ast}[u,\vec v]x^+(\g^{-1}z)-x^+(\g^{-1}z)\Phi_{\vec\l}^{(n,m)\ast}[u,\vec v]=\psi^-(\g^{-1/2}z)\left[\hat\rho_{\vec v}^{(0,m)}(x^+(z))\cdot\Phi_{\vec\l}^{(n,m)\ast}[u,\vec v]\right],\\
&\Phi_{\vec\l}^{(n,m)\ast}[u,\vec v]x^-(z)-\g^{-m}\Psi_{\vec\l}(z)x^-(z)\Phi_{\vec\l}^{(n,m)\ast}[u,\vec v]=\hat\rho_{\vec v}^{(0,m)}(x^-(z))\cdot\Phi_{\vec\l}^{(n,m)\ast}[u,\vec v],
\end{split}
\end{align}
where the dot denotes the action in the vertical space. The AFS lemmas are represented on the figure \refOld{fig_AFS}, in a very simplified manner. The insertion of the symbol $\maltese$ denotes the action of DIM generators $e\in\CE$, and two insertions the action of the coproduct.

\subsection{Gaiotto state}
The q-deformed Gaitto state is a Whittaker state for the q-Virasoro (or q-W) algebra \cite{Awata2009,Awata2010,Taki2014,Awata2011a,feigin2011}, it is a deformed version of the original Gaiotto state defined for the Virasoro algebra \cite{Gaiotto:2009ma,Marshakov:2009gn,Kanno:2011fw,Kanno:2012xt}. The intertwiners $\Phi$ and $\Phi^\ast$ defined in the previous section can be interpreted as an uplift of the Gaiotto state to the horizontal representation space. Indeed, the Gaiotto state can be recovered by taking the vacuum expectation value in the horizontal spaces,
\begin{align}
\begin{split}\label{def_Gaiotto}
&|G,\vec v\rangle\rangle=_{\ (1,n^\ast)_{u^\ast}\!\!\!\!}\langle\vac|\Phi^{(n^\ast,m)\ast}[u^\ast,\vec v]|\vac\rangle_{(1,n^\ast+m)_{u^{\ast\prime}}}=\sum_{\vec\l}a_{\vec\l}\ \la\Phi^{(n^\ast,m)\ast}_{\vec\l}[u^\ast,\vec v]\ra\ |\vec\l,\vec v\rangle\rangle,\\
&\langle\langle G,\vec v|=_{\ (1,n+m)_{u'}\!\!\!\!}\langle\vac|\Phi^{(n,m)}[u,\vec v]|\vac\rangle_{(1,n)_{u}}=\sum_{\vec\l}a_{\vec\l}\ \la\Phi^{(n,m)}_{\vec\l}[u,\vec v]\ra\langle\langle\vec\l,\vec v|,
\end{split}
\end{align}
where we have used the definitions
\begin{equation}\label{def_Phi_c}
\Phi^{(n^\ast,m)\ast}_{\vec\l}[u^\ast,\vec v]=\langle\langle\vec v,\vec\l|\Phi^{(n^\ast,m)\ast}[u^\ast,\vec v],\quad\Phi^{(n,m)}_{\vec\l}[u,\vec v]=\Phi^{(n,m)}[u,\vec v]|\vec v,\vec\l\rangle\rangle
\end{equation} 
and introduced the closure relation in the vertical space
\begin{equation}\label{closure}
\mathbbm{1}=\sum_{\vec\l} a_{\vec\l}\ |\vec v,\vec\l\rangle\rangle\langle\langle\vec v,\vec\l|.
\end{equation} 

The characterization of the Gaiotto state under the transformation $x^\pm(z)$ given in \cite{Bourgine2016} can be recovered by taking the vev of the AFS lemmas, evaluating the horizontal action of the generators using the formulas \ref{braiding_xpm} given in appendix. As an example, we consider the action of $x^+(z)$ on the Gaiotto state. Using the property \ref{dual_vert_repres}, the vertical action on the coordinate $\Phi_{\vec\l}^\ast$ becomes the dual action on the Gaiotto state, and the vev of the third relation in \ref{AFS_xpm} can be written in the form
\begin{equation}
\rho_{\vec v}^{(0,m)}(x^+(z))|G,\vec v\rangle\rangle = u^\ast\g^{n}z^{-n}\left(\CY^-(zq_3^{-1})-\CY^+(zq_3^{-1})\right)|G,\vec v\rangle\rangle.
\end{equation} 
where the operators $\CY^\pm(z)$ were defined as the diagonal operators in the basis $|\vec\l,\vec v\rangle\rangle$ with the eigenvalues $[\CY_{\vec\l}(z)]_\pm$ consisting of power series in $z^{\mp1}$. In order to compare with the reference \cite{Bourgine2016}, we need to change the normalization and define $e(z)=z^m\rho_{\vec v}^{(0,m)}(x^+(z))$. The operator $e_+(z)$ is the defined as the projection of $e(z)$ on the strictly negative powers of $z$. Assuming $n\leq 0$, the operator $z^{-n+m}\CY^-(zq_3^{-1})$ does not contribute to the projection as its eigenvalues generate only positive powers, and we recover
\begin{equation}
e_+(z)|G,\vec v\rangle\rangle = -u^\ast\g^{n}\pr_\infty^-\left[z^{-n+m}\CY^+(zq_3^{-1})\right]|G,\vec v\rangle\rangle.
\end{equation} 
where $\pr_\infty^-$ as been introduced in \cite{Bourgine2016} as the projection on the strictly negative powers of an expansion at $z\to\infty$. Similarly, $e_-(z)$ is defined as the projection of $e(z)$ on positive powers of $z$. Assuming $n>m$, this time the operator $z^{-n+m}\CY^+(zq_3^{-1})$ does not contribute, so that\footnote{When comparing with the results of reference \cite{Bourgine2016}, one should keep in mind the difference of notations: $\CY^+_\text{there}(z)=\CY^+_\text{here}(z)$ but $\CY^-_\text{there}(z)=\nu z^m\CY^-_\text{here}(zq_3^{-1})$.}
\begin{equation}
e_-(z)|G,\vec v\rangle\rangle = u^\ast\g^{n}\pr_0^+\left[z^{-n+m}\CY^-(zq_3^{-1})\right]|G,\vec v\rangle\rangle.
\end{equation} 
where $\pr_0^+$ denotes the projection on positive powers of an expansion around $z=0$.

Hence, up to minor differences in states normalization, we recover here the results previously obtained on the transformation of the Gaiotto state under the DIM symmetry generators. This identification provides an interpretation for the index $\k_R$ and $\k_L$ associated to the Gaiotto state (and its dual respectively). These two ad hoc parameters were introduced in \cite{Bourgine2016} in order to reproduce the Chern-Simons coupling $\k=\k_R-\k_L$ of the gauge theory, despite the lack of an interpretation for themselves. From our observation, it is clear that these parameters are identified with the horizontal level of the intertwiners, $\k_R=n^\ast$ and $\k_L=n$.

\subsection{Vertical intertwiner}
\begin{figure}
\begin{center}
\begin{tikzpicture}
\draw (0,0) -- (1,0) -- (2,1) -- (3,1);
\draw (1,0) -- (1,-1);
\draw (2,1) -- (2,2);
\node[below,scale=0.7] at (1,-1) {$(0,m_1)_{\vec v_1}$};
\node[left,scale=0.7] at (0,0) {$(1,n_1)_{u_1}$};
\node[below right,scale=0.7] at (1,0) {$\Phi^{(n_1,m_1)}[u_1,\vec v_1]$};
\node[right,scale=0.7] at (1.3,0.3) {$(1,n_1+m_1)_{u_1'}=(1,n_2+m_2)_{u'_2}$};
\node[left,scale=0.7] at (2,1) {$\Phi^{(n_2,m_2)\ast}[u_2,\vec v_2]$};
\node[right,scale=0.7] at (3,1) {$(1,n_2)_{u_2}$};
\node[above,scale=0.7] at (2,2) {$(0,m_2)_{\vec v_2}$};
\end{tikzpicture}
\hspace{1cm}
\begin{tikzpicture}
\draw (-0.7,-0.7) -- (0,0) -- (2,0) -- (2.7,0.7);
\draw (0,0) -- (0,1);
\draw (2,0) -- (2,-1);
\node[below,scale=0.7] at (2,-1) {$(0,m_1)_{\vec v_1}$};
\node[above,scale=0.7] at (1,0) {$(1,n)_{u}$};
\node[below right,scale=0.7] at (2,0) {$\Phi^{(n,m_1)}[u,\vec v_1]$};
\node[above right,scale=0.7] at (2.7,0.7) {$(1,n+m_1)_{u_1'}$};
\node[left,scale=0.7] at (0,0) {$\Phi^{(n,m_2)\ast}[u,\vec v_2]$};
\node[below left,scale=0.7] at (-0.7,-0.7) {$(1,n+m_2)_{u'_2}$};
\node[above,scale=0.7] at (0,1) {$(0,m_2)_{\vec v_2}$};
\end{tikzpicture}
\end{center}
\caption{$\Pi^{(n_1,n_2,m_1,m_2)}[u_1,\vec v_1,u_2,\vec v_2]$ and $\Pi^{(n,m_1,m_2)\ast}[u_1,\vec v_1,\vec v_2]$}
\label{fig_Pi}
\end{figure}
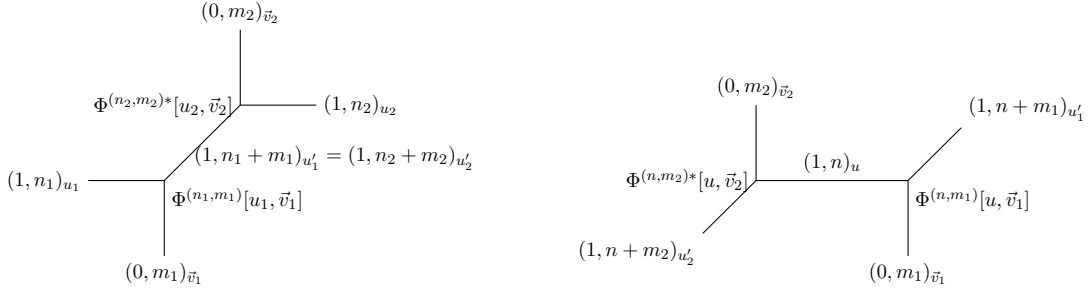

The contraction of the intertwiner $\Phi_{\vec\l_1}^{(n_1,m_1)}[u_1,\vec v_1]$ with the dual intertwiner $\Phi_{\vec\l_2}^{(n_2,m_2)\ast}[u_2,\vec v_2]$ in the horizontal channel is realized as a product in the $q$-boson space. This product is possible only if the two representations in the intermediate space coincide, e.g. if $u_1=u_2$ and $n_1=n_2$ for the product $\Phi_1\Phi_2^\ast$. In the case of the opposite product $\Phi_2^\ast\Phi_1$, the condition is $n_1+m_1=n_2+m_2$ and $u'_1=u'_2$. Both are represented on the figure \refOld{fig_Pi}. Using the q-commutation relations from appendix \refOld{AppB}, the two products can be normal-ordered and reproduce the bifundamental contribution \ref{def_Zbf}:\footnote{A priori, this result allows us to write the following formal relation between $\Pi$ and $\Pi^\ast$,
\begin{equation}
\Pi_{\vec\l_1,\vec\l_2}^{(n_1,n_2,m_1,m_2)}[u_1,\vec v_1,u_2,\vec v_2]=\left(\dfrac{\g^{m_1}u_1'}{u_1}\right)^{|\vec\l_2|}\left(\dfrac{\g^{m_2}u_2'}{u_2}\right)^{-|\vec\l_1|}\prod_{x\in\vec\l_1}\chi_x^{m_2}\prod_{x\in\vec\l_2}\chi_x^{-m_1}\dfrac{\prod_{l=1}^{m_1}\prod_{l'=1}^{m_2}\CG(v_{l'}^{(2)}/\g v_l^{(1)})}{\prod_{l=1}^{m_1}\prod_{l'=1}^{m_2}\CG(v_l^{(1)}/\g v_{l'}^{(2)})}\Pi_{\vec\l_1,\vec\l_2}^{(n,m_1,m_2)\ast}[u,\vec v_1,\vec v_2],
\end{equation}
that could be interpreted as the q-commutation of creation and annihilation operators for D5 branes. However, this relation does not make sense in the representation formalism of DIM algebra due to the mismatch of the representation spaces. It only makes sense if all the horizontal representation spaces are identified as the Fock space for the free bosonic modes. We will not use this relation further in this paper.}
\begin{align}
\begin{split}
&\Pi_{\vec\l_1,\vec\l_2}^{(n_1,n_2,m_1,m_2)}[u_1,\vec v_1,u_2,\vec v_2]=\dfrac{\Zbf(\vec v_1,\vec \l_1,\vec v_2,\vec \l_2|\g^{-1})}{\prod_{l=1}^{m_1}\prod_{l'=1}^{m_2}\CG(v_l^{(1)}/\g v_{l'}^{(2)})}:\Phi_{\vec\l_2}^{(n_2,m_2)\ast}[u_2,\vec v_2]\Phi_{\vec\l_1}^{(n_1,m_1)}[u_1,\vec v_1]:\\
&\Pi_{\vec\l_1,\vec\l_2}^{(n,m_1,m_2)\ast}[u,\vec v_1,\vec v_2]=\dfrac{\Zbf(\vec v_2,\vec \l_2,\vec v_1,\vec \l_1|\g^{-1})}{\prod_{l=1}^{m_1}\prod_{l'=1}^{m_2}\CG(v_{l'}^{(2)}/\g v_l^{(1)})}:\Phi_{\vec\l_2}^{(n,m_2)\ast}[u,\vec v_2]\Phi_{\vec\l_1}^{(n,m_1)}[u,\vec v_1]:,
\end{split}
\end{align}
where we have introduced the notations
\begin{align}
\begin{split}
&\Pi_{\vec\l_1,\vec\l_2}^{(n_1,n_2,m_1,m_2)}[u_1,\vec v_1,u_2,\vec v_2]=\Phi_{\vec\l_2}^{(n_2,m_2)\ast}[u_2,\vec v_2]\Phi_{\vec\l_1}^{(n_1,m_1)}[u_1,\vec v_1]:(1,n_1)_{u_1}\to(1,n_2)_{u_2},\\
&\Pi_{\vec\l_1,\vec\l_2}^{(n,m_1,m_2)\ast}[u,\vec v_1,\vec v_2]=\Phi_{\vec\l_1}^{(n,m_1)}[u,\vec v_1]\Phi_{\vec\l_2}^{(n,m_2)\ast}[u,\vec v_2]:(1,n+m_2)_{u'_2}\to(1,n+m_1)_{u'_1}.
\end{split}
\end{align}
In fact, the quantity $\Pi_{\vec\l_1,\vec\l_2}^{(n_1,n_2,m_1,m_2)}[u_1,\vec v_1,u_2,\vec v_2]$ with $n_\a=m_\a=1$ has already been studied in \cite{Awata2016} where it has been interpreted as a T-matrix, and it was further shown that it obeys RTT-relations with the R-matrix canonically associated to the coproduct $\D$ defined in \ref{AFS_coproduct}. In this paper, we will not refer to this interpretation anymore, although we expect it to hold beyond the restricted case studied in \cite{Awata2016}.

Due to the associativity of the product in the q-boson Fock space, the horizontal channel contraction satisfies $(\Phi_\l e)\Phi_\l^\ast=\Phi_\l(e\Phi_\l^{\ast})$ for any element $e$ in the appropriate representation. This property is depicted on the figure \refOld{fig_invar} (right). When combined with the AFS lemmas, it produces some important q-commutation relations:
\begin{align}\small
\begin{split}\label{AFS_Pi}
&x^+(z)\Pi_{\vec\l_1,\vec\l_2}-\g^{-m_1}\Psi_{\vec\l_1}(z)\Pi_{\vec\l_1,\vec\l_2}x^+(z)=\ \Phi_{\vec\l_2}^\ast\left(\rho_{\vec v_1}^{(0,m_1)}(x^+(z))\cdot\Phi_{\vec\l_1}\right)-\psi^-(\g^{1/2}z)\left(\hat\rho_{\vec v_2}^{(0,m_2)}(x^+(\g z))\cdot\Phi_{\vec\l_2}^\ast\right)\Phi_{\vec\l_1}\\
&\g^{-m_2}\Psi_{\vec\l_2}(z)x^-(z)\Pi_{\vec\l_1,\vec\l_2}-\Pi_{\vec\l_1,\vec\l_2}x^-(z)=\ \Phi_{\vec\l_2}^\ast\left(\rho_{\vec v_1}^{(0,m_1)}(x^-(\g z))\cdot\Phi_{\vec\l_1}\right)\psi^+(\g^{1/2}z)-\left(\hat\rho_{\vec v_2}^{(0,m_2)}(x^-(z))\cdot\Phi_{\vec\l_2}^\ast\right)\Phi_{\vec\l_1}\\
&\g^{-m_2}\Psi_{\vec\l_2}(z)x^-(z)\Pi_{\vec\l_1,\vec\l_2}^\ast-\Pi_{\vec\l_1,\vec\l_2}^\ast x^-(z)=\ \left(\rho_{\vec v_1}^{(0,m_1)}(x^-(\g z))\cdot\Phi_{\vec\l_1}\right)\Phi_{\vec\l_2}^\ast\psi^+(\g^{1/2}z)-\Phi_{\vec\l_1}\left(\hat\rho_{\vec v_2}^{(0,m_2)}(x^-(z))\cdot\Phi_{\vec\l_2}^\ast\right),\\
&x^+(z)\Pi_{\vec\l_1,\vec\l_2}^\ast-\g^{-m_1}\Psi_{\vec\l_1}(z)\Pi_{\vec\l_1,\vec\l_2}^\ast x^+(z)=\ \left(\rho_{\vec v_1}^{(0,m_1)}(x^+(z))\cdot\Phi_{\vec\l_1}\right)\Phi_{\vec\l_2}^\ast-\psi^-(\g^{1/2}z)\Phi_{\vec\l_1}\left(\hat\rho_{\vec v_2}^{(0,m_2)}(x^+(\g z))\cdot\Phi_{\vec\l_2}^\ast\right),\\
\end{split}
\end{align}
where arguments $n_\a$, $m_\a$, $u_\a$, and $\vec v_\a$ have been omitted for a better readability.

In \cite{carlsson2008exts}, Carlsson and Okounkov have introduced an operator intertwining between two vertical $(0,1)$ representations in order to describe the matter multiplet in the bifundamental representation of $U(1)\times U(1)$. This operator was later generalized to $U(m_1)\times U(m_2)$ bifundamental representations in \cite{Bourgine2016}, and related to the Gaiotto state in the formal limit $m_2\to0$. It is defined as an intertwiner between $(0,m_1)$ and $(0,m_2)$ vertical representation spaces, obtained as the horizontal vev of the operator $\Pi$,
\begin{equation}\label{def_V12}
V_{12}^{(m_1,m_2)} = \prod_{l=1}^{m_1}\prod_{l'=1}^{m_2}\CG(v_{l'}^{(2)}/\g v_{l}^{(1)})\ \la\Phi^{(n_1,m_1)\ast}[u_1,\vec v_1]\Phi^{(n_2,m_2)}[u_2,\vec v_2]\ra.
\end{equation} 
This operator $V_{12}^{(m_1,m_2)}$ will be called the \textit{vertical interwiner} to distinguish it from the AFS intertwiners. In Toda field theory, $V_{12}^{(m,m)}$ is identified with the degenerate 
primary field \cite{Kanno:2013aha}  multiplied by the Carlsson-Okounkov $U(1)$ factor.
By definition, its matrix elements reproduce the bifundamental contribution,
\begin{align}
\begin{split}
\langle\langle \vec v_1,\vec\l_1|V_{12}^{(m_1,m_2)}|\vec v_2,\vec\l_2\rangle\rangle &= \prod_{l=1}^{m_1}\prod_{l'=1}^{m_2}\CG(v_{l'}^{(2)}/\g v_{l}^{(1)})\ \la\Phi_{\vec\l_1}^{(n_1,m_1)\ast}[u_1,\vec v_1]\Phi^{(n_2,m_2)}_{\vec\l_2}[u_2,\vec v_2]\ra\\
&=t_{n_1,m_1}^\ast(\vec\l_1,u_1,\vec v_1)t_{n_2,m_2}(\vec\l_2,u_2,\vec v_2)\Zbf(\vec v_2,\vec \l_2,\vec v_1,\vec \l_1|\g^{-1}).
\end{split}
\end{align}
Because of a different choice of conventions here, the role of the two nodes is exchanged. This is a feature we will encounter again later when discussing qq-characters. We also observe the presence of the extra prefactor of the form $\la\Phi_1^\ast\ra\la\Phi_2\ra$. Since the dependence on the two nodes is factorized, this term can be absorbed in a renormalization of the basis. Note however that the presence of this prefactor slightly modifies the commutation relations with the action of DIM generators.

The constraint $n_1+m_1=n_2+m_2$ coming from the horizontal contraction has already been observed in \cite{Bourgine2016}. Furthermore, the q-commutation relations with DIM generators that were obtained there can be recovered by taking the vev of the relations \ref{AFS_Pi}. For instance, focusing on the first relation in \ref{AFS_Pi} that involves $x^+(z)$, we find
\begin{align}
\begin{split}
&V_{12}^{(m_1,m_2)}\hat\rho_{\vec v_2}^{(0,m_2)}(x^+(z))-\g^{n_1}\rho_{\vec v_1}^{(0,m_1)}(x^+(\g z))V_{12}^{(m_1,m_2)}\\
=&u_1\g^{n_1}z^{-n_1}\left(\CY^+(z\g^{-1})V_{12}^{(m_1,m_2)}\dfrac1{\CY^+(z)}-\CY^-(z\g^{-1})V_{12}^{(m_1,m_2)}\dfrac1{\CY^-(z)}\right).
\end{split}
\end{align}
Upon projection, we recover here the relations obtained previously in \cite{Bourgine2016}, up to the difference in normalization of states and operators.

\section{Quantum Weyl reflection and qq-characters}
\subsection{Horizontal intertwiner and qq-character for the $A_1$ quiver}
\begin{figure}
\begin{center}
\begin{tikzpicture}
\draw (-0.7,-0.7) -- (0,0) -- (0,2) -- (0.7,2.7);
\draw (-1,2) -- (0,2);
\draw (0,0) -- (1,0);
\node[below left,scale=0.7] at (-0.7,-0.7) {$(1,n^\ast+m)_{u'^\ast}$};
\node[left,scale=0.7] at (-1,2) {$(1,n)_{u}$};
\node[below right,scale=0.7] at (0,2) {$\Phi^{(n,m)}[u,\vec v]$};
\node[above right,scale=0.7] at (0.7,2.7) {$(1,n+m)_{u'}$};
\node[left,scale=0.7] at (0,0) {$\Phi^{(n^\ast,m)\ast}[u^\ast,\vec v]$};
\node[right,scale=0.7] at (1,0) {$(1,n^\ast)_{u^\ast}$};
\node[right,scale=0.7] at (0,1) {$(0,m)_{\vec v}$};
\end{tikzpicture}
\caption{$\CT_{U(m)}$ operator for $U(m)$ instanton partition function}
\label{fig_Um}
\end{center}
\end{figure}
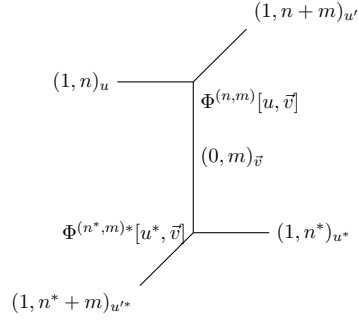

\subsubsection{Definition of the horizontal intertwiner}
In contrast with the horizontal contraction, the vertical contraction is defined as a scalar product between the intertwiner and its dual in a shared vertical representation space. Because of this shared vertical space, the weights $\vec v$ and the label $m$ of the two intertwiners must be equal. As a result, we obtain a tensor product of operators acting in two different Fock spaces, traced over the basis of the common vertical space,
\begin{align}
\begin{split}\label{def_TUm}
\CT_{U(m)}&=\Phi^{(n,m)}[u,\vec v]\cdot\Phi^{(n^\ast,m)\ast}[u^\ast,\vec v]=\sum_{\vec\l}a_{\vec\l}\ \Phi_{\vec\l}^{(n^\ast,m)\ast}[u^\ast,\vec v]\otimes\Phi_{\vec\l}^{(n,m)}[u,\vec v]\\
&:(1,n^\ast+m)_{u^{\ast\prime}}\otimes(1,n)_u\to(1,n^\ast)_{u^\ast}\otimes(1,n+m)_{u'}
\end{split}
\end{align}
Here the second equality has been obtained using the definitions \ref{def_Phi_c} after introduction of the closure relation \ref{closure} between the two intertwiners. For later convenience, the two Fock spaces have been exchanged.
Since it intertwines two horizontal spaces, it will be referred to as the {\it horizontal intertwiner}.

The vertical contraction channel is represented on the figure \refOld{fig_Um}. It defines an operator $\CT_{U(m)}$ associated to the pure $U(m)$ $A_1$ gauge theory. Indeed, the instanton partition function of the gauge theory is equal to the vev of this operator $\CT_{U(m)}$ in the tensored horizontal spaces,
\begin{align}
\begin{split}
\Zinst[A_1]&=\left(_{\ (1,n^\ast)_{u^\ast}\!}\langle\vac|\otimes_{\ (1,n+m)_{u'}\!\!}\langle\vac|\right)\CT_{U(m)}\left(|\vac\rangle_{(1,n^\ast+m)_{u^{\ast\prime}}}\otimes|\vac\rangle_{(1,n)_u}\right)\\
&=\sum_{\vec\l}\qf^{|\vec\l|}\Zv(\vec v,\vec \l)\ZCS(\k,\vec \l),
\end{split}
\end{align}
under the following identification of the Chern-Simons level and the exponentiated gauge coupling
\begin{equation}
\k=n^\ast-n,\quad \qf=\g^{-\k-m}\dfrac{u}{u^\ast}.
\end{equation}
This identification confirms the relation between the horizontal level of representations and the Chern-Simons level. It also leads to interpret the weights $u$ and $u^\ast$ of the two horizontal channels as the position of the NS5 branes (dressed by coinciding D5 branes).

The standard expression of the instanton partition function as a scalar product of Gaiotto states can be recovered from the definition \ref{def_TUm} by introducing the definition \ref{def_Gaiotto} of the Gaiotto states,
\begin{equation}
\Zinst[A_1]=\la\Phi^{(n,m)}[u,\vec v]\cdot\Phi^{(n^\ast,m)\ast}[u^\ast,\vec v]\ra=\langle\langle G,\vec v|G,\vec v\rangle\rangle.
\end{equation} 
Contrary to the expression given in \cite{Bourgine2016}, there is no need to insert an extra gauge coupling operator $\qf^D$ in this formula. The reason being that here the gauge coupling dependence is naturally shared between the two Gaiotto states as it appears from the ratio $u/u^\ast$ of the horizontal weights entering in the intertwiners normalization.

At first sight, it might seem artificial to inject the vector contribution into the coefficients $a_{\vec\l}$ inside the trace in the definition of the operator $\CT_{U(m)}$. However, it should be noted that the form of the vertical representations (and the dual ones) is mostly determined from the constraints of the AFS lemmas, once the horizontal representations and the intertwiners are defined. Then, the equivalence of vertical representations \ref{dual_vert_repres} imposes the constraints \ref{prop_al} on the coefficients $a_{\vec\l}$ (defined as the inverse norm of the states). These recursion relations look very similar to the discrete Ward identities satisfied by the vector contribution. Indeed, up to the few extra factors present in \ref{def_al}, the vector contribution solves these constraints. It thus enters naturally in the definition of the $\CT$-operator of the $U(m)$ gauge theory through the scalar product. In summary, our construction is very rigid, and the form of this operators is already determined from the choice of horizontal representations and normalization of intertwiners.

\paragraph{R-matrix:} In \cite{Awata2011}, partition functions of $U(m)$ $\CN=1$ gauge theories have been constructed in a similar way if we take into account the decomposition \ref{Phi_decomposed} and \ref{Phi_ast_decomposed} of the generalized intertwiners, and the property of the coefficients $a_{\vec\l}$,
\begin{align}
\begin{split}
&\prod_{l=1}^ma_{\l_l}\ \Phi_{\l_m}^{(n^\ast)\ast}[u_m^\ast,v_m]\cdots\Phi_{\l_1}^{(n^\ast+m-1)\ast}[u_1^\ast,v_1]\otimes\Phi_{\l_m}^{(n+m-1)}[u_m,v_m]\cdots\Phi_{\l_1}^{(n)}[u_1,v_1]\\
=&\prod_{\superp{l,l'=1}{l>l'}}^m\CG(v_{l'}/v_l)\CG(v_{l'}/\g^2v_l)\times a_{\vec\l}\ \Phi_{\vec\l}^{(n^\ast,m)}[u^\ast,\vec v]\otimes\Phi_{\vec\l}^{(n,m)}[u,\vec v].
\end{split}
\end{align}
In the end, the only difference with the operator used in \cite{Awata2011} is the harmless prefactor of $\CG$-functions depending only on the ratio of the vertical weights (Coulomb branch vevs). The normal ordering of operators involved in the definition of generalized intertwiners brings extra Nekrasov factors that have been interpreted as elements of a diagonal R-matrix in \cite{Awata2016}. Here, these factors are absorbed in a change of states normalization, in agreement with the definition \ref{def_vert} of the $(0,m)$ vertical representation. More precisely, they are compensated by the replacement of the product over $a_{\l_l}$ by the inverse norm coefficient $a_{\vec\l}$.

\paragraph{Remark on the reflection symmetry:} The intertwiner is exchanged with its dual under the $\s_5$ reflection symmetry, up to a normalization factor that can be absorbed in the transformation of the coefficients $a_{\vec \l}$. The operator $\CT_{U(m)}=\Phi\cdot\Phi^\ast$ becomes $\Phi^\ast\cdot\Phi$. The exchange of the horizontal spaces corresponds to a reflection of the representation web with $x^5$ axis. Since the tensor product is commutative, the vev of the operator $\CT_{U(m)}$, and so the instanton partition function, remains invariant. However, in the process $n$ and $n^\ast$ are exchanged, which has the effect to flip the sign of the Chern-Simons level $\k$.

\subsubsection{Horizontal intertwiner as screening current and fundamental qq-character}
\begin{figure}
\begin{center}
\begin{tikzpicture}
\draw (-0.7,-0.7) -- (0,0) -- (0,2) -- (0.7,2.7);
\draw (-1,2) -- (0,2);
\draw (0,0) -- (1,0);
\node[below left,scale=0.7] at (0,2) {$\Phi$};
\node[below right,scale=0.7] at (0,0) {$\Phi^\ast$};
\node[scale=0.7] at (0,0.5) {$\maltese$};
\node[scale=0.7] at (1.5,1) {$=$};
\end{tikzpicture}
\begin{tikzpicture}
\draw (-0.7,-0.7) -- (0,0) -- (0,2) -- (0.7,2.7);
\draw (-1,2) -- (0,2);
\draw (0,0) -- (1,0);
\node[below left,scale=0.7] at (0,2) {$\Phi$};
\node[below right,scale=0.7] at (0,0) {$\Phi^\ast$};
\node[scale=0.7] at (0,1.5) {$\maltese$};
\end{tikzpicture}
\hspace{2cm}
\begin{tikzpicture}
\draw (-0.7,-0.7) -- (0,0) -- (2,0) -- (2.7,0.7);
\draw (0,0) -- (0,1);
\draw (2,0) -- (2,-1);
\node[above,scale=0.7] at (2,0) {$\Phi$};
\node[below,scale=0.7] at (0,0) {$\Phi^\ast$};
\node[scale=0.7] at (0.5,0) {$\maltese$};
\node[scale=0.7] at (3,0) {$=$};
\node[scale=0.7] at (0,-2) {$\quad$};
\end{tikzpicture}
\begin{tikzpicture}
\draw (-0.7,-0.7) -- (0,0) -- (2,0) -- (2.7,0.7);
\draw (0,0) -- (0,1);
\draw (2,0) -- (2,-1);
\node[above,scale=0.7] at (2,0) {$\Phi$};
\node[below,scale=0.7] at (0,0) {$\Phi^\ast$};
\node[scale=0.7] at (1.5,0) {$\maltese$};
\node[scale=0.7] at (0,-2) {$\quad$};
\end{tikzpicture}
\caption{Commutation relations in the vertical and horizontal channels (respectively) for the insertion of DIM operators, here denoted by the symbol $\maltese$}
\label{fig_invar}
\end{center}
\end{figure}
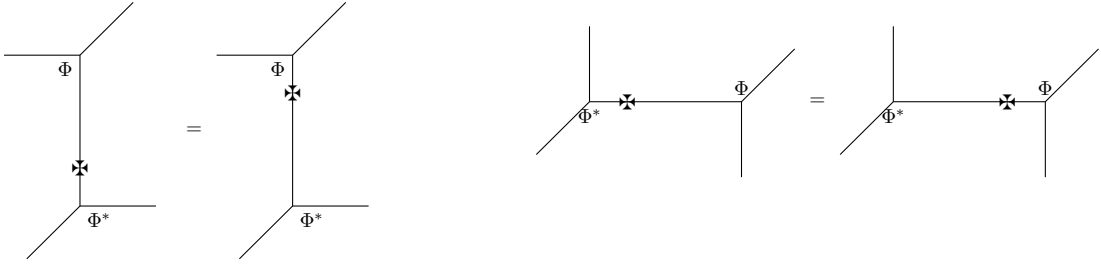

\begin{figure}
\begin{center}
\begin{tikzpicture}
\draw (-0.7,-0.7) -- (0,0) -- (0,2) -- (0.7,2.7);
\draw (-1,2) -- (0,2);
\draw (0,0) -- (1,0);
\node[below left,scale=0.7] at (0,2) {$\Phi$};
\node[below right,scale=0.7] at (0,0) {$\Phi^\ast$};
\node[scale=0.7] at (-0.35,-0.35) {$\maltese$};
\node[scale=0.7] at (-0.5,2) {$\maltese$};
\node[scale=0.7] at (1.5,1) {$=$};
\end{tikzpicture}
\begin{tikzpicture}
\draw (-0.7,-0.7) -- (0,0) -- (0,2) -- (0.7,2.7);
\draw (-1,2) -- (0,2);
\draw (0,0) -- (1,0);
\node[below left,scale=0.7] at (0,2) {$\Phi$};
\node[below right,scale=0.7] at (0,0) {$\Phi^\ast$};
\node[scale=0.7] at (0,0.5) {$\maltese$};
\node[scale=0.7] at (0.5,0) {$\maltese$};
\node[scale=0.7] at (-0.5,2) {$\maltese$};
\node[scale=0.7] at (1.5,1) {$=$};
\end{tikzpicture}
\begin{tikzpicture}
\draw (-0.7,-0.7) -- (0,0) -- (0,2) -- (0.7,2.7);
\draw (-1,2) -- (0,2);
\draw (0,0) -- (1,0);
\node[below left,scale=0.7] at (0,2) {$\Phi$};
\node[below right,scale=0.7] at (0,0) {$\Phi^\ast$};
\node[scale=0.7] at (0,1.5) {$\maltese$};
\node[scale=0.7] at (0.5,0) {$\maltese$};
\node[scale=0.7] at (-0.5,2) {$\maltese$};
\node[scale=0.7] at (1.5,1) {$=$};
\end{tikzpicture}
\begin{tikzpicture}
\draw (-0.7,-0.7) -- (0,0) -- (0,2) -- (0.7,2.7);
\draw (-1,2) -- (0,2);
\draw (0,0) -- (1,0);
\node[below left,scale=0.7] at (0,2) {$\Phi$};
\node[below right,scale=0.7] at (0,0) {$\Phi^\ast$};
\node[scale=0.7] at (0.5,0) {$\maltese$};
\node[scale=0.7] at (0.35,2.35) {$\maltese$};
\end{tikzpicture}
\caption{Heuristic proof of the invariance of the qq-character for a $U(m)$ gauge theory}
\label{fig_qq_char}
\end{center}
\end{figure}
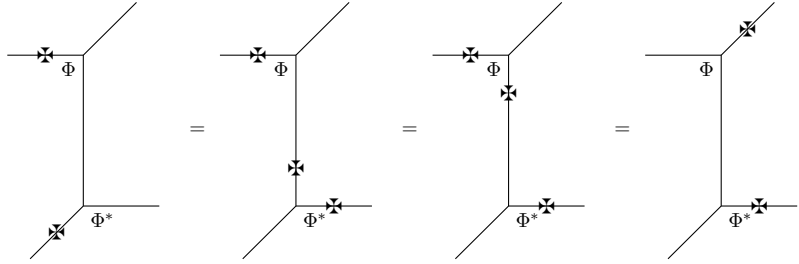

The contraction in the vertical channel commutes with the action of the DIM generators $x^\pm(z)$,
\begin{equation}\label{com_Phi_e}
\left(1\otimes\rho_{\vec v}^{(0,m)}(e)\right)\cdot\CT_{U(m)}=\left(\hat\rho_{\vec v}^{(0,m)}(e)\otimes 1\right)\cdot\CT_{U(m)}.
\end{equation} 
Here, the product notation $\cdot$ has been introduced to emphasize the action in vertical representation spaces. Explicitly, this identity reads
\begin{equation}
\sum_{\vec\l}a_{\vec\l}\ \Phi_{\vec\l}^{(n^\ast,m)}[u^\ast,\vec v]\otimes\left(\rho_{\vec v}^{(0,m)}(e)\cdot\Phi_{\vec\l}^{(n,m)}[u,\vec v]\right)=\sum_{\vec\l}a_{\vec\l}\ \left(\hat\rho_{\vec v}^{(0,m)}(e)\cdot\Phi_{\vec\l}^{(n^\ast,m)}[u^\ast,\vec v]\right)\otimes\Phi_{\vec\l}^{(n,m)}[u,\vec v].
\end{equation}
It expresses the fact that the action on the vertical ket states of arbitrary algebra element should be equated with the action on the bra states. This relation can be easily deduced from the properties \ref{prop_al} obeyed by the coefficients $a_{\vec\l}$. The invariance property of the vertical contraction is represented on the figure \refOld{fig_invar} (left), it is the non-trivial equivalent of the associativity property $(\Phi_\l e)\Phi_\l^\ast=\Phi_\l(e\Phi_\l^{\ast})$ for the horizontal channel. The commutation of $x^\pm(z)$ in the vertical representation has been exploited in \cite{Bourgine2015c,Bourgine2016} in order to establish the regularity property of qq-characters. Here instead, it will prove more convenient to work in the horizontal channel.

Combining this vertical commutation property with the AFS lemmas  \ref{AFS} provides the commutation of the coproduct of DIM elements in the tensored horizontal channels:
\begin{equation}\label{com_D_e}
\left(\rho_{u^\ast}^{(1,n^\ast)}\otimes\rho_{u'}^{(1,n+m)}\right)\D(e)\ \CT_{U(m)} = \CT_{U(m)} \left( \rho_{u'^\ast}^{(1,n^\ast+m)}\otimes\rho_{u}^{(1,n)}\right)\D(e).
\end{equation}
The proof is sketched on the figure \refOld{fig_qq_char}, it follows from the application of the AFS lemma on the dual intertwiner, then the use of formula \ref{com_Phi_e} to commute the vertical actions, and finally the application of AFS lemma again but on the intertwiner $\Phi$. This relation generalizes the results obtained in \cite{Awata2016a} to $m>1$. Since the operator $\CT$ commutes with (the coproduct of) every generator of DIM, it can be interpreted as a screening operator.

In order to build the qq-characters, we define the operator $\CX_{\sAbox}^\pm(z)$ as the coproduct of $x^\pm(z)$,
\begin{equation}
\CX_{\sAbox}^\pm(z)=\D(x^\pm(z)).
\end{equation}
It results from the commutation relation \ref{com_D_e} that the quantities defined as the vev in the horizontal spaces of the products $\CX_{\sAbox}^\pm(z)\CT_{U(m)}$ are polynomials, up to a possible multiplication by a negative power of $z$. This follows from the radial ordering of operators in the q-boson Fock spaces. Indeed, the correlators $\la\CX_{\sAbox}^\pm(z)\CT_{U(m)}\ra$ are well-defined for $|z|>|\chi_x|$, so that they have to be expanded for $z$ near infinity. Similarly, the correlators $\la\CT_{U(m)}\CX_{\sAbox}^\pm(z)\ra$ have to be expanded near $z=0$. The non-trivial equality between the two expansions implies that the correlators are polynomials multiplied by a power of $z$. Just like the equality of the two expansions of any holomorphic function $f(z)$,
\begin{equation}
f(z)\superpsim{\sim}{\infty}\sum_{k=-p}^\infty f_kz^{-k},\quad f(z)\superpsim{\sim}{0}\sum_{k=-q}^\infty \tilde{f}_kz^k,
\end{equation} 
implies the restriction of the summations to a finite extent, $f(z)=\sum_{k=-q}^p f_k z^k=z^{-q}\times\text{Poly}(z)$. These polynomial quantities are the qq-characters introduced by Nekrasov in \cite{Nekrasov2015,Nekrasov2016}. Explicitly, the quantity defined as 
\begin{equation}\label{def_chi_sAbox}
\chi_{\sAbox}^+(z)=\dfrac{\nu z^{n^\ast+m}}{u^\ast\g^{2n^\ast}}\dfrac{\la\CX_{\sAbox}^+(z\g^{-1})\CT_{U(m)}\ra}{\la\CT_{U(m)}\ra},\quad \nu^{-1}=\prod_{l=1}^m(-q_3v_l),
\end{equation} 
can be evaluated in the horizontal representation spaces by exploiting the commutation relation of q-bosonic modes and their action on the vacuum state. Doing so, we recover the expression of the fundamental qq-character already obtained in \cite{Bourgine2016},
\begin{align}
\begin{split}\label{chi_sAbox}
\chi_{\sAbox}^+(z)&=\la\nu z^m\CY_{\vec\l}(zq_3^{-1})+\qf\dfrac{z^\k}{\CY_{\vec\l}(z)}\rag\\
&=\dfrac1{\Zinst[A_1]}\sum_{\{\vec\l\}}\qf^{|\vec\l|}\Zv(\vec v,\vec \l)\ZCS(\k,\vec \l)\left(\nu z^m\CY_{\vec\l}(zq_3^{-1})+\qf\dfrac{z^\k}{\CY_{\vec\l}(z)}\right).
\end{split}
\end{align}
Note that the average of operators in the gauge theory defined in \ref{def_vev_gauge} is in fact a weighted sum over the $m$-tuple Young diagrams $\vec\l$ so that the LHS is effectively independent of $\vec\l$. Due to the asymptotic properties of the function $\CY_{\vec\l}(z)$, namely $\CY_{\vec\l}(z)\superpsim{\sim}{\infty} 1$ and $\CY_{\vec\l}(z)\superpsim{\sim}{0} (u'/u)z^{-m}$, the correlator in the RHS of \ref{chi_sAbox} is a polynomial of degree $m$ in the physical range $|\k|\leq m$ of the Chern-Simons level. This asymptotic behavior explains retrospectively the power of $z$ chosen for the prefactor in the definition \ref{def_chi_sAbox}. Note that an equivalent qq-character $\chi_{\sAbox}^-(z)$ would have be obtained by considering the commutation relations of $\D(x^-(z))$. On the other hand, the commutations of $\D(\psi^\pm(z))$ only provide trivial identities.

Although the derivation of the qq-character has been performed in the horizontal representation, exploiting the Fock space structure, the final expression coincides with the one obtained in the vertical representation. In fact, this expression is closer to the vertical direction, in the sense that it can be expressed as the expectation value of the operators $\CY^\pm(z)$ in the Gaiotto state (either plus or minus since the resulting expression is a polynomial):
\begin{equation}
\chi_{\sAbox}^+(z)=\dfrac{\langle\langle G,\vec v|\nu z^m\CY^\pm(zq_3^{-1})+\qf z^\k\CY^\pm(z)^{-1}|G,\vec v\rangle\rangle}{\langle\langle G,\vec v|G,\vec v\rangle\rangle}.
\end{equation} 
The reason for the equivalence of the two derivation is to be found in the AFS lemma that translates between vertical and horizontal actions. The vertical derivation is closer to the AGT dual presentation \cite{Awata2009} in terms of q-Virasoro (q-Toda) theory \cite{Shiraishi1995}. On the other hand, the main advantage of the horizontal derivation lies in the simplification of the effective calculations since it relies only on q-bosonic modes commutations. It will allow us to propose the expression of higher qq-characters in the following sections.

In this paper, our construction is parallel to the one given by Kimura and Pestun in \cite{Kimura2015}. The precise connection between the two algebraic approaches is discussed in the appendix \refOld{App_KP}. In their work, a partition function state is obtained by application of screening operators to the vacuum. A similar state can be constructed here by applying the $\CT$-operator to the tensored horizontal vacuum $|\vac\rangle\otimes|\vac\rangle$. However, a major difference lies in the absence of the (Kadomtsev-Petviashvili) times deformation introduced in \cite{Kimura2015} and identified with the positive q-bosonic modes of their Fock space. Nevertheless, both approaches consist in constructing a screening operator, together with an operator commuting with it. It is shown in appendix \refOld{App_KP} that the latter coincide with the operator $\CX_{\sAbox}^+(z)$ used here, up to a q-Heisenberg ``$U(1)$'' factor. In both cases, the qq-character is obtained by taking the vacuum expectation value of the product of the two commuting operators.

\paragraph{Remark on a useful identity:} The coproduct symmetrizes the insertion of DIM generators in one of the tensor spaces, so that the resulting expression commutes with the operator $\CT_{U(m)}$. In fact, it is also possible to define an operator with two insertions of DIM generators such as
\begin{equation}\label{def_CX++}
\CX^{++}(z)=\psi^-(\hg_{(1)}^{-1/2}z)x^+(\hg_{(1)}z)\otimes x^+(z),
\end{equation} 
so that it again commutes with $\CT_{U(m)}$. This property will play an important role in the construction of qq-characters for linear quivers of higher rank. It can be obtained by normal ordering the products $\CX^{++}\CT$ and $\CT\CX^{++}$ independently. The resulting expressions have no extra poles appart from the points $z=0$ and $z=\infty$, so that they can be analytically continued to the whole complex plane in which they are equal. However, due to the absence of any extra singularity, taking the vev of the commutation relation only provides a trivial identity, and no non-trivial qq-character can be associated to this operator.

\subsubsection{Higher qq-characters}
In our formalism, higher qq-characters are simply obtained by taking products of coproducts of DIM elements. Since these coproducts commute with $\CT_{U(m)}$ individually, so will their product. As an illustration, consider the qq-character associated to the symmetric representation $\AAbox$ that can be obtained using 
\begin{equation}
\CX^+_{\sAAbox}(z,w)=\D(x^+(z)x^+(w))=\D(x^+(z))\D(x^+(w)).
\end{equation} 
Explicitly, 
\begin{align}
\begin{split}
\CX^+_{\sAAbox}(z,w)=&x^+(z)x^+(w)\otimes 1+\psi^-(\hg_{(1)}^{1/2}z)x^+(w)\otimes x^+(\hg_{(1)}z)\\
&+x^+(z)\psi^-(\hg_{(1)}^{1/2}w)\otimes x^+(\hg_{(1)}w)+\psi^-(\hg_{(1)}^{1/2}z)\psi^-(\hg_{(1)}^{1/2}w)\otimes x^+(\hg_{(1)}z)x^+(\hg_{(1)}w).
\end{split}
\end{align}
Evaluating the vev in the horizontal Fock spaces
\begin{equation}
\chi_{\sAAbox}^+(z,w)=S(w/z)\left(\dfrac{\nu}{u^\ast\g^{2n^\ast}}\right)^2(zw)^{n^\ast+m}\dfrac{\la\CX_{\sAAbox}^+(z\g^{-1},w\g^{-1})\CT_{U(m)}\ra}{\la\CT_{U(m)}\ra},
\end{equation} 
gives
\begin{align}
\begin{split}
\chi_{\sAAbox}^+(z,w)
&=\la\left(\nu z^m\CY_{\vec\l}(zq_3^{-1})+\qf\dfrac{z^\k}{\CY_{\vec\l}(z)}\right)\left(\nu w^m\CY_{\vec\l}(wq_3^{-1})+\qf\dfrac{w^\k}{\CY_{\vec\l}(w)}\right)\rag\\
&+\nu\qf\dfrac{(1-q_1)(1-q_2)zw}{z-w}\la\dfrac{z^m w^\k\CY_{\vec\l}(zq_3^{-1})}{(zq_3^{-1}-w)\CY_{\vec\l}(w)}-\dfrac{z^\k w^m\CY_{\vec\l}(wq_3^{-1})}{(wq_3^{-1}-z)\CY_{\vec\l}(z)}\rag,
\end{split}
\end{align}
where extra scattering factors arise from normal ordering products of DIM operators. Writing $z=z_1\z$, $w=z_2\z$, it is shown that the quantity $\chi_{\sAAbox}^+(z,w)$ is a polynomial in $\z$ as a consequence of the commutation between $\CX_{\sAAbox}^+$ and $\CT_{U(m)}$. In fact, it is possible to prove a stronger result: the ratio of $\chi_{\sAAbox}^+(z,w)$ by the scattering factor $S(w/z)$ is a polynomial in both variables $z$ and $w$.

\subsection{Quantum Weyl reflection}
\subsubsection{Horizontal intertwiner for the $A_2$ quiver}
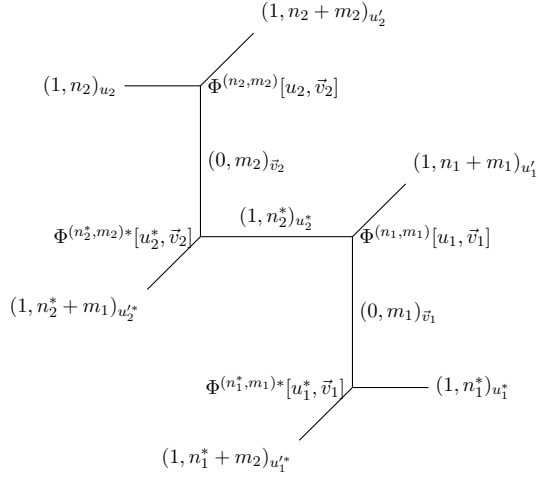
\begin{figure}
\begin{center}
\begin{tikzpicture}
\draw (0,0) -- (1,0) -- (1,-2) -- (3,-2) -- (3,-4) -- (4,-4);
\draw (1,0) -- (1.7,0.7);
\draw (0.3,-2.7) -- (1,-2);
\draw (3,-2) -- (3.7,-1.3);
\draw (2.3,-4.7) -- (3,-4);
\node[above right,scale=0.7] at (1.7,0.7) {$(1,n_2+m_2)_{u'_2}$};
\node[above right,scale=0.7] at (3.7,-1.3) {$(1,n_1+m_1)_{u'_1}$};
\node[below left,scale=0.7] at (0.3,-2.7) {$(1,n_2^\ast+m_1)_{u'^\ast_2}$};
\node[below left,scale=0.7] at (2.3,-4.7) {$(1,n_1^\ast+m_2)_{u'^\ast_1}$};
\node[left,scale=0.7] at (0,0) {$(1,n_2)_{u_2}$};
\node[right,scale=0.7] at (4,-4) {$(1,n_1^\ast)_{u_1^\ast}$};
\node[above,scale=0.7] at (2,-2) {$(1,n_2^\ast)_{u_2^\ast}$};
\node[right,scale=0.7] at (1,-1) {$(0,m_2)_{\vec v_2}$};
\node[right,scale=0.7] at (3,-3) {$(0,m_1)_{\vec v_1}$};
\node[right,scale=0.7] at (1,0) {$\Phi^{(n_2,m_2)}[u_2,\vec v_2]$};
\node[left,scale=0.7] at (1,-2) {$\Phi^{(n_2^\ast,m_2)\ast}[u_2^\ast,\vec v_2]$};
\node[right,scale=0.7] at (3,-2) {$\Phi^{(n_1,m_1)}[u_1,\vec v_1]$};
\node[left,scale=0.7] at (3,-4) {$\Phi^{(n_1^\ast,m_1)\ast}[u_1^\ast,\vec v_1]$};
\end{tikzpicture}
\caption{Configuration relevant to the $U(m_2)\times U(m_1)$ gauge theory}
\label{fig_UmUm}
\end{center}
\end{figure}

The $A_2$ quiver gauge theory, with gauge group $U(m_2)\times U(m_1)$ is described by an operator $\CT_{U(m_2)\times U(m_1)}$ involving two vertical contractions (for the vector multiplet contributions), and a single horizontal contraction (the bifundamental contribution):
\begin{align}
\begin{split}
\CT_{U(m_2)\times U(m_1)}&=\sum_{\vec\l_1,\vec\l_2}a_{\vec\l_1}a_{\vec\l_2}\ \Phi_{\vec\l_1}^{(n_1^\ast,m_1)\ast}[u_1^\ast,\vec v_1]\otimes\Phi_{\l_1}^{(n_1,m_1)}[u_1,\vec v_1]\Phi_{\l_2}^{(n_2^\ast,m_2)\ast}[u_2^\ast,\vec v_2]\otimes\Phi_{\vec\l_2}^{(n_2,m_2)}[u_2,\vec v_2]\\
&:(1,n_1^\ast+m_1)_{u_1'^\ast}\otimes(1,n_2^\ast+m_2)_{u'_2}\otimes(1,n_2)_{u_2}\to(1,n_1^\ast)_{u_1^\ast}\otimes(1,n_1+m_1)_{u'_1}\otimes(1,n_2+m_2)_{u'_2},
\end{split}
\end{align}
with the constraints $n_2^\ast=n_1$ and $u_2^\ast=u_1$ in order to match the representation levels and weights in the horizontal channel. Actually, this operator can be obtained from the combination of the operators $\CT_{U(m_1)}$ and $\CT_{U(m_2)}$ associated to the two gauge groups, using a new product $\circ$ to represent the horizontal contraction, $\CT_{U(m_2)\times U(m_1)}=\CT_{U(m_1)}\circ\CT_{U(m_2)}$. The new product corresponds to the concatenation of two chains of tensor products of respective length $r$ and $s$, in order to form a chain of length $r+s-1$:
\begin{equation}\label{def_circ}
(a_1\otimes\cdots\otimes a_r)\circ(b_1\otimes\cdots\otimes b_s)=a_1\otimes\cdots \otimes a_{r-1}\otimes a_rb_1\otimes b_2\cdots\otimes b_s.
\end{equation} 

As before, the instanton partition function of the gauge theory corresponds to the vev in the horizontal spaces, up to a factor of $\CG$-functions depending only on the Coulomb branch vevs,
\begin{align}
\begin{split}
\Zinst[A_2]&=\prod_{l=1}^{m_1}\prod_{l'=1}^{m_2}\CG(v_{l'}^{(2)}/\g v_{l}^{(1)})\times\la\CT_{U(m_2)\times U(m_1)}\ra\\
&=\sum_{\vec\l_1,\vec \l_2}\qf_1^{|\vec\l_1|}\qf_2^{|\vec\l_2|}\Zv(\vec v_1,\vec\l_1)\ZCS(\k_1,\vec\l_1)\Zv(\vec v_2,\vec\l_2)\ZCS(\k_2,\vec\l_2)\Zbf(\vec v_2,\vec\l_2,\vec v_1,\vec\l_1|\g^{-1}),
\end{split}
\end{align}
with the identification $\qf_\a=\g^{-\k_\a-m_\a}u_\a/u_\a^\ast$ and $\k_\a=n_\a^\ast-n_\a$ for the gauge coupling and the Chern-Simons level associated to the two gauge groups $\a=1,2$.

\subsubsection{Fundamental qq-character for the $A_2$ quiver}
A different qq-character is associated to each node of the quiver diagram, in correspondence with antisymmetric representations of the Lie algebra. In the case of the $A_2$ quiver, the two relevant representations are the fundamental $\Abox$ and the fully antisymmetric $\Bbox$ ones.  We focus first on the construction of the fundamental qq-character $\chi_{\sAbox}^+(z)$, leaving the antisymmetric representation for the next section. Within our conventions, the fundamental representation is associated to the action of $x^+(z)$ on the first node.

The commutation property \ref{com_Phi_e} of the vertical channel is valid at each node in the form
\begin{align}
\begin{split}
&\left(\hat\rho_{\vec v_1}^{(0,m_1)}(e)\otimes1\otimes1\right)\cdot\CT_{U(m_2)\times U(m_1)}=\left(1\otimes\rho_{\vec v_1}^{(0,m_1)}(e)\otimes 1\right)\cdot\CT_{U(m_2)\times U(m_1)}\\
&\left(1\otimes\hat\rho_{\vec v_2}^{(0,m_2)}(e)\otimes1\right)\cdot\CT_{U(m_2)\times U(m_1)}=\left(1\otimes1\otimes\rho_{\vec v_2}^{(0,m_2)}(e)\right)\cdot\CT_{U(m_2)\times U(m_1)}.
\end{split}
\end{align}
The operator leading to the fundamental qq-character can be obtained using the double coproduct
\begin{equation}
\D^{\sAbox}=(\D\otimes1)\D=(1\otimes\D)\D.
\end{equation}
The action of this double co-product on the DIM generators reads
\begin{align}
\begin{split}
\D^{\sAbox}(x^+(z))=&x^+(z)\otimes1\otimes1+\psi^-(\hg^{1/2}\otimes1\otimes 1\ z)\otimes x^+(\hg\otimes1\otimes1\ z)\otimes1\\
&+\psi^-(\hg^{1/2}\otimes1\otimes1\ z)\otimes\psi^-(\hg\otimes\hg^{1/2}\otimes1\ z)\otimes x^+(\hg\otimes\hg\otimes1\ z)\\
\D^{\sAbox}(x^-(z))=&1\otimes1\otimes x^-(z)+1\otimes x^-(1\otimes1\otimes\hg\ z)\otimes\psi^+(1\otimes1\otimes\hg^{1/2}\ z)\\
&+x^-(1\otimes\hg\otimes\hg\ z)\otimes\psi^+(1\otimes\hg^{1/2}\otimes\hg\ z)\otimes\psi^+(1\otimes1\otimes\hg^{1/2}\ z)\\
\D^{\sAbox}(\psi^\pm(z))=&\psi^\pm(1\otimes\hg^{\pm1/2}\otimes\hg^{\pm1/2}\ z)\otimes\psi^\pm(\hg^{\mp1/2}\otimes1\otimes\hg^{\pm1/2}\ z)\otimes\psi^\pm(\hg^{\mp1/2}\otimes\hg^{\mp1/2}\otimes1\ z).
\end{split}
\end{align}
Combining the general commutation properties of the vertical and horizontal contractions, together with the AFS lemmas \ref{AFS}, it is possible to show that the horizontal action of the generators commutes with the $\CT$-operator:
\begin{align}
\begin{split}\label{com_Phi_UmUm}
&\left(\rho_{u_1^\ast}^{(1,n_1^\ast)}\otimes\rho_{u'_1}^{(1,n_1+m_1)}\otimes\rho_{u'_2}^{(1,n_2+m_2)}\right)\D^{\sAbox}(e)\ \CT_{U(m_2)\times U(m_1)}\\
=&\ \CT_{U(m_2)\times U(m_1)}\ \left(\rho_{u_1'^\ast}^{(1,n_1^\ast+m_1)}\otimes\rho_{u'_2}^{(1,n_2^\ast+m_2)}\otimes\rho_{u_2}^{(1,n_2)}\right)\D^{\sAbox}(e).
\end{split}
\end{align}
The proof is a tedious but straightforward calculation, using the same method as in the single node case.

Introducing the operator $\CX_{\sAbox}^+(z)=\D^{\sAbox}(x^+(z))$ that commutes with $\CT_{U(m_2)\times U(m_1)}$, the fundamental qq-character can be written (omitting the horizontal representations):
\begin{equation}
\chi_{\sAbox}^+(z)=\dfrac{\nu_1}{u_1^\ast\g^{2n_1^\ast}}z^{n_1^\ast+m_1}\dfrac{\la\CX_{\sAbox}^+(z\g^{-1})\CT_{U(m_2)\times U(m_1)}\ra}{\la\CT_{U(m_2)\times U(m_1)}\ra}
\end{equation} 
with $\nu_\a=u_\a/(q_3^{m_\a}u'_\a)$. Again, this quantity is a polynomial in $z$ because $\CX_{\sAbox}^+$ commutes with $\CT$. The power of $z$ in the prefactor is fixed by consideration of the asymptotic behavior. Evaluating the correlators in the q-bosonic Fock spaces, we recover the expression given in \cite{Bourgine2016} for the fundamental $A_2$ qq-character with a bifundamental mass $\mu=\g^{-1}$
in the vertical channel:\footnote{To simplify formulas, the labels corresponding to the two nodes have been exchanged here with respect to the conventions employed in \cite{Bourgine2016}.}
\begin{equation}
\chi_{\sAbox}^+(z)=\la\nu_1z^{m_1}\CY_{\vec\l_1}(z\g^{-2})+\qf_1 z^{\k_1}\dfrac{\CY_{\vec\l_2}(z\g^{-1})}{\CY_{\vec\l_1}(z)}+\qf_1\qf_2\dfrac{\nu_1}{\nu_2}\g^{\k_2+2m_1-m_2}\dfrac{z^{\k_1+\k_2+m_1-m_2}}{\CY_{\vec\l_2}(\g z)}\rag.
\end{equation}

\subsubsection{Quantum Weyl reflection and the second qq-character}
There are two ways to obtain the second qq-character. The simplest one is to consider the insertion of the operator $\CX_{\sAbox}^-(z)=\D^{\sAbox}(x^-(z\g^{-1}))$ that also commutes with the $\CT$-operator. Defining
\begin{equation}
\chi_{\sAbox}^-(z)=\nu_2u'_2\g^{2(n_2+m_2)}z^{-n_2}\dfrac{\la\CX_{\sAbox}^-(z\g^{-1})\CT_{U(m_2)\times U(m_1)}\ra}{\la\CT_{U(m_2)\times U(m_1)}\ra}
\end{equation}
gives after evaluation of each horizontal actions,
\begin{equation}\label{chi_sAbox_m}
\chi_{\sAbox}^-(z)=\la\nu_2 z^{m_2}\CY_{\vec\l_2}(z\g^{-2})+\qf_2\nu_1\g^{m_1}z^{m_1+\k_2}\dfrac{\CY_{\vec\l_1}(z\g^{-1})}{\CY_{\vec\l_2}(z)}+\qf_1\qf_2\g^{\k_1}\dfrac{z^{\k_1+\k_2}}{\CY_{\vec\l_1}(\g z)}\rag,
\end{equation} 
which is indeed the second qq-character of the $A_2$ quiver found in \cite{Bourgine2016}. In fact, because of the reflection symmetry obeyed by the $A_2$ quiver, we have $\chi_{\sAbox}^-(z)=\chi_{\sBbox}^+(z)$ which explains why we have obtained the correct qq-character. This property seems to be a consequence of the $\s_5$ symmetry of the representation web combined with the $\CS^2$ rotation of the DIM algebra.

There is a more natural way to derive the second qq-character, which is to start from insertions of $x^+(z)$ in two of the three horizontal lines at the end of the diagram, with spectral parameters fine-tuned to obtain the commutation with the $\CT$-operator. Indeed, the first term of the qq-character, that is proportional to $\CY_{\vec\l_2}(z\g^{-2})$, can be obtained from the insertion of the operator
\begin{equation}\label{op_x_x}
\psi^-(\hg^{-1/2}\otimes 1\otimes 1z)x^+(\hg\otimes1\otimes1\ z)\otimes x^+(z)\otimes1
\end{equation} 
on the left of the representation web. The other terms entering the expression of the qq-characters are known to be obtained by acting with the Weyl reflection on the first term \cite{Kimura2015}.

The Weyl reflection for the $A_2$ quiver diagram sends the co-weight $w_2$ attached to the second node to
\begin{equation}
w_2\xrightarrow{\a_2} w_1-w_2\xrightarrow{\a_1} -w_1\equivalent \CY_{\vec\l_2}\xrightarrow{\a_2}\dfrac{\CY_{\vec\l_1}}{\CY_{\vec\l_2}}\xrightarrow{\a_1}\dfrac1{\CY_{\vec\l_1}},
\end{equation} 
where $\a_i$ denote the roots of the Lie algebra with the Dynkin diagram $A_2$. To each intermediate expression has been associated a term of the qq-character. A similar transformation can be defined on the DIM generator $x^+(z)$ involved in tensorial expressions like \ref{op_x_x}. It will be called the \textit{quantum Weyl reflection}. The quantum Weyl reflection with respect to the root $\a_i$ consists in replacing the insertion of $x^+$ in the $i$th tensor space with an insertion of $\psi^-$ in the $i$th space and $x^+$ in the $(i+1)$th space, together with the appropriate shifts of the spectral parameters:\footnote{The quantum Weyl transformation can be defined in a similar manner on the generator $x^-(z)$, i.e. in such a way that it reproduces the coproduct in the case of the fundamental representation:
\begin{equation}
\cdots\otimes1\otimes_i x^-(z)\otimes \cdots\xrightarrow{\a_{i}}\cdots\otimes x^-(\hg_{(i+1)}z)\otimes_i\psi^+(\hg_{(i+1)}^{1/2}z)\otimes\cdots
\end{equation}}
\begin{equation}
\cdots\otimes x^+(z)\otimes_i1\otimes \cdots\xrightarrow{\a_{i}}\cdots\otimes\psi^-(\hg_{(i)}^{1/2}z)\otimes_i x^+(\hg_{(i)}z)\otimes\cdots\\
\end{equation}
where $\hg_{(i)}=(1\otimes)^{i-1}\otimes \hg(\otimes1)^{r+1-i}$. The transformation is forbidden if two $x^+$ were to collide in the same space. In a sense, the generators $x^+$ obey a fermionic statistics in the tensor spaces. It is further assumed that the operators $\psi^-$ are ordered on the left of operators $x^+$, although this fact does not modify the derivation of the qq-characters.

Before applying the quantum Weyl reflection to obtain the operator relevant to the second node of the $A_2$ quiver, let us review how this transformation works in the known cases. The $A_1$ quiver is described by two horizontal spaces, it has a single weight $w\xrightarrow{\a}-w$ and the qq-character $\chi^+_{\sAbox}(z)$ has only two terms. The application of the quantum Weyl reflection on $x^+(z)\otimes1$ leads to the coproduct $\D(x^+(z))$.

Turning to the $A_2$ quiver, the quantum Weyl reflections of 
\begin{align}
\begin{split}
&x^+(z)\otimes1\otimes 1\xrightarrow{\a_1} \psi^-(\hg^{1/2}\otimes 1\otimes 1\ z)\otimes x^+(\hg\otimes 1\otimes 1\ z)\otimes 1\\
&\xrightarrow{\a_2} \psi^-(\hg^{1/2}\otimes 1\otimes 1\ z)\otimes\psi^-(\hg\otimes\hg^{1/2}\otimes 1\ z)\otimes x^+(\hg\otimes \hg\otimes 1\ z)
\end{split}
\end{align}
reproduces the three terms in the expression of the operator $\CX^+_{\sAbox}(z)$ constructed from the application of the squared coproduct $\D^{\sAbox}$. Thus, the quantum Weyl reflection defines a generalization of the coproduct that implements the action of an operator into three copies of the initial space.

Now, we apply the Weyl reflection to the operator \ref{op_x_x} of the $A_2$ quiver, and sum over the three terms in order to define
\begin{align}
\begin{split}
&\D^{\sBbox}(x^+(z))=\psi^-(\hg^{-1/2}\otimes 1\otimes1\ z)x^+(\hg\otimes1\otimes1\ z)\otimes x^+(z)\otimes1\\
&+\psi^-(\hg^{-1/2}\otimes 1\otimes 1\ z)x^+(\hg\otimes1\otimes1\ z)\otimes \psi^-(1\otimes \hg^{1/2}\otimes1\ z)\otimes x^+(1\otimes\hg\otimes1\ z)\\
&+\psi^-(\hg^{-1/2}\otimes 1\otimes 1\ z)\psi^-(\hg^{3/2}\otimes 1\otimes 1\ z)\otimes \psi^-(1\otimes \hg^{1/2}\otimes1\ z)x^+(\hg^2\otimes1\otimes1\ z)\otimes x^+(1\otimes\hg\otimes1\ z).
\end{split}
\end{align}
After a tedious but straightforward computation, it can be shown that the operator $\CX^+_{\sBbox}(z)=\D^{\sBbox}(x^+(z))$ commutes with $\CT$. The commutation of the operator defined in \ref{def_CX++} with $\CT_{U(m)}$ (seen here as a subdiagram) is essential for the various cancellations to occur. The corresponding qq-character is defined as
\begin{equation}
\chi_{\sBbox}^+(z)=\dfrac{\nu_2}{u_1^\ast u'_1}\g^{-2(n_1+n_1^\ast+m_1)}z^{n_1+n_1^\ast+m_1+m_2}\dfrac{\la\CX_{\sBbox}^-(z\g^{-1})\CT_{U(m_2)\times U(m_1)}\ra}{\la\CT_{U(m_2)\times U(m_1)}\ra}.
\end{equation} 
The evaluation of the vev in the horizontal spaces reproduces the expression of $\chi_{\sAbox}^-(z)$ given in \ref{chi_sAbox_m}, showing that indeed $\chi_{\sBbox}^+(z)=\chi_{\sAbox}^-(z)$.

\subsubsection{Generalization to the $A_r$ quivers}
The results obtained for the $A_2$ quiver are easily generalized to linear quivers with an arbitrary number of nodes $r$. The $\CT$-operator is constructed using $r-1$ contractions in the horizontal channel, rendered by the product $\circ$ defined in \ref{def_circ},
\begin{equation}
\CT_{U(m_r)\times\cdots\times U(m_1)}=\CT_{U(m_1)}\circ\cdots\circ\CT_{U(m_r)},
\end{equation}
where the order of indices labeling gauge groups has been reversed for convenience. Schematically, it reads
\begin{align}
\begin{split}
\CT_{U(m_r)\times\cdots\times U(1)}=\sum_{\vec\l_1,\cdots\vec\l_r}\prod_{s=1}^ra_{\vec\l_r}\ \Phi_{\l_1}^\ast\otimes\Phi_{\l_1}\Phi_{\l_2}^\ast\otimes\cdots\otimes\Phi_{\l_r-1}\Phi_{\l_r}^\ast\otimes\Phi_{\l_r},
\end{split}
\end{align}
where we have omitted all the weights and level parameters. This expression implies the constraints $n_s=n_{s+1}^\ast$ and $u_s=u_{s+1}^\ast$ in order to match the representation spaces in horizontal channels. Then, up to a prefactor of $\CG$-functions, the vev reproduces the instanton partition function \ref{def_Zinst} for the quiver $\G=A_r$ under the identification $\k_s=n_s^\ast-n_s$ for the Chern-Simons levels, and $\qf_s=\g^{-\k_s-m_s}u_s/u_s^\ast$ for the gauge couplings.

As before, the fundamental qq-character, attached to the first node, can be obtained by multiple applications of the coproduct. The $A_r$ fundamental coproduct $\D^{\sAbox}$ is defined recursively as
\begin{equation}
\D^{\sAbox}=(\D(\otimes1)^{r-1})\cdot(\D(\otimes1)^{r-2})\cdots \D,
\end{equation} 
and acts on the DIM generators as follows:
\begin{align}
\begin{split}
&\D^{\sAbox}(x^+(z))=\sum_{s=1}^{r+1}\psi^-(\hg_{(1)}^{1/2}\ z)\otimes\cdots\otimes\psi^-(\hg_{(s-2)!}\hg_{(s-1)}^{1/2}\ z)\otimes x^+(\hg_{(s-1)!}\ z)\ (\otimes1)^{r+1-s},\\
&\D^{\sAbox}(x^-(z))=\sum_{s=1}^{r+1}(1\otimes)^{r+1-s}x^-(\hg_{(s-1)!}^\top\ z)\otimes\psi^+((\hg_{(s-2)!}\hg_{s-1}^{1/2})^\top\ z)\otimes\cdots\otimes\psi^+(\hg_{(r+1)}^{1/2}\ z),\\
&\D^{\sAbox}(\psi^\pm(z))=\psi^\pm(1(\otimes\hg^{\pm1/2})^r\  z)\otimes\psi^\pm(\hg^{\mp1/2}\otimes1(\otimes\hg^{\pm1/2})^{r-1}\ z)\otimes\cdots\otimes\psi^\pm((\hg^{\mp1/2}\otimes)^r1\ z),
\end{split}
\end{align}
with $\hg_{(s)!}=\hg_{(1)}\hg_{(2)}\cdots\hg_{(s)}=(\hat\g\otimes)^{s}1(\otimes1)^{r-s}$, and the tensorial transpose defined as $(a_1\otimes a_2\otimes\cdots\otimes a_{r+1})^\top=a_{r+1}\otimes a_r\otimes\cdots\otimes a_1$.

The fundamental coproduct of DIM generators commutes with the $\CT$-operator,
\begin{equation}
\D^{\sAbox}(e)\cdot \CT_{U(m_r)\times\cdots\times U(1)}=\CT_{U(m_r)\times\cdots\times U(1)}\cdot\D^{\sAbox}(e),
\end{equation} 
where we have omitted to indicate the horizontal representations. As a result, the fundamental qq-character defined as
\begin{equation}
\chi_{\sAbox}^+(z) = \dfrac{\nu_1}{u_1^\ast}\g^{-2n_1^\ast}z^{n_1^\ast+m_1}\dfrac{\la \CX_{\sAbox}^+(z\g^{-1})\CT_{U(m_r)\times\cdots\times U(1)}\ra}{\la \CT_{U(m_r)\times\cdots\times U(1)}\ra},\quad \CX_{\sAbox}^+(z)=\D^{\sAbox}(x^+(z)),
\end{equation}
is a polynomial. Explicit evaluation of the correlators for each horizontal space provides the formula
\begin{align}
\begin{split}
\chi_{\sAbox}^+(z)=&\Bigg\langle\nu_1z^m_1\CY_{\vec\l_1}(z\g^{-2})+\sum_{s=1}^{r-1}\dfrac{\nu_1}{\nu_s}\left(\prod_{i=1}^s \qf_i z^{\k_i}\g^{2m_i-m_s+\sum_{j=i+1}^s\k_j}\right)z^{m_1-m_s}\g^{-m_s}\dfrac{\CY_{\vec\l_{s+1}}(z\g^{s-2})}{\CY_{\vec\l_s}(z\g^{s-1})}\\
&\quad+\dfrac{\nu_1}{\nu_r}\left(\prod_{i=1}^r \qf_i z^{\k_i}\g^{2m_i-m_r+\sum_{j=i+1}^r\k_j}\right)z^{m_1-m_r}\g^{-m_r}\dfrac1{\CY_{\vec\l_r}(z\g^{r-1})}\Bigg\rangle_\text{gauge}.
\end{split}
\end{align}


The qq-character associated to the $s$th node corresponds to the antisymmetric representation denoted by the Young diagram $(s)$ with $s$ boxes, all in the first column. The corresponding operator $\CX_{(s)}^+(z)$ is obtained by application of quantum Weyl reflections on the operator
\begin{align}
\begin{split}
&\psi_{[1]}^-(z)x^+(\hg_{(s-1)!}z)\otimes\psi_{[2]}^-(z)\ x^+(\hg_{(s-2)!}z)\otimes\cdots\otimes \psi_{[s-1]}^-(z)x^+(\hg_{(1)}z)\otimes x^+(z) (\otimes1)^{r+1-s},\\
&\text{with:}\quad \psi^-_{[i]}(z)=\prod_{j=1}^{s-i}\psi^-\left(\hg_{(j-1)!}\hg_{(i)}^{-1/2}\prod_{k=i+1}^{s-j}\hg_{(k)}^{-1}\right).
\end{split}
\end{align}
Note that since the operator is evaluated in horizontal representations, the position of the central element $\hg$ in the arguments of operators is somewhat arbitrary here. As an illustration, the $A_3$ quiver is treated in details in the appendix \refOld{App_A3}. This construction can also be applied to the generator $x^-(z)$. Because of the reflection symmetry of the quiver diagram, the corresponding qq-characters are expected to obey the relation $\chi_{(s)}^-(z)=\chi_{(r+1-s)}^+(z)$.

In fact, it is possible to define qq-characters associated to arbitrary representations of the Lie algebra. To a representation labeled by a Young diagram $\l$ is associated the operator $\CX_{\l}^+(\vec z)$ obtained by taking the product over the columns $\l_i$ of the operators $\CX_{(\l_i)}^+(z_i)$ defined previously. This construction works if the first column of the Young diagram contains at most $r$ boxes. By construction, these operators commute with the $\CT$-operator of the gauge theory, and the vev $\la\CX_{\l}^+(\vec z)\CT\ra$ is a polynomial up to multiplication by a monomial of the variables $z_i$.

\subsubsection{Inclusion of fundamental/antifundamental matter fields}

\begin{figure}
	\begin{center}
		\begin{tikzpicture}
		\hspace{-1cm}
		\draw (-1, 1) -- (-1,0) -- (1,0) -- (1,-2) -- (2,-2);
		\draw (1,0) -- (1.7,0.7);
		\draw (-1.7, -0.7) -- (-1, 0); 
		\draw (0.3,-2.7) -- (1,-2);
		\node[above right,scale=0.7] at (1.7,0.7) {$(1,n+m)_{u'}$};
		\node[below left,scale=0.7] at (0.3,-2.7) {$(1,n^\ast+m)_{u'^\ast}$};
		\node[below left,scale=0.7] at (-1.7,-0.7) {$(1,n+f)_{u''}$};
		\node[left,scale=0.7] at (0.7,0.3) {$(1,n_3^\ast)_{u_3^\ast}$};
		\node[above,scale=0.7] at (2,-2) {$(1,n^\ast)_{u^\ast}$};
		\node[right,scale=0.7] at (1,-1) {$(0,m)_{\vec v}$};
		\node[above,scale=0.7] at (-1,1) {$(0,\tf)_{\vec\mu^{(\text{af})}}$};
		\node[right,scale=0.7] at (1,0) {$\Phi^{(n,m)}[u,\vec v]$};
		\node[left,scale=0.7] at (1,-2) {$\Phi^{(n^\ast,m)\ast}[u^\ast,\vec v]$};
		\node[left,scale=0.7] at (-1,0) {$\Phi_{\vec\vac}^{(n,\tf)\ast}[u,\vec\mu^{(\text{af})}]$};
		\end{tikzpicture}
		\caption{Representation web of the $A_1$ quiver with antifundamental matter}
		\label{fig_A1matter}
	\end{center}
\end{figure}

Matter fields are introduced by semi-infinite D5 branes that are vertical edges in the representation web. These can be inserted either in the bottom or top part of the diagram, leading to fundamental ($\Phi$) or antifundamental ($\Phi^\ast$) matter respectively. It is well-known in gauge theory that such matter fields can be obtained by introducing extra gauge groups, sending the corresponding gauge coupling $\qf$ to zero. This constraints the Young diagrams $\vec\l$ associated to this gauge group in the partition function expansion \ref{def_Zinst} to be empty, hence generating the contributions
\begin{equation}
\Zf(\g^{-1}\vec\mu^{(\text{f})},\vec \l)=\Zbf(\vec v,\vec\l,\vec\mu^{(\text{f})},\vec\vac|\g^{-1}),\quad  \Zaf(\g\vec \mu^{(\text{af})},\vec \l)=\Zbf(\vec\mu^{(\text{af})},\vec\vac,\vec v,\vec\l|\g^{-1}).
\end{equation} 

In this spirit, we can regard the massive $A_1$ quiver as the limit of the $A_3$ quiver as $\qf_1,\qf_3\to0$. This procedure corresponds to send two NS5 branes at infinity. Taking the formula from appendix \refOld{App_A3} for the  $A_3$ qq-character $\chi_{\sBbox}^+(z)$ associated to the second node, and sending the gauge couplings $\qf_1,\qf_3\to0$ while $\qf_2=\qf$ is held fixed, we indeed recover the massive $A_1$ qq-character $\chi_{\sAbox}^+(z)$ obtained in \cite{Bourgine2016},
\begin{equation}\label{qq_massive}
\chi_{\sAbox}^+(z)=\Bigg\langle\nu z^{m}\CY_{\vec\l}(z\g^{-2})+\qf z^{\k}\dfrac{\pf(z\g^{-2})\paf(z)}{\CY_{\vec\l}(z)}\Bigg\rangle_{\text{gauge}},
\end{equation} 
since
\begin{equation}
\nu_1\g^{m_1}z^{m_1}\CY_{\vec\vac_1}(z\g^{-1})=\prod_{l=1}^{f}(1-z\g^{-1}(\mu_l^{(\text{f})})^{-1})=\pf(z\g^{-2}),\quad \CY_{\vec\vac_3}(z\g^{-1})=\prod_{l=1}^{\tilde{f}}(1-\g\mu_l^{(\text{af})}z^{-1})=\paf(z),
\end{equation} 
and we have dropped the label $2$ of the middle node.

In our formalism, the gauge coupling $\qf$ is obtained as a ratio of horizontal weights $u/u^\ast$. The limiting procedure $\qf\to0$ corresponds to send either $u$ to zero for some intertwiner $\Phi_\l$, or $u^\ast$ to infinity for the dual intertwiner $\Phi_\l^\ast$. In either case, the normalization coefficients, $t_{n,m}$ or $t_{n,m}^\ast$, vanishes except when the Young diagrams $\vec\l$ are empty. The case of the antifundamental matter is the easiest one to consider. Indeed, it is observed from the AFS lemma that since $R(\vec\vac)=\vac$, the vacuum intertwiner $\Phi_{\vac}^\ast$ commutes with the action of $x^+(z)$. As a result, an additional horizontal contraction with this operator, as represented on the figure \refOld{fig_A1matter}, does not spoil the commutation with the operator $\CX_{\sAbox}^+(z)$. In this case, the $\CT$-operator is simplified as the extra $\Phi_{\vac}$ can be decoupled,\footnote{Said it otherwise, the action of $x^+(z)$ on $\Phi_{\vac}$ being proportional to $u\to0$, the extra horizontal channel can be dropped.}
\begin{equation}
\CT_{U(m)}^{(\text{af})}=\sum_{\vec\l}a_{\vec\l}\ \Phi_{\vec\l}^{(n^\ast,m)\ast}[u^\ast,\vec v]\otimes\Phi_{\vec\l}^{(n,m)}[u,\vec v]\Phi_{\vec\vac}^{(n,\tf)\ast}[u,\vec\mu^{(\text{af})}].
\end{equation} 

On the other hand, in the case of fundamental matter, it does not seem possible to fully decouple the extra horizontal channel, and we are forced to define the $\CT$-operator within three different Fock spaces,
\begin{equation}\label{CT_Um_af}
\CT_{U(m)}^{(\text{af})}=\sum_{\vec\l}a_{\vec\l}\ \Phi_{\vec\vac}^{(n,f)\ast}[u,\vec\mu^{(\text{f})}]\otimes\Phi_{\vec\vac}^{(n,f)}[u,\vec\mu^{(\text{f})}]\Phi_{\vec\l}^{(n^\ast,m)\ast}[u^\ast,\vec v]\otimes\Phi_{\vec\l}^{(n,m)}[u,\vec v],
\end{equation}
in order to observe the commutation relation with the operator $\CX_{\sAbox}^+(z)$.\footnote{Taking the limit $u_1^\ast\to\infty$ in the $A_3$ operator $\CX_{\sBbox}^+(z)$, the dominant terms are those with a $x^+(z)$ generator inserted in the first space. They are of order $\sim u_1^\ast$ and reproduce the two terms in the massive qq-character \ref{qq_massive}.} This problem is related to the non-commutation of $\Phi_{\vec\vac}$ with $x^+(z)$, it can be solved by considering the commutation with the operator $\CX_{\sAbox}^-(z)$ instead. However, the problem persists if both fundamental and antifundamental matter are introduced, in which case the only solution is to consider a third horizontal channel with a trivial vertical contraction as in \ref{CT_Um_af}.

The treatment of fundamental matter here is rather different from the usual brane description. In particular, we do not observe a limitation on the number of fields in this algebraic construction, which may be an effect of the presence of Chern-Simons terms. It would be advisable to achieve a deeper understanding of the precise difference between the two constructions.

Since our understanding of fundamental matter is based on gauging the flavor group, the generalization of these results to all linear quivers would require to construct arbitrary quiver theories, which is way beyond the scope of our paper. However, we hope to be able to address this issue in a near future.

\section{Perspectives}
We have proposed an algebraic method to derive qq-characters of linear quiver $\CN=1$ gauge theories with $U(m)$ gauge groups. It is based on the insertion of DIM generators in a tensored horizontal representation, symmetrized in order to define an operator commuting with the $\CT$-operator of the gauge theory. This method provides an efficient way to derive the explicit expression of the qq-characters as correlators in the gauge theory.

There are several directions in which this work can be extended. The most natural one is the treatment of DE-type quivers. In the case of D-type quivers, the brane construction of Kapustin \cite{Kapustin1998} involving an orientifold brane seems relevant. Progress along this direction will be reported elsewhere. Affine quivers could also be considered. There, the extra compact dimension seems to impose the consideration of a ring of tensor spaces in the horizontal representations in which an infinite number of quantum Weyl transformations can be applied. A much harder problem would consist in studying gauge theories with DE-type gauge groups, i.e. $Sp(m)$ or $SO(m)$ groups. The recent construction of Hayashi and Ohmori \cite{Hayashi2017} could be helpful in this context.

We hope that the generalized intertwiners introduced here will also be useful in the description of the underlying integrability, leading to a generalization of the $\CR$-matrix construction \cite{Awata2016,Awata2016b}.

%
Finally, the action of a similar quantum algebra has been observed in the context of higher spins \cite{Gaberdiel2017}, and it would be interesting to investigate the role played by these fundamental objects that are interwiners and qq-characters.

\section*{Acknowledgments}

J.-E.B. would like to thank A. Sciarappa and Joonho Kim for discussions. In the early stages of this project, he has been supported by an I.N.F.N. post-doctoral fellowship within the grant GAST, and the UniTo-SanPaolo research  grant Nr TO-Call3-2012-0088 {\it ``Modern Applications of String Theory'' (MAST)}, the ESF Network {\it ``Holographic methods for strongly coupled systems'' (HoloGrav)} (09-RNP-092 (PESC)) and the MPNS--COST Action MP1210. He also wishes to thank Tokyo University for their generous financial support during his stay.
YM is partially supported by Grants-in-Aid for Scientific Research (Kakenhi \#25400246) from MEXT, Japan. 
RZ is supported by JSPS fellowship and he is also grateful for the hospitality during his stay in KIAS. 

Part of the results of the paper were announced in the workshop "Progress in Quantum Field Theory and String Theory II" (March 27-31, 2017 Osaka City University).  We would like to thank the participants of the workshop, especially H. Awata, H. Itoyama, H. Kanno, Y. Zenkevich with whom very useful discussion was made.
\appendix

\section{Different expressions for the vertical representation}\label{AppA}
In \cite{Bourgine2016}, a different-looking vertical representation has been employed. At the level $(0,1)$, it reads \footnote{\label{fn13} Here the generators have been multiplied by a constant factor without altering the commutation relations as follows:
	\begin{equation}
	e(z)\to z^{-1}\sqrt{(1-q_3)v} e(z),\quad f(z)\to z\sqrt{(1-q_3)v}f(z),\quad \psi^\pm(z)\to(1-q_3)v\psi^\pm(z).
	\end{equation}
	The definition of the function $\Psi_\l(z)$ has also been modified in order to reflect this change of normalization.}
\begin{align}
\begin{split}
&e(z)|v,\l\rangle =z^{-1} \sum_{x\in A(\l)}\d(z/\chi_x)\L_x(\l)|v,\l+x\rangle\\
&f(z)|v,\l\rangle = \sum_{x\in R(\l)}\d(z/\chi_x)\L_x(\l)|v,\l-x\rangle,\\
&\psi^\pm(z)|v,\l\rangle=-\g_1\left[\Psi_\l(z)\right]_\pm|v,\l\rangle,\quad \g_1=(1-q_1)(1-q_2)(1-q_3),
\end{split}
\end{align}
with function $\Psi_\l(z)$ defined in \ref{def_CY}, and the coefficients being the square root of the residues $\L_x(\l)^2=\pm\res_{z=\chi_x}\Psi_\l(z)$. However, the normalization of the states can be modified by an arbitrary factor: letting $|v,\l\rangle\rangle=\CN(\l)|v,\l\rangle$, we have in general
\begin{align}
\begin{split}
&e(z)|v,\l\rangle\rangle = z^{-1}\sum_{x\in A(\l)}\d(z/\chi_x)\L_x(\l)\dfrac{\CN(\l)}{\CN(\l+x)}|v,\l+x\rangle\rangle,\\ &f(z)|v,\l\rangle\rangle = \sum_{x\in R(\l)}\d(z/\chi_x)\L_x(\l)\dfrac{\CN(\l)}{\CN(\l-x)}|v,\l-x\rangle\rangle,
\end{split}
\end{align}
Note that the action of the Cartan is not modified since they are diagonal in this basis. Choosing the normalization factor as\footnote{The recursive property is inherited from the discrete Ward identity obeyed by the vector contribution \cite{Bourgine2016},
	\begin{equation}
	\dfrac{\Zv(v,\l+x)}{\Zv(v,\l)}=\dfrac{(1-q_3)^2v}{\g_1\chi_x^2}\dfrac1{\CY_\l(\chi_xq_3^{-1})}\res_{z=\chi_x}\dfrac1{\CY_\l(z)}=\dfrac{(1-q_3)^2v}{\g_1\chi_x^2}\dfrac{\L_x(\l)^2}{\CY_\l(\chi_xq_3^{-1})^2}.
	\end{equation}}
\begin{equation}
\CN(\l)=\dfrac1{\sqrt{\Zv(v,\l)}}\prod_{x\in\l}\left(\dfrac{(1-q_3)v^{1/2}}{\g_1^{1/2}\chi_x}\right),\implies \dfrac{\CN(\l)}{\CN(\l+x)}=\dfrac{\L_x(\l)}{\CY_\l(q_3^{-1}\chi_x)},
\end{equation}
using the fact that
\begin{equation}
\res_{z=\chi_{x\in A(\l)}}\Psi_\l(z)=\CY_\l(\chi_x q_3^{-1})\res_{z=\chi_{x\in A(\l)}}\dfrac1{\CY_\l(z)},\quad\res_{z=\chi_{x\in R(\l)}}\Psi_\l(z)=\dfrac1{\CY_\l(\chi_x)}\res_{z=\chi_{x\in R(\l)}}\CY_\l(zq_3^{-1}),
\end{equation}
and the property $\L_x(\l)=\L_x(\l-x)$, the new representation can be written
\begin{align}
\begin{split}
&e(z)|v,\l\rangle\rangle = \sum_{x\in A(\l)}\d(z/\chi_x)\res_{z=\chi_x}\dfrac1{z\CY_\l(z)}|v,\l+x\rangle\rangle,\\
&f(z)|v,\l\rangle\rangle = \sum_{x\in R(\l)}\d(z/\chi_x)\CY_{\l-x}(q_3^{-1}\chi_x)|v,\l-x\rangle\rangle.
\end{split}
\end{align}
The second relation simplifies after a careful treatment of the limit $z\to\chi_x$ in the expression $\CY_{\l-x}(zq_3^{-1})=\CY_\l(zq_3^{-1})/S(q_3\chi_x/z)$,
\begin{equation}
f(z)|v,\l\rangle\rangle = \dfrac{(1-q_3)^2}{\g_1 q_3}\sum_{x\in R(\l)}\d(z/\chi_x)\res_{z=\chi_x}z^{-1}\CY_{\l}(zq_3^{-1})|v,\l-x\rangle\rangle.
\end{equation}
Finally, we notice that the coefficient in front of the commutator $[e,f]$ is different from the one in \ref{def_DIM} for $[x^+,x^-]$. In order to recover the same convention, we need to multiply
\begin{equation}
f(z)\to \dfrac{\g_1q_3}{(1-q_3)^2}f(z),\quad \psi^\pm(z)\to-\g_1^{-1}\psi^\pm(z).
\end{equation}
Under the identification of the renormalized $f(z)$ with $x^-(z)$, and $e(z)$ with $x^+(z)$, we end up with the vertical representation \ref{def_vert}. In addition, an extra cosmetic factor of $\g^{-1}$ has been added in front of $x^-(z)$ and $\psi^\pm(z)$ in order to simplify some expressions. It is important to stress that our renormalized vertical representation here does not coincide with the one used in AFS's paper in which the normalization of the intertwiners $\Phi_\l$ and $\Phi_\l^\ast$ is also different.

\section{Useful formulas for the horizontal representation}\label{AppB}
\subsection{$q$-bosons vertex operators}
The vertex operators $\eta$, $\xi$ and $\vphi^\pm$ satisfy the relations
\begin{align}
\begin{split}
&\eta(z)\eta(w)=S(w/z)^{-1}:\eta(z)\eta(w):,\quad \xi(z)\xi(w)=S(z/w)^{-1}:\xi(z)\xi(w):,\\
&\eta(z)\xi(w)=S(\g w/z):\eta(z)\xi(w):,\quad \xi(w)\eta(z)=S(\g z/w) :\eta(z)\xi(w):,\\
&\vphi^+(\g^{-1/2}z)\eta(w)=\dfrac{S(z/w)}{S(w/z)}:\vphi^+(\g^{-1/2}z)\eta(w):,\quad \eta(w)\vphi^+(\g^{-1/2}z)=:\vphi^+(\g^{-1/2}z)\eta(w):,\\
&\vphi^+(\g^{1/2}z)\xi(w)=\dfrac{S(w/z)}{S(z/w)}:\vphi^+(\g^{1/2}z)\xi(w):,\quad \xi(w)\vphi^+(\g^{1/2}z)=:\vphi^+(\g^{1/2}z)\xi(w):,\\
&\vphi^-(\g^{1/2}z)\eta(w)=:\vphi^-(\g^{1/2}z)\eta(w):,\quad \eta(w)\vphi^-(\g^{1/2}z)=\dfrac{S(w/z)}{S(z/w)}:\vphi^-(\g^{1/2}z)\eta(w):,\\
&\vphi^-(\g^{-1/2}z)\xi(w)=:\vphi^-(\g^{-1/2}z)\xi(w):,\quad \xi(w)\vphi^-(\g^{-1/2}z)=\dfrac{S(z/w)}{S(w/z)}:\vphi^-(\g^{-1/2}z)\xi(w):.\\
\end{split}
\end{align}

Explicitly, the vacuum intertwiners read
\begin{align}
\begin{split}
&\Phi_\vac(v)=\exp\left(-\sum_{k=1}^\infty\dfrac1k\dfrac1{1-q^k}v^ka_{-k}\right)\exp\left(\sum_{k=1}^\infty\dfrac1{k}\dfrac1{1-q^{-k}}v^{-k}a_k\right),\\
&\Phi^\ast_\vac(v)=\exp\left(\sum_{k=1}^\infty\dfrac1k\dfrac1{1-q^k}\g^kv^na_{-k}\right)\exp\left(-\sum_{k=1}^\infty\dfrac1{k}\dfrac1{1-q^{-k}}\g^kv^{-k}a_k\right),
\end{split}
\end{align}
they obey the relations
\begin{align}
\begin{split}
&\eta(z)\Phi_\vac(w)=\dfrac1{1-w/z}:\eta(z)\Phi_\vac(w):,\quad \Phi_\vac(w)\eta(z)=\dfrac1{1-z/(\g^2w)}:\eta(z)\Phi_\vac(w):,\\
&\xi(z)\Phi_\vac(w)=\left(1-\g w/z\right):\xi(z)\Phi_\vac(w):,\quad \Phi_\vac(w)\xi(z)=\left(1-z/(\g w)\right):\xi(z)\Phi_\vac(w):\\
&\vphi^+(\g^{-1/2}z)\Phi_\vac(w)=\dfrac{1-\g^2w/z}{1-w/z}:\vphi^+(\g^{-1/2}z)\Phi_\vac(w):,\quad \Phi_\vac(w)\vphi^+(\g^{-1/2}z)=:\vphi^+(\g^{-1/2}z)\Phi_\vac(w):,\\
&\vphi^-(\g^{1/2}z)\Phi_\vac(w)=:\vphi^-(\g^{1/2}z)\Phi_\vac(w):,\quad \Phi_\vac(w)\vphi^-(\g^{1/2}z)=\dfrac{1-z/w}{1-z/(\g^2 w)}:\vphi^-(\g^{1/2}z)\Phi_\vac(w):,\\
&\eta(z)\Phi_\vac^\ast(w)=\left(1-\g w/z\right) :\eta(z)\Phi_\vac^\ast(w):,\quad \Phi_\vac^\ast(w)\eta(z)=\left(1-z/(\g w)\right) :\eta(z)\Phi_\vac^\ast(w):,\\
&\xi(z)\Phi_\vac^\ast(w)=\dfrac1{1-\g^2w/z}:\xi(z)\Phi_\vac^\ast(w):,\quad \Phi_\vac^\ast(w)\xi(z)=\dfrac1{1-z/w} :\xi(z)\Phi_\vac^\ast(w):,\\
&\vphi^+(\g^{1/2}z)\Phi_\vac^\ast(w)=\dfrac{1-w/z}{1-\g^2w/z}:\vphi^+(\g^{1/2}z)\Phi^\ast_\vac(w):,\quad \Phi^\ast_\vac(w)\vphi^+(\g^{1/2}z)=:\vphi^+(\g^{1/2}z)\Phi^\ast_\vac(w):,\\
&\vphi^-(\g^{-1/2}z)\Phi^\ast_\vac(w)=:\vphi^-(\g^{-1/2}z)\Phi^\ast_\vac(w):,\quad \Phi^\ast_\vac(w)\vphi^-(\g^{-1/2}z)=\dfrac{1-z/(\g^2 w)}{1-z/w}:\vphi^-(\g^{-1/2}z)\Phi^\ast_\vac(w):.\\
\end{split}
\end{align}
Note also the properties
\begin{align}
\begin{split}\label{def_CG}
&\Phi_\vac(z)\Phi_\vac(w)=\CG(w/\g^2 z):\Phi_\vac(z)\Phi_\vac(w):,\quad\Phi_\vac^\ast(z)\Phi_\vac^\ast(w)=\CG(w/z):\Phi^\ast_\vac(z)\Phi_\vac^\ast(w):,\\
&\Phi_\vac(z)\Phi_\vac^\ast(w)=\CG(w/(\g z))^{-1} :\Phi_\vac(z)\Phi_\vac^\ast(w):,\quad\Phi_\vac^\ast(w) \Phi_\vac(z)=\CG(z/(\g w))^{-1} :\Phi_\vac(z)\Phi_\vac^\ast(w):,\\
&\text{with }\quad \CG(z)=\exp\left(-\sum_{k=1}^\infty\dfrac1k\dfrac{z^k}{(1-q^{k})(1-t^{-k})}\right)=\prod_{i,j=1}^\infty\left(1-zq_1^{i-1}q_2^{j-1}\right),
\end{split}
\end{align}
and the fact that
\begin{equation}\label{prop_phi}
\vphi^+(\g^{1/2}z)=:\xi(z)\eta(\g z):,\quad \vphi^-(\g^{-1/2}z)=:\xi(z)\eta(\g^{-1}z):.
\end{equation}

\subsection{Commutation relations in horizontal representations}
The simplest relations are the commutations between the operators $\psi^\pm$ and the intertwiners, they can be derived easily by combining the properties given previously and the formula \ref{def_CY}:\footnote{Operators are supposed to be radially ordered.}
\begin{align}
\begin{split}\label{braiding_psipm}
|z|>|\chi_x|:\qquad &\psi^+(\g^{-1/2}z)\Phi_{\vec\l}^{(n,m)}[u,\vec v]=\g^{-(n+m)}\Psi_{\vec\l}(z):\vphi^+(\g^{-1/2}z)\Phi_{\vec\l}^{(n,m)}[u,\vec v]:\\
&\psi^-(\g^{1/2}z)\Phi_{\vec\l}^{(n,m)}[u,\vec v]=\g^{n+m}:\vphi^-(\g^{1/2}z)\Phi_{\vec\l}^{(n,m)}[u,\vec v]:\\
&\psi^+(\g^{1/2}z)\Phi_{\vec\l}^{(n,m)\ast}[u,\vec v]=\g^{-n}\Psi_{\vec\l}(z)^{-1}:\vphi^+(\g^{1/2}z)\Phi_{\vec\l}^{(n,m)\ast}[u,\vec v]:\\
&\psi^-(\g^{-1/2}z)\Phi_{\vec\l}^{(n,m)\ast}[u,\vec v]=\g^n:\vphi^-(\g^{-1/2}z)\Phi_{\vec\l}^{(n,m)\ast}[u,\vec v]:\\
|z|<|\chi_x|:\qquad & \Phi_{\vec\l}^{(n,m)}[u,\vec v]\psi^+(\g^{-1/2}z)=\g^{-n}:\vphi^+(\g^{-1/2}z)\Phi_{\vec\l}^{(n,m)}[u,\vec v]:\\
&\Phi_{\vec\l}^{(n,m)}[u,\vec v]\psi^-(\g^{1/2}z)=\g^{n+2m}\Psi_{\vec\l}(z)^{-1}:\vphi^-(\g^{1/2}z)\Phi_{\vec\l}^{(n,m)}[u,\vec v]:\\
&\Phi_{\vec\l}^{(n,m)\ast}[u,\vec v]\psi^+(\g^{1/2}z)=\g^{-n-m}:\vphi^+(\g^{1/2}z)\Phi_{\vec\l}^{(n,m)\ast}[u,\vec v]:\\
&\Phi_{\vec\l}^{(n,m)\ast}[u,\vec v]\psi^-(\g^{-1/2}z)=\g^{n-m}\Psi_{\vec\l}(z):\vphi^-(\g^{-1/2}z)\Phi_{\vec\l}^{(n,m)\ast}[u,\vec v]:\\
\end{split}
\end{align}
In these expressions, the representation $(1,n+m)_{u'}$ of the DIM generator is understood (but omitted) on the left of the operator $\Phi_{\vec\l}^{(n,m)}$, while the representation on the right is $(1,n)_u$. The two representations are exchanged for the dual operator: $(1,n)_u$ is on the left of $\Phi_{\vec\l}^{(n,m)\ast}$ while $(1,n+m)_{u'}$ is on the right. Similar expressions can be derived for $x^\pm$:
\begin{align}
\begin{split}\label{braiding_xpm}
|z|>|\chi_x|:\qquad & x^+(z)\Phi_{\vec\l}^{(n,m)}[u,\vec v]=\dfrac{\g^{n+m}u'}{z^{n+m}\CY_{\vec\l}(z)}\ :\eta(z)\Phi_{\vec\l}^{(n,m)}[u,\vec v]:\\
& x^-(z)\Phi_{\vec\l}^{(n,m)}[u,\vec v] = \dfrac{z^{n+m}}{u'\g^{n+m}}\CY_{\vec\l}(z\g^{-1}) :\xi(z)\Phi_{\vec\l}^{(n,m)}[u,\vec v]:\\
& x^+(z)\Phi_{\vec\l}^{(n,m)\ast}[u,\vec v] = u\g^n z^{-n}\CY_{\vec\l}(z\g^{-1}):\eta(z)\Phi_{\vec\l}^{(n,m)\ast}[u,\vec v]:\\
& x^-(z)\Phi_{\vec\l}^{(n,m)\ast}[u,\vec v]= \dfrac{z^n}{u\g^n\CY_{\vec\l}(zq_3^{-1})}:\xi(z)\Phi_{\vec\l}^{(n,m)\ast}[u,\vec v]:\\
|z|<|\chi_x|:\qquad & \Phi_{\vec\l}^{(n,m)}[u,\vec v] x^+(z) = \g^m\Psi_{\vec\l}(z)^{-1}\dfrac{u'\g^{n+m} }{z^{n+m}\CY_{\vec\l}(z)}\ :\eta(z)\Phi_{\vec\l}^{(n,m)}[u,\vec v]:\\
& \Phi_{\vec\l}^{(n,m)}[u,\vec v] x^-(z) = \dfrac{z^{n+m}}{u'\g^{n+m} }\CY_{\vec\l}(z\g^{-1}) :\xi(z)\Phi_{\vec\l}^{(n,m)}[u,\vec v]:\\
& \Phi_{\vec\l}^{(n,m)\ast}[u,\vec v] x^+(z) = u\g^nz^{-n}\CY_{\vec\l}(z\g^{-1}):\eta(z)\Phi_{\vec\l}^{(n,m)\ast}[u,\vec v]:\\
& \Phi_{\vec\l}^{(n,m)\ast}[u,\vec v]x^-(z) = \g^{-m} \Psi_{\vec\l}(z)\ \dfrac{z^n}{u\g^n\CY_\l(zq_3^{-1})}:\xi(z)\Phi_{\vec\l}^{(n,m)\ast}[u,\vec v]:
\end{split}
\end{align}

\section{Derivation of the AFS lemmas}\label{AppC}
The proof of the relations involving $\psi^\pm(z)$ is a matter of writing the commutation relations \ref{braiding_psipm}. Hence the focus here is on the generators $x^\pm(z)$. We first examine the product of $x^+(z)$ and $\Phi_{\vec\l}^{(n,m)}[u,\vec v]$, the proof is based on the following decomposition for the function
\begin{equation}
\dfrac1{z\CY_{\vec\l}(z)}=\sum_{x\in A(\vec\l)}\dfrac1{z-\chi_x}\res_{z=\chi_x}\dfrac1{z\CY_{\vec\l}(z)}.
\end{equation}
As a consequence, we can write the right product of $x^+(z)$ on $\Phi_\l$ in \ref{braiding_xpm} as
\begin{align}
\begin{split}
x^+(z) \Phi_{\vec\l}^{(n,m)}[u,\vec v] =& u'\g^{n+m}z^{-n-m+1}\sum_{x\in A(\vec\l)}\dfrac1{z-\chi_x}\res_{z=\chi_x}\dfrac1{z\CY_{\vec\l}(z)}\ :\eta(z)\Phi_{\vec\l}^{(n,m)}[u,\vec v]:\\
=& u'\g^{n+m}z^{-n-m+1}\sum_{x\in A(\vec\l)}\dfrac1{z-\chi_x}\res_{z=\chi_x}\dfrac1{z\CY_{\vec\l}(z)}\ :\eta(\chi_x)\Phi_{\vec\l}^{(n,m)}[u,\vec v]:\\
&+ u'\g^{n+m}z^{-n-m+1}\sum_{x\in A(\vec\l)}\dfrac{:\eta(z)\Phi_{\vec\l}^{(n,m)}[u,\vec v]:-:\eta(\chi_x)\Phi_{\vec\l}^{(n,m)}[u,\vec v]:}{z-\chi_x}\res_{z=\chi_x}\dfrac1{z\CY_{\vec\l}(z)}.
\end{split}
\end{align}
This expression is valid for $|z|>|\chi_x|$, however the second line of the last equality has no pole at $z=\chi_x$ and can be analytically continued to $|z|<|\chi_x|$. This is not true for the first line, and the fraction should be expanded in positive powers of $z$. A similar expression can be obtained for $|z|<|\chi_x|$ by considering the left product of $x^+(z)$ on $\Phi_\l$:
\begin{align}
\begin{split}
\g^{-m}\Psi_{\vec\l}(z) \Phi_{\vec\l}^{(n,m)}[u,\vec v] x^+(z) =& u'\g^{n+m}z^{-n-m+1}\sum_{x\in A(\vec\l)}\dfrac1{z-\chi_x}\res_{z=\chi_x}\dfrac1{z\CY_{\vec\l}(z)}\ :\eta(z)\Phi_{\vec\l}^{(n,m)}[u,\vec v]:\\
=&u'\g^{n+m}z^{-n-m+1} \sum_{x\in A(\vec\l)}\dfrac1{z-\chi_x}\res_{z=\chi_x}\dfrac1{z\CY_{\vec\l}(z)}\ :\eta(\chi_x)\Phi_{\vec\l}^{(n,m)}[u,\vec v]:\\
&+ u'\g^{n+m}z^{-n-m+1}\sum_{x\in A(\vec\l)}\dfrac{:\eta(z)\Phi_{\vec\l}^{(n,m)}[u,\vec v]:-:\eta(\chi_x)\Phi_{\vec\l}^{(n,m)}[u,\vec v]:}{z-\chi_x}\res_{z=\chi_x}\dfrac1{z\CY_{\vec\l}(z)}.
\end{split}
\end{align}
Taking the difference of the two, the terms with no singularity cancel each-other. The remaining expression is a difference of expansions in powers of $z$ and $z^{-1}$ that forms a delta function,
\begin{equation}
x^+(z)\Phi_{\vec\l}^{(n,m)}[u,\vec v]-\g^{-m}\Psi_{\vec\l}(z)\Phi_{\vec\l}^{(n,m)}[u,\vec v]x^+(z)= u'\g^{n+m}z^{-n-m}\sum_{x\in A(\vec\l)}\d(z/\chi_x)\res_{z=\chi_x}\dfrac1{z\CY_{\vec\l}(z)}\ :\eta(\chi_x)\Phi_{\vec\l}^{(n,m)}[u,\vec v]:
\end{equation}
Then, since $\Phi_{\vec\l}^{(n,m)}$ is built as a product of operators $\eta(\chi_x)$ for all $x\in\l$, the vertex operator $:\eta(\chi_x)\Phi_{\vec\l}^{(n,m)}:$ can be written as $\Phi^{(n,m)}_{\vec\l+x}$. Taking into account the prefactor
\begin{equation}\label{prop_t}
\dfrac{t_n(\vec \l,u,v)}{t_n(\vec \l+x,u,v)}=\dfrac{\chi_x^{n+1}}{u'\g^{n+1}},
\end{equation}
we recover the AFS lemma in the form \ref{AFS_Phi}.

A similar argument can be employed to treat the action of $x^-(z)$, with the poles located at the points $z=\g^{-1}\chi_{x\in R(\vec\l)}$, and the operator :$\xi(z)\Phi_{\vec\l}^{(n,m)}$: simplified using the property \ref{prop_phi} of the appendix \refOld{AppB}.\footnote{The following property is useful here,
\begin{equation}\label{prop_res}
\res_{z=\g^{-1}\a}f(z)=\g^{-1}\res_{z=\a}f(z\g^{-1}).
\end{equation}} However, in this case, a more elegant proof is also possible. It is based on the formula for the commutation relation between the modes $x_k^-$ and $\Phi_\l^{(n)}$ that can be found in \cite{DiFrancesco1997} (formula (6.15)). By definition, we have
\begin{equation}
x_k^-=\oint_0\dfrac{dz}{2i\pi}z^{k-1}x^-(z)
\end{equation}
so that
\begin{align}
\begin{split}
[x_k^-,\Phi_{\vec\l}^{(n,m)}[u,\vec v]]&=\oint_{\superp{z=0}{|z|>|\chi_x|}}\dfrac{dz}{2i\pi}z^{k-1}x^-(z)\Phi_{\vec\l}^{(n,m)}[u,\vec v]-\oint_{\superp{z=0}{|z|<|\chi_x|}}\dfrac{dz}{2i\pi}z^{k-1}\Phi_{\vec\l}^{(n,m)}[u,\vec v]x^-(z)\\
&=\sum_{x\in R(\vec\l)}\oint_{\chi_x\g^{-1}}\dfrac{dz}{2i\pi}\dfrac{z^{k+n+m-1}}{u'\g^{n+m}}\CY_{\vec\l}(z\g^{-1})\ :\xi(z)\Phi_{\vec\l}^{(n,m)}[u,\vec v]:
\end{split}
\end{align}
The second equality is the consequence of several cancellations between poles, such that only the poles of $\CY_{\vec\l}(z\g^{-1})$ will contribute. The expression for the product of operators is taken from \ref{braiding_xpm}. The contour integral can be reduced to the residue contributions of the integrand, which simplifies thanks to the properties \ref{prop_phi} and \ref{prop_res} to give
\begin{equation}
[x_k^-,\Phi_{\vec\l}^{(n,m)}[u,\vec v]]=\sum_{x\in R(\vec\l)}\chi_x^{k+m-2}\g^{-2m-k+1}\res_{z=\chi_x}\CY_{\vec\l}(zq_3^{-1})\ \Phi_{\vec\l-x}^{(n,m)}[u,\vec v]\psi^+(\g^{-1/2}\chi_x).
\end{equation}
Summing over the index $k$ with the spectral parameter at the power $z^{-k}$, we recover the AFS lemma \ref{AFS_Phi}. This short computation gives some insight on the interpretation of the AFS lemma: it is valid for each power of $z$ in a formal expansion.

\section{Connection with quiver W-algebras}\label{App_KP}
In \cite{Kimura2015,Kimura2016a}, Kimura and Pestun have introduced quantum W-algebras based on the Dynkin diagram $\G$ of simple Lie algebras of ADE type. These algebras are constructed upon a set of q-bosonic modes $s_k^{(i)}$ with $k\in\mathbb{Z}$ and $i\in\G$ that obey the commutation relations
\begin{equation}\label{com_sk}
[s_k^{(i)},s_{-k'}^{(i')}]=-\dfrac1k\dfrac{1-q^k}{1-t^k}\d_{k,k'}c_{ii'}^{[k]},\quad k>0,
\end{equation} 
where $c_{ii'}^{[k]}$ denotes the $k$-th Adams operation applied to the mass-deformed Cartan matrix. For instance, in the case of the $A_3$ quiver with bifundamental masses $\mu_{ii'}=\g^{-1}$, this matrix reads
\begin{equation}
c_{ii'}^{[k]}=\left(
\begin{array}{ccc}
1+q_3^k & -\g^k & 0\\
-\g^k & 1+q_3^k & -\g^k\\
0 & -\g^k & 1+q_3^k\\
\end{array}
\right).
\end{equation} 
Since this algebra is also acting on Nekrasov partition functions, it should be related to the DIM algebra considered in our paper. The aim of this appendix is to highlight this connection. It is based on the decomposition of the tensor product of two $(1,0)$ DIM representations into q-Heisenberg$\oplus$q-Virasoro algebras. This decomposition has been described by Mironov, Morozov and Zenkevich in \cite{Mironov2016}, and this appendix is just a reformulation of their results in our notations.

In order to simplify the discussion, we will neglect the role of zero modes, $\Phi_{\vac}$,... We will also restrict ourselves to $U(1)$ gauge groups at each node of the quiver diagram. It is an easy exercise to extend the argument to more general cases. We first focus on the $A_1$ quiver for which the $\CT$-operator is built as a vertical contraction of two intertwiners, $\CT_{U(1)}=\tr \Phi^\ast\otimes\Phi$. Since two horizontal spaces are involved, we need two copies of the q-bosonic modes in order to represent the horizontal action of the intertwiner and its dual. We denote these modes $a_k^{(i)}$ with $i=1,2$. By definition, modes with a different value of the label $i$ commute, while modes with the same label obey the commutation relation \ref{com_an}:
\begin{equation}
[a_k^{(i)},a_{-k'}^{(i')}]=k\dfrac{1-q^k}{1-t^k}\d_{k,k'}\d_{i,i'},\quad k>0.
\end{equation}
The operator $\CT$ involves a trace over Young diagram realizations of a product over the box content of the diagram. Each factor contains the following operator evaluated at $z=\chi_x$ for some $x\in\l$,
\begin{equation}\label{op_xi_eta}
\xi(z)\otimes\eta(z)=\exp\left(-\sum_{k=1}^\infty\dfrac{1-t^{-k}}{k}z^k\left(a_{-k}^{(1)}\g^k-a_{-k}^{(2)}\right)\right)\exp\left(\sum_{k=1}^\infty\dfrac{1-t^{k}}{k}z^{-k}\left(a_{k}^{(1)}\g^k-a_{k}^{(2)}\right)\right).
\end{equation} 
It leads to identify the modes $ks_k=-\g^{|k|}a_k^{(1)}+a_k^{(2)}$. They indeed obey the commutation relation \ref{com_sk} with the deformed $A_1$ Cartan matrix $c^{[k]}=1+q_3^k$. As a result, the operator \ref{op_xi_eta} can be expressed in terms of the screening operator defined in \cite{Kimura2015},
\begin{equation}
\xi(z)\otimes\eta(z)\simeq:S(z)^{-1}S(q_2z):,\quad S(z)=:\exp\left(\sum_{k\in\mathbb{Z}}z^ks_{-k}\right):.
\end{equation} 
Taking the product over the boxes $x\in\l$, several cancellations occur, and the final result is expressed in terms of a product over each column $i$ of height $\l_i$,
\begin{equation}
:\prod_{x\in\l}\xi(\chi_x)\otimes\eta(\chi_x):\ \simeq\ :\prod_i S(vq_1^{i-1}q_2^{\l_i}):
\end{equation} 
where we have neglected the boundary terms $S(vq_1^{i-1})$ that can be taken care of using zero modes. In the RHS, the product is taken over the elements of the set $\CX$ defined in \cite{Kimura2015}, and we can formally identify the state $|Z_T\rangle$ representing the partition function with the action of $\CT_{U(m)}$ over the horizontal vacuum states:
\begin{equation}
|Z_T\rangle\simeq\sum_\l :\prod_i S(vq_1^{i-1}q_2^{\l_i}): |\vac\rangle\simeq\CT_{U(1)}\left(|\vac\rangle\otimes|\vac\rangle\right).
\end{equation} 

The modes $s_k$ can be used to build the stress-energy tensor of the q-Virasoro algebra. The orthogonal combination $kb_k=a_k^{(1)}+\g^{|k|}a_k^{(2)}$, which by definition commutes with $s_k$, obeys the q-bosonic commutation relation
\begin{equation}
[b_k,b_{-k'}]=-\dfrac1k\dfrac{1-q^k}{1-t^k}\d_{k,k'}(1+q_3^{k})\quad k>0.
\end{equation} 
Thus, we have obtained the formal decomposition
\begin{equation}
(1,0)\otimes(1,0)=\text{q-Heisenberg}\oplus\text{q-Virasoro}.
\end{equation} 

It is also interesting to rewrite the coproduct of $x^+(z)$ in terms of the modes $b_k,s_k$:
\begin{equation}
:\D(x^+(z)):\ \simeq\ :\exp\left(-\sum_{k\in\mathbb{Z}}\dfrac{1-t^k}{1+q_3^{|k|}}z^{-k}b_k\right):\left[Y(z\g^{-1})+:Y(z\g)^{-1}:\right]
\end{equation} 
where, following Kimura and Pestun, we have introduced the operator
\begin{equation}
Y(z)=:\exp\left(\sum_{k\in\mathbb{Z}}z^{-k}y_k\right):,\quad y_k=\dfrac{1-t^k}{1+q_3^k}s_k.
\end{equation} 
Hence, up to a $U(1)$ factor, we recover in $\D(x^+(z))$ the operator $T$ of Kimura and Pestun, identified with the fundamental current (stress-energy tensor) of q-Virasoro \cite{Shiraishi1995}.\footnote{Note that we have chosen to denote the Q-operator of Kimura and Pestun as $\CT$ since the partition function is obtained as the vev of this operator. On the other hand, their T-operator has been denoted $\CX^\pm$ to emphasize the fact that it comes from the generators $x^\pm$ of the DIM algebra, and that the TQ-relation only holds if we forget about the difference of representations.}

For a general linear quiver diagram $A_r$, the modes $s_k^{(i)}$ $i=1\cdots r$ are associated to the nodes of the diagram. On the other hand, the $\CT$-operator is written as an $(r+1)$th tensorial product
\begin{align}
\begin{split}
\CT_{U(m_r)\times\cdots\times U(m_1)}&=\tr_{12\cdots r}\Phi_1^\ast\otimes\Phi_1\Phi_2^\ast\otimes\cdots\otimes\Phi_{r-1}\Phi_r^\ast\otimes\Phi_r\\
&\simeq\sum_{\l^{(1)},\cdots, \l^{(r)}}\left(\prod_{x\in\l^{(1)}}\xi(\chi_x)\otimes\eta(\chi_x)\right)\circ\left(\prod_{x\in\l^{(2)}}\xi(\chi_x)\otimes\eta(\chi_x)\right)\circ\cdots\circ \left(\prod_{x\in\l^{(r)}}\xi(\chi_x)\otimes\eta(\chi_x)\right).
\end{split}
\end{align}
In this expression, a different set of modes $a_k^{(i)}$ is attached to each tensor space, with $i=1\cdots r+1$. It leads to identify the modes as follows:
\begin{equation}
ks_k^{(i)}=-\g^{|k|}a_k^{(i)}+a_k^{(i+1)},\quad kb_k=\sum_{i=1}^{r+1}\g^{|k|i}a_k^{(i)}.
\end{equation} 
Under this identification, the modes $s_k^{(i)}$ reproduce the commutation relation \ref{com_sk} with the deformed Cartan matrix of the $A_r$ Dynkin diagram. In addition, they all commute with the modes $b_k$. Thus, for a general linear quiver, we have the formal decomposition
\begin{equation}
(1,0)^{\otimes(r+1)}=\text{q-Heisenberg}\oplus W_r.
\end{equation}

\section{Derivation of the qq-characters for the $A_3$ quiver}\label{App_A3}
\begin{figure}
	\begin{center}
		\begin{tikzpicture}
		\draw (8, 2) -- (9, 2) -- (9,0) -- (11,0) -- (11,-2) -- (13,-2) -- (13,-4) -- (14,-4);
		\draw (11,0) -- (11.7,0.7);
		\draw (9, 2) -- (9.7, 2.7);
		\draw (8.3, -0.7) -- (9, 0); 
		\draw (10.3,-2.7) -- (11,-2);
		\draw (13,-2) -- (13.7,-1.3);
		\draw (12.3,-4.7) -- (13,-4);
		\node[above right,scale=0.7] at (11.7,0.7) {$(1,n+m)_{u'}$};
		\node[above right,scale=0.7] at (9.7,2.7) {$(1,n_3+m_3)_{u'_3}$};
		\node[above right,scale=0.7] at (13.7,-1.3) {$(1,n_1+m_1)_{u'_1}$};
		\node[below left,scale=0.7] at (10.3,-2.7) {$(1,n^\ast+m)_{u'^\ast}$};
		\node[below left,scale=0.7] at (8.3,-0.7) {$(1,n_3^\ast+m_3)_{u'^\ast_3}$};
		\node[below left,scale=0.7] at (12.3,-4.7) {$(1,n_1^\ast+m_1)_{u'^\ast_1}$};
		\node[left,scale=0.7] at (10.7,0.3) {$(1,n_3^\ast)_{u_3^\ast}$};
		\node[right,scale=0.7] at (14,-4) {$(1,n_1^\ast)_{u_1^\ast}$};
		\node[left,scale=0.7] at (8.3,2) {$(1,n_3)_{u_3}$};
		\node[above,scale=0.7] at (12,-2) {$(1,n^\ast)_{u^\ast}$};
		\node[right,scale=0.7] at (11,-1) {$(0,m)_{\vec v}$};
		\node[right,scale=0.7] at (9,1) {$(0,m_3)_{\vec v_3}$};
		\node[right,scale=0.7] at (13,-3) {$(0,m_1)_{\vec v_1}$};
		\node[right,scale=0.7] at (11,0) {$\Phi^{(n,m)}[u,\vec v]$};
		\node[right,scale=0.7] at (9,2) {$\Phi^{(n_3,m_3)}[u_3,\vec v_3]$};
		\node[left,scale=0.7] at (11,-2) {$\Phi^{(n^\ast,m)\ast}[u^\ast,\vec v]$};
		\node[left,scale=0.7] at (9,0) {$\Phi^{(n_3^\ast,m_3)\ast}[u_3^\ast,\vec v_3]$};
		\node[right,scale=0.7] at (13,-2) {$\Phi^{(n_1,m_1)}[u_1,\vec v_1]$};
		\node[left,scale=0.7] at (13,-4) {$\Phi^{(n_1^\ast,m_1)\ast}[u_1^\ast,\vec v_1]$};
		\end{tikzpicture}
		\caption{Representation web of the $A_3$ quiver}
		\label{fig_A3}
	\end{center}
\end{figure}
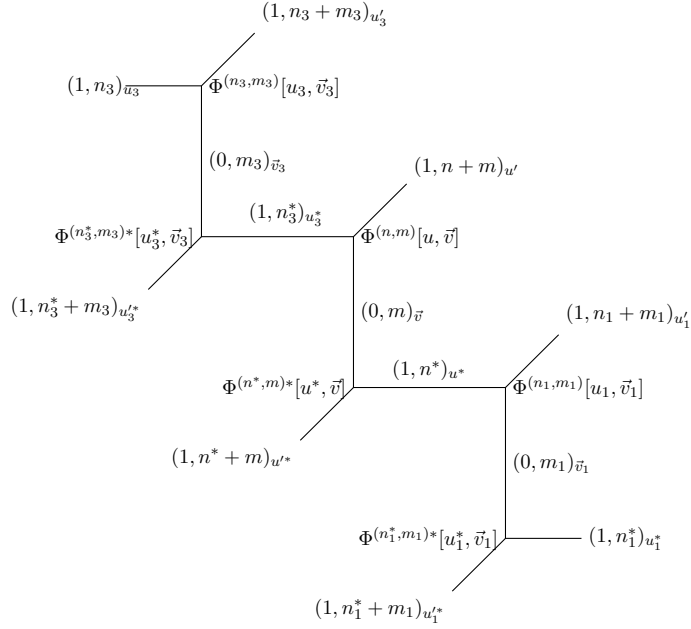

The qq-characters associated to the three nodes of the $A_3$ quiver are labeled by the Young diagrams $\Abox$, $\Bbox$ and $\Cbox$. They can be constructed from the following operators:
\begin{align}
\begin{split}
\D^{\sAbox}(x^+(z))=&x_{:0}^+\otimes1\otimes1\otimes1+\psi^-_{:1/2}\otimes x_{:1}^+\otimes 1\otimes 1+\psi^-_{:1/2}\otimes\psi^-_{:3/2}\otimes x_{:2}^+\otimes 1\\
&+\psi^-_{:1/2}\otimes\psi^-_{:3/2}\otimes\psi^-_{:5/2}\otimes x_{:3}^+\\
\D^{\sBbox}(x^+(z))=&\psi^-_{:-1/2}x_{:1}^+\otimes x_{:0}^+\otimes1\otimes1+\psi^-_{:-1/2}x_{:1}^+\otimes\psi^-_{:1/2} \otimes x_{:1}^+\otimes1+\psi^-_{:-1/2}\psi^-_{:3/2}\otimes\psi^-_{:1/2}x_{:2}^+\otimes x_{:1}^+\otimes1\\
&+\psi^-_{:-1/2}x_{:1}^+\otimes\psi^-_{:1/2} \otimes \psi^-_{:3/2} \otimes x_{:2}^++\psi^-_{:-1/2}\psi_{:3/2}^-\otimes\psi^-_{:1/2} x_{:2}^+\otimes \psi^-_{:3/2} \otimes x_{:2}^+\\
&+\psi^-_{:-1/2}\psi_{:3/2}^-\otimes\psi^-_{:1/2} \psi^-_{:5/2}\otimes \psi^-_{:3/2}x_{:3}^+ \otimes x_{:2}^+\\
\D^{\sCbox}(x^+(z))=&\psi^-_{:-3/2}\psi^-_{:1/2}x_{:2}^+\otimes\psi^-_{:-1/2}x_{:1}^+\otimes x_{:0}^+\otimes1+\psi^-_{:-3/2}\psi^-_{:1/2}x_{:2}^+\otimes\psi^-_{:-1/2}x_{:1}^+\otimes \psi_{:1/2}^-\otimes x_{:1}^+\\
&+\psi^-_{:-3/2}\psi^-_{:1/2}x_{:2}^+\otimes\psi^-_{:-1/2}\psi_{:3/2}^-\otimes \psi_{:1/2}^-x_{:2}^+\otimes x_{:1}^+\\
&+\psi^-_{:-3/2}\psi^-_{:1/2}\psi^-_{5/2}\otimes\psi^-_{:-1/2}\psi_{:3/2}^-x_{:3}^+\otimes \psi_{:1/2}^-x_{:2}^+\otimes x_{:1}^+
\end{split}
\end{align}
where we have introduced the shortcut notations $x_{:k}^+=x^+(\g^k z)$, $\psi^-_{:k}=\psi^-(\g^k z)$. Note that the argument of operators has been simplified taking advantage of the fact that they act in the horizontal representations where $\hg$ becomes $\g$. After a long and tedious computation, it is possible to show that these expressions do commute with the operator $\CT_{U(m_3)\times U(m_2)\times U(m_1)}$. In practice, we have used a short program in Python to perform the algebraic manipulations.

Defining the qq-characters as
\begin{align}
\begin{split}
&\chi_{\sAbox}^+(z)=\dfrac{\nu_1}{u_1^\ast\g^{2n_1^\ast}}z^{n_1^\ast+m_1}\dfrac{\la\D^{\sAbox}(x^+(z\g^{-1}))\CT_{U(m_3)\times U(m_2)\times U(m_1)}\ra}{\la\CT_{U(m_3)\times U(m_2)\times U(m_1)}\ra}\\
&\chi_{\sBbox}^+(z)=\dfrac{\nu_2}{u_1'u_1^\ast\g^{2n_1^\ast+2n_1+2m_1}}z^{n_1^\ast+n_1+m_1+m_2}\dfrac{\la\D^{\sBbox}(x^+(z\g^{-1}))\CT_{U(m_3)\times U(m_2)\times U(m_1)}\ra}{\la\CT_{U(m_3)\times U(m_2)\times U(m_1)}\ra}\\
&\chi_{\sCbox}^+(z)=\dfrac{\nu_3}{u_1'u_1^\ast u_2'\g^{2n_1^\ast+2n_1+2m_1+2n_2+2m_2}}z^{n_1^\ast+n_1+n_2+m_1+m_2+m_3}\dfrac{\la\D^{\sCbox}(x^+(z\g^{-1}))\CT_{U(m_3)\times U(m_2)\times U(m_1)}\ra}{\la\CT_{U(m_3)\times U(m_2)\times U(m_1)}\ra},
\end{split}
\end{align}
we find the following expressions after evaluation in the four independent Fock spaces,
\begin{align}
\begin{split}
\chi_{\sAbox}^+(z)=&\Bigg\langle\nu_1 z^{m_1}\CY_{\vec\l_1}(z\g^{-2})+\qf_1z^{\k_1}\dfrac{\CY_{\vec\l_2}(z\g^{-1})}{\CY_{\vec\l_1}(z)}+\qf_1\qf_2\dfrac{\nu_1}{\nu_2}\g^{\k_2+2m_1-m_2}z^{\k_1+\k_2+m_1-m_2}\dfrac{\CY_{\vec\l_3}(z)}{\CY_{\vec\l_2}(z\g)}\\
&+\qf_1\qf_2\qf_3\dfrac{\nu_1}{\nu_3}\g^{\k_2+2\k_3+2m_1+2m_2-2m_3}\dfrac{z^{\k_1+\k_2+\k_3+m_1-m_3}}{\CY_{\vec\l_3}(z\g^2)}\Bigg\rangle_\text{gauge}\\
\chi_{\sBbox}^+(z)=&\Bigg\langle\nu_2z^{m_2}\CY_{\vec\l_2}(z\g^{-2})+\qf_2\nu_1\g^{m_1}z^{\k_2+m_1}\dfrac{\CY_{\vec\l_1}(z\g^{-1})\CY_{\vec\l_3}(z\g^{-1})}{\CY_{\vec\l_2}(z)}+\qf_1\qf_2\g^{\k_1}z^{\k_1+\k_2}\dfrac{\CY_{\vec\l_3}(z\g^{-1})}{\CY_{\vec\l_1}(z\g)}\\
&+\qf_2\qf_3\dfrac{\nu_1\nu_2}{\nu_3}\g^{-\k_2+m_2}(z\g)^{\k_2+\k_3+m_1+m_2-m_3}\dfrac{\CY_{\vec\l_1}(z\g^{-1})}{\CY_{\vec\l_3}(z\g)}\\
&+\qf_1\qf_2\qf_3\dfrac{\nu_2}{\nu_3}\g^{-\k_2+m_2}(z\g)^{\k_1+\k_2+\k_3+m_2-m_3}\dfrac{\CY_{\vec\l_2}(z)}{\CY_{\vec\l_1}(z\g)\CY_{\vec\l_3}(z\g)}+\qf_1\qf_2^2\qf_3\dfrac{\nu_1}{\nu_3}\g^{2m_1}\dfrac{(z\g)^{\k_1+2\k_2+\k_3+m_1-m_3}}{\CY_{\vec\l_2}(z\g^2)}\Bigg\rangle_{\text{gauge}}\\
\chi_{\sCbox}^+(z)=&\Bigg\langle\nu_3z^{m_3}\CY_{\vec\l_3}(z\g^{-2})+\qf_3\n_2\g^{m_2}z^{\k_3+m_2}\dfrac{\CY_{\vec\l_2}(z\g^{-1})}{\CY_{\vec\l_3}(z)}\\
&+\qf_2\qf_3\nu_1\g^{\k_2+2m_1}z^{\k_2+\k_3+m_1}\dfrac{\CY_{\vec\l_1}(z)}{\CY_{\vec\l_2}(z\g)}+\qf_1\qf_2\qf_3\g^{2\k_1+\k_2}\dfrac{z^{\k_1+\k_2+\k_3}}{\CY_{\vec\l_1}(z\g^2)}\Bigg\rangle_\text{gauge}.
\end{split}
\end{align}

\bibliographystyle{unsrt}

\begin{thebibliography}{10}

\bibitem{Hanany1996}
A.~Hanany and E.~Witten.
\newblock {Type IIB superstrings, BPS monopoles, and three-dimensional gauge
  dynamics}.
\newblock {\em Nucl. Phys.}, B492:152--190, 1997.

\bibitem{Aharony1997}
O.~Aharony and A.~Hanany.
\newblock {Branes, superpotentials and superconformal fixed points}.
\newblock {\em Nucl. Phys.}, B504:239--271, 1997.

\bibitem{Aharony1997a}
O.~Aharony, A.~Hanany, and B.~Kol.
\newblock {Webs of (p,q) five-branes, five-dimensional field theories and grid
  diagrams}.
\newblock {\em JHEP}, 01:002, 1998.

\bibitem{aganagic2005topological}
M.~Aganagic, A.~Klemm, M.~Marino, and C.~Vafa.
\newblock {The topological vertex}.
\newblock {\em Communications in mathematical physics}, 254(2):425--478, 2005.

\bibitem{Leung1997}
N.~C. Leung and C.~Vafa.
\newblock {Branes and toric geometry}.
\newblock {\em Adv. Theor. Math. Phys.}, 2:91--118, 1998.

\bibitem{Nekrasov2004}
N.~Nekrasov.
\newblock {Seiberg-Witten prepotential from instanton counting}.
\newblock {\em Adv.Theor.Math.Phys.}, 7:831--864, 2004.

\bibitem{iqbal2009refined}
A.~Iqbal, C.~Kozcaz, and C.~Vafa.
\newblock {The refined topological vertex}.
\newblock {\em Journal of High Energy Physics}, 2009(10):069, 2009.

\bibitem{Morozov2015}
A.~Morozov and Y.~Zenkevich.
\newblock {Decomposing Nekrasov Decomposition}.
\newblock {\em JHEP}, 02:098, 2016.

\bibitem{Mironov2016a}
A.~Mironov, A.~Morozov, and Y.~Zenkevich.
\newblock {Spectral duality in elliptic systems, six-dimensional gauge theories
  and topological strings}.
\newblock {\em JHEP}, 05:121, 2016.

\bibitem{Bourgine2015c}
J.-E. Bourgine, Y.~Matsuo, and Hong Zhang.
\newblock {Holomorphic field realization of SH$^c$ and quantum geometry of
  quiver gauge theories}.
\newblock {\em JHEP}, 04:167, 2016.

\bibitem{Bourgine2016}
J.-E. Bourgine, M.~Fukuda, Y.~Matsuo, Hong Zhang, and Rui-Dong Zhu.
\newblock {Coherent states in quantum $\mathcal{W}_{1+\infty}$ algebra and
  qq-character for 5d Super Yang-Mills}.
\newblock {\em PTEP}, 2016(12):123B05, 2016.

\bibitem{SHc}
O.~Schiffmann and E.~Vasserot.
\newblock {Cherednik algebras, W-algebras and the equivariant cohomology of the
  moduli space of instantons on A 2}.
\newblock {\em Publications math{\'e}matiques de l'IH{\'E}S}, 118(1):213--342,
  2013.

\bibitem{Ding1997}
J.~Ding and K.~Iohara.
\newblock {Generalization of Drinfeld Quantum Affine Algebras}.
\newblock {\em Letters in Mathematical Physics}, 41(2):181--193, 1997.

\bibitem{Miki2007}
Kei Miki.
\newblock {A (q, $\gamma$) analog of the $W_{1+\infty}$ algebra}.
\newblock {\em Journal of Mathematical Physics}, 48(12):3520, 2007.

\bibitem{Feigin2009}
B.~{Feigin} and A.~{Tsymbaliuk}.
\newblock {Heisenberg action in the equivariant K-theory of Hilbert schemes via
  Shuffle Algebra}.
\newblock {\em ArXiv e-prints}, April 2009.

\bibitem{feigin2011quantum}
B.~Feigin, E.~Feigin, M.~Jimbo, T.~Miwa, and E.~Mukhin.
\newblock {Quantum continuous $\mathfrak{gl}_\infty$ : Semi-infinite
  construction of representations}.
\newblock {\em Kyoto Journal of Mathematics}, 51(2):337--364, 2011.

\bibitem{Feigin:2010qea}
B.~Feigin, E.~Feigin, M.~Jimbo, T.~Miwa, and E.~Mukhin.
\newblock {Quantum continuous $gl_\infty$: Tensor products of Fock modules and
  $W_n$ characters}.
\newblock 2010.

\bibitem{fukuda2015sh}
M.~Fukuda, S.~Nakamura, Y.~Matsuo, and Rui-Dong Zhu.
\newblock {SH$^c$ Realization of Minimal Model CFT: Triality, Poset and Burge
  Condition}.
\newblock {\em JHEP}, 2015(11):168, 2015.

\bibitem{Maulik2012}
D.~Maulik and A.~Okounkov.
\newblock {Quantum Groups and Quantum Cohomology}, 2012.

\bibitem{Smirnov2013}
A.~Smirnov.
\newblock {On the Instanton R-matrix}, 2013.

\bibitem{Awata2016}
H.~Awata, H.~Kanno, A.~Mironov, Al. Morozov, A.~Morozov, Y.~Ohkubo, and
  Y.~Zenkevich.
\newblock {Toric Calabi-Yau threefolds as quantum integrable systems. R-matrix
  and RTT relations}.
\newblock 2016.

\bibitem{Awata2016b}
H.~Awata, H.~Kanno, A.~Mironov, Al. Morozov, A.~Morozov, Y.~Ohkubo, and
  Y.~Zenkevich.
\newblock {Anomaly in RTT relation for DIM algebra and network matrix models}.
\newblock 2016.

\bibitem{NPS}
N.~Nekrasov, V.~Pestun, and S.~Shatashvili.
\newblock {Quantum geometry and quiver gauge theories}.
\newblock 2013.

\bibitem{Nekrasov2015}
N.~Nekrasov.
\newblock {BPS/CFT correspondence: non-perturbative Dyson-Schwinger equations
  and qq-characters}.

\bibitem{Nekrasov2016}
N.~Nekrasov.
\newblock {BPS/CFT correspondence II: Instantons at crossroads, Moduli and
  Compactness Theorem}.
\newblock 2016.

\bibitem{Nekrasov2016b}
N.~Nekrasov.
\newblock {BPS/CFT Correspondence III: Gauge Origami partition function and
  qq-characters}.
\newblock 2016.

\bibitem{Nekrasov2016a}
N.~Nekrasov and N.~S. Prabhakar.
\newblock {Spiked Instantons from Intersecting D-branes}.
\newblock {\em Nucl. Phys.}, B914:257--300, 2017.

\bibitem{Kim2016}
Hee-Cheol Kim.
\newblock {Line defects and 5d instanton partition functions}.
\newblock {\em JHEP}, 03:199, 2016.

\bibitem{Knight1995}
H.~Knight.
\newblock {Spectra of Tensor Products of Finite Dimensional Representations of
  Yangians}.
\newblock {\em Journal of Algebra}, 174(1):187--196, 1995.

\bibitem{Frenkel1998}
E.~Frenkel and N.~Reshetikhin.
\newblock {The q-characters of representations of quantum affine algebras and
  deformations of W-algebras}.
\newblock 1998.

\bibitem{Kimura2015}
T.~Kimura and V.~Pestun.
\newblock {Quiver W-algebras}.
\newblock 2015.

\bibitem{Kimura2016a}
T.~Kimura and V.~Pestun.
\newblock {Quiver elliptic W-algebras}.
\newblock 2016.

\bibitem{Mironov2016}
A.~Mironov, A.~Morozov, and Y.~Zenkevich.
\newblock {Ding--Iohara--Miki symmetry of network matrix models}.
\newblock {\em Phys. Lett.}, B762:196--208, 2016.

\bibitem{Awata2016a}
H.~Awata, H.~Kanno, T.~Matsumoto, A.~Mironov, Al. Morozov, A.~Morozov,
  Y.~Ohkubo, and Y.~Zenkevich.
\newblock {Explicit examples of DIM constraints for network matrix models}.
\newblock {\em JHEP}, 07:103, 2016.

\bibitem{Awata2011}
H.~Awata, B.~Feigin, and J.~Shiraishi.
\newblock {Quantum Algebraic Approach to Refined Topological Vertex}.
\newblock {\em JHEP}, 03:041, 2012.

\bibitem{Feigin2009a}
B.~{Feigin}, K.~{Hashizume}, A.~{Hoshino}, J.~{Shiraishi}, and S.~{Yanagida}.
\newblock {A commutative algebra on degenerate CP$^{1}$ and Macdonald
  polynomials}.

\bibitem{rieffel1981c}
M.~A. Rieffel.
\newblock {$C^*$-algebras associated with irrational rotations}.
\newblock {\em Pacific Journal of Mathematics}, 93(2):415--429, 1981.

\bibitem{rieffel1988projective}
M.~A. Rieffel.
\newblock {Projective modules over higher-dimensional noncommutative tori}.
\newblock {\em Canad. J. Math}, 40(2):257--338, 1988.

\bibitem{grassi2016topological}
A.~Grassi, Y.~Hatsuda, and M.~Marino.
\newblock {Topological strings from quantum mechanics}.
\newblock In {\em {Annales Henri Poincar{\'e}}}, volume~17, pages 3177--3235.
  Springer, 2016.

\bibitem{nakajima2005instanton}
H.~Nakajima and K.~Yoshioka.
\newblock {Instanton counting on blowup. I. 4-dimensional pure gauge theory}.

\bibitem{Kanno:2013aha}
S.~Kanno, Y.~Matsuo, and Hong Zhang.
\newblock {Extended Conformal Symmetry and Recursion Formulae for Nekrasov
  Partition Function}.
\newblock {\em JHEP}, 1308:028, 2013.

\bibitem{Gaberdiel:2012ku}
M.~R. Gaberdiel and R.~Gopakumar.
\newblock {Triality in Minimal Model Holography}.
\newblock {\em JHEP}, 07:127, 2012.

\bibitem{Altschuler:1990th}
D.~Altschuler, M.~Bauer, and H.~Saleur.
\newblock {Level rank duality in nonunitary coset theories}.
\newblock {\em J. Phys.}, A23:L789--L794, 1990.

\bibitem{kuniba1991ferro}
A.~Kuniba, T.~Nakanishi, and J.~Suzuki.
\newblock {Ferro-and antiferro-magnetizations in RSOS models}.
\newblock {\em Nuclear Physics B}, 356(3):750--774, 1991.

\bibitem{Macdonald1995}
I.~G. Macdonald.
\newblock {\em {Symmetric functions and Hall polynomials}}.
\newblock {Oxford mathematical monographs}. Clarendon Press New York, Oxford,
  1995.
\newblock 1{\`e}re Impression broch{\'e}e 1998. Retirage 2002, 2003, 2004.

\bibitem{Bourgine2017}
J.-E. Bourgine and D.~Fioravanti.
\newblock {Omega-deformed Seiberg-Witten relations}.
\newblock {\em To appear}.

\bibitem{Bourgine2017a}
J.-E. Bourgine and D.~Fioravanti.
\newblock {Quantum integrability in the Nekrasov-Shatashvili limit}.
\newblock {\em To appear}.

\bibitem{Okounkov2003a}
A.~Okounkov, N.~Reshetikhin, and C.~Vafa.
\newblock {Quantum Calabi-Yau and classical crystals}.

\bibitem{Awata2008}
H.~Awata and H.~Kanno.
\newblock {Refined BPS state counting from Nekrasov's formula and Macdonald
  functions}.
\newblock {\em Int. J. Mod. Phys.}, A24:2253--2306, 2009.

\bibitem{Awata2009}
H.~Awata and Y.~Yamada.
\newblock {Five-dimensional AGT Conjecture and the Deformed Virasoro Algebra}.
\newblock {\em JHEP}, 01:125, 2010.

\bibitem{Awata2010}
H.~Awata and Y.~Yamada.
\newblock {Five-dimensional AGT Relation and the Deformed beta-ensemble}.
\newblock {\em Prog. Theor. Phys.}, 124:227--262, 2010.

\bibitem{Taki2014}
M.~Taki.
\newblock {On AGT-W Conjecture and q-Deformed W-Algebra}.
\newblock 2014.

\bibitem{Awata2011a}
H.~Awata, B.~Feigin, A.~Hoshino, M.~Kanai, J.~Shiraishi, and S.~Yanagida.
\newblock {Notes on Ding-Iohara algebra and AGT conjecture}.
\newblock 2011.

\bibitem{feigin2011}
B.~L. Feigin and A.~I. Tsymbaliuk.
\newblock {Equivariant $K$ -theory of Hilbert schemes via shuffle algebra}.
\newblock {\em Kyoto J. Math.}, 51(4):831--854, 2011.

\bibitem{Gaiotto:2009ma}
D.~Gaiotto.
\newblock {Asymptotically free N=2 theories and irregular conformal blocks}.
\newblock 2009.

\bibitem{Marshakov:2009gn}
A.~Marshakov, A.~Mironov, and A.~Morozov.
\newblock {On non-conformal limit of the AGT relations}.
\newblock {\em Phys. Lett.}, B682:125--129, 2009.

\bibitem{Kanno:2011fw}
H.~Kanno and Y.~Tachikawa.
\newblock {Instanton counting with a surface operator and the chain-saw
  quiver}.
\newblock {\em JHEP}, 06:119, 2011.

\bibitem{Kanno:2012xt}
H.~Kanno and M.~Taki.
\newblock {Generalized Whittaker states for instanton counting with fundamental
  hypermultiplets}.
\newblock {\em JHEP}, 05:052, 2012.

\bibitem{carlsson2008exts}
E.~Carlsson and A.~Okounkov.
\newblock {Exts and vertex operators}.
\newblock {\em arXiv preprint arXiv:0801.2565}, 2008.

\bibitem{Shiraishi1995}
J.~Shiraishi, H.~Kubo, H.~Awata, and S.~Odake.
\newblock {A Quantum deformation of the Virasoro algebra and the Macdonald
  symmetric functions}.
\newblock {\em Lett. Math. Phys.}, 38:33--51, 1996.

\bibitem{Kapustin1998}
A.~Kapustin.
\newblock {D(n) quivers from branes}.
\newblock {\em JHEP}, 12:015, 1998.

\bibitem{Hayashi2017}
H.~Hayashi and K.~Ohmori.
\newblock {5d/6d DE instantons from trivalent gluing of web diagrams}.
\newblock 2017.

\bibitem{Gaberdiel2017}
M.~R. Gaberdiel, R.~Gopakumar, Wei Li, and Cheng Peng.
\newblock {Higher Spins and Yangian Symmetries}.
\newblock 2017.

\bibitem{DiFrancesco1997}
P.~{Di Francesco}, P.~Mathieu, and D.~Senechal.
\newblock {\em {Conformal field theory}}.
\newblock 1997.
\newblock New York, USA: Springer (1997) 890 p.

\end{thebibliography}

\end{document}